\documentclass[
  a4paper,
 superscriptaddress,
 amsmath,amssymb, twocolumn,
 aps,
 prl,
 longbibliography,
 floatfix,
 notitlepage,
 showpacs,
 noeprint,
]{revtex4-2}

\usepackage[colorlinks=true,citecolor=blue]{hyperref}
\usepackage[normalem]{ulem}

\usepackage[english]{babel}

\usepackage{xcolor}
\usepackage{graphicx}
\graphicspath{.}

\newcommand{\noteFPM}[1]%
{\begingroup{\color{red} FPM: #1}\endgroup}

\begin{document}
\title{Non-hermitian topology and entanglement in an optomechanical superlattice}

\author{Wojciech Brzezicki}
\affiliation{Institute of Theoretical Physics, Jagiellonian University, ulica S.
  \L{}ojasiewicza 11, PL-30348 Krak\'ow, Poland}
\affiliation{International Research Centre MagTop, Institute of
  Physics, Polish Academy of Sciences, Aleja Lotnik\'ow 32/46, PL-02668 Warsaw,
  Poland}

\author{Timo Hyart}
\affiliation{Computational Physics Laboratory, Physics Unit, Faculty of
  Engineering and Natural Sciences,
  Tampere University, FI-33014 Tampere, Finland}
 \affiliation{Department of Applied Physics, Aalto University, 00076 Aalto, Espoo, Finland}

\author{Francesco Massel}
\affiliation{Department of Science and Industry Systems, University of
  South-Eastern Norway, PO Box 235, Kongsberg, Norway}

\begin{abstract}
  The interplay between topology, dissipation and nonlinearities can give rise
  to a wealth of new phenomena and pave the way for novel topological lasers, sensors and other quantum devices.
  Along these lines, we propose here an optomechanical setup in which the
  concomitant presence of a spatially modulated external drive and dissipation
  gives rise to a topologically nontrivial state for mechanical and optical
  excitations. We are able to show that the one-dimensional system considered
  here exhibits topologically protected end states for which mechanical and
  optical degrees of  freedom are entangled. We show such entanglement to be
  robust with respect to  the presence of nonzero-temperature baths and we propose a protocol for experimental observation of the entanglement.
\end{abstract}

\maketitle


\paragraph{Introduction.}

Since the seminal work by Thouless, Kohomoto, Nightingale and den Nijs
\cite{Thouless.1982}, describing the quantization of the Hall conductance in
terms of topological invariant of their Hamiltonian, topology has
played a central role in the determination of properties of condensed matter
systems. The classification of Hamiltonians belonging to  Altland-Zirnbauer
symmetry classes~\cite{Altland.1997}, characterized by particle-hole, time
reversal and chiral symmetries, has led to the systematic understanding of the
relationship between topological invariants and symmetries in various
dimensional systems -- the tenfold way~\cite{Schnyder.2008, Kitaev09, Ryu_2010}. Over the years, it has been realized that the role of
topology in determining the properties of physical systems extends beyond the
realm of solid-state electron systems, with ramifications, for instance,
into the physics of optical and optomechanical systems~\cite{Haldane.2008,Raghu.2008czi, Xu.201614,
  El-Ganainy.2018, Bandres.2018lml, Mittal.2018, Blanco-Redondo.2018,
  Miri.2019,Ozawa.2019, Pino.2022,Patil.2022, Dai.2024, Guria.2024, Metelmann.2015,Bernier.2017,Verhagen.2017,Malz.2018, Youssefi.2022, Peano.2015,Sanavio.2020,Ren.2022, Budich20, Koch22,wu_manipulating_2024}, where the combination of topological and non-hermitian physics has recently lead to the realization of topological lasers~\cite{Bandres.2018lml} and topologically protected entangled states~\cite{Mittal.2018,Blanco-Redondo.2018}. Upon broadening the scope of the topological considerations to include the non-hermitian effects present in open systems \cite{Bergholtz21}, it was discovered that the generalization of the topological classification
 to non-Hermitian systems  implies an extension of the tenfold way to 38 symmetry classes~\cite{Kawabata.2019}.

\begin{figure}[ht]
  \includegraphics[width=0.95\columnwidth]{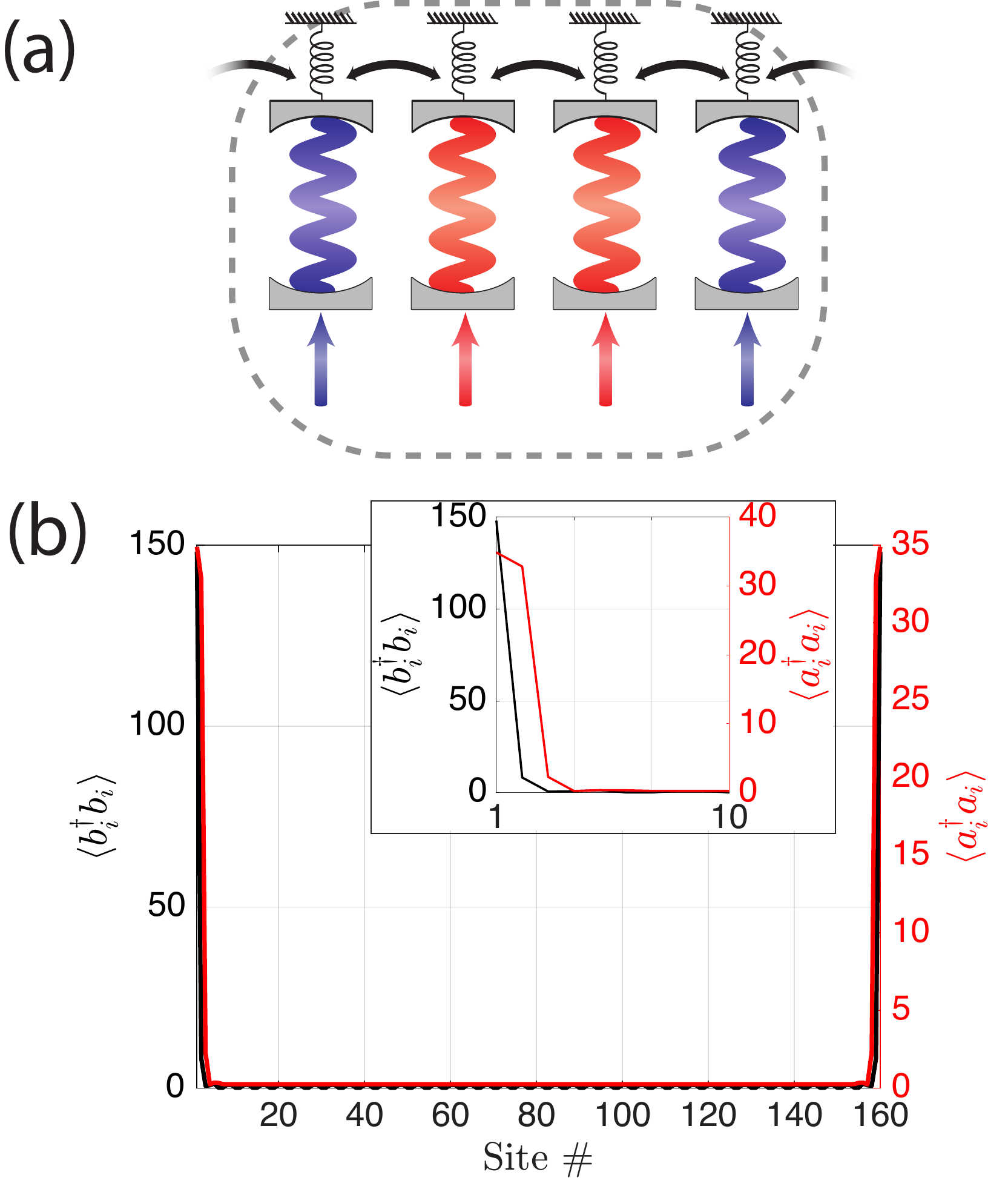}
  \caption{(a) Cartoon depiction of the superlattice unit cell. Each
    optomechanical system is driven either on the blue ($G_+$, first and fourth
    sites) or the red ($G_-$, second and third) mechanical sideband. The
    coupling between sites arises from a hopping term ($J$) between the
    mechanical modes. (b) Mechanical (black) and optical (red) equilibrium
    population of the chain for $G_+=0.242$, $G_-=1$, $J=0.5$, $\gamma=10^{-4}$, and $n_\mathrm{c}=n_\mathrm{m}=0$. For
    this value of $G_+$ the system is approaching the instability for the end
    states. Energies are in units of $\kappa$.} \label{fig:1}
\end{figure}
In the present work, we propose a new approach to topology in optomechanical
system, investigating the role of spatially-modulated drive and dissipation in
determining the topological properties of an optomechanical chain, taking into
account the role played by quantum fluctuations. We show that an
external drive can induce a topologically nontrivial phase, leading, for open
boundary conditions, to topologically protected end  states. Importantly, when the system is in the topologically nontrivial phase we obtain robust stationary-state
entanglement between the mechanical and optical end  modes.

The generation of entanglement between optical and mechanical modes in
optomechanical systems has been experimentally demonstrated in a pulsed
regime~\cite{Palomaki.2013}, but earlier theoretical
proposals~\cite{Genes.2008wrq} of stationary-state entanglement require the
system to be in the so-called ultrastrong coupling regime, which is difficult to achieve experimentally (see~\cite{Peterson.2019}
for a relatively recent demonstration of an optomechanical system in the
ultrastrong coupling regime).
Conversely, in our system, for realistic experimental
parameters, we can achieve stationary-state entanglement that
survives in the presence of significant thermal noise for the mechanical degrees
of freedom, and tolerates moderate amounts of optical noise. Furthermore, since the end states are topologically protected, our results are expected to remain valid also in the presence of significant disorder in all model parameters describing the optical cavity modes and mechanical oscillators of the optomechanical chain. Finally, our analysis suggests that,
beyond the instability threshold for end modes, topological lasing of entangled
optomechanical modes is possible.

\paragraph{The system.} Our setup is constituted by a one-dimensional lattice
of conventional  optomechanical cavities, each of which is driven by a strong coherent tone.
On each site, we choose to detune the drive from the optical cavity resonance by
an amount corresponding to the mechanical resonant frequency $\omega_{\rm m}$.
The detuning can either be negative ($\omega_{\rm R} = \omega_{\rm c} - \omega_{\rm  m}$, red-detuned drive) or positive ($\omega_{\rm B} = \omega_{\rm c} +\omega_{\rm m}$, blue-detuned drive). Echoing the analysis performed in Ref.~\cite{Brzezicki.2019, Brzezicki.2023}, where the properties of a non-hermitian one-dimensional chain were modulated in a dissipative
superlattice pattern according to a ABBA pattern, we choose to alternate the
drive of the optomechanical chain in a blue-red-red-blue (BRRB) driving pattern
(see Fig.~\ref{fig:1}). Our choice reflects the idea that, within the framework
of linear optomechanics~\cite{Aspelmeyer.2014,Bowen.2015}, in the presence of a
strong driving coherent tone, an optical cavity coupled through radiation
pressure to a mechanical oscillator can be analyzed in terms of two
linearly-coupled harmonic oscillators describing the dynamics of the
electromagnetic field and mechanical fluctuations. In the so-called
sideband-resolved regime, a drive at a frequency  $\omega_{\rm B}$ will induce an effective parametric-down-converter ($a^\dagger
b^\dagger + {\rm h.c.}$) coupling between electromagnetic ($a$) and mechanical
($b$) degrees of freedom, while a drive at $\omega_{\rm R}$ will lead to a coherent beam-splitter process ($a^\dagger b +
{\rm h.c.}$) between mechanical and electromagnetic fluctuations. For an
isolated optomechanical system, a beam-splitter coupling can lead to ground
state cooling of the mechanical mode ~\cite{Marquardt.2007, Teufel.2011ju},
while the former implies amplification and, ultimately, instability and lasing
of the mechanical motion~\cite{Ludwig.2008,Massel.2011}. These two effects can
be understood in terms of an effective increase (respectively, decrease) of the
mechanical damping rate.

In addition to the on-site coherent drives, we also
introduce a  hopping term between mechanical excitations, which can be
realized, for instance, in an optomechanical crystal
\cite{Safavi-Naeini.2010} combining, in a single structure, the properties of
photonic~\cite{Painter.1999,Vučković.2002} and phononic~\cite{Vasseur.2007,III.2009}
crystals. Combining these elements, the Hamiltonian describing the coherent part of the
dynamics for the mechanical and optical fluctuations in a chain comprising $4N$ sites is given by
\begin{eqnarray}
  \label{eq:1}
    H &=&  \sum^{N-1}_{\rm j=0} \big[ G_+ (a_{\rm 4 j+1}^\dagger b_{\rm 4 j+1 }^\dagger + a_{\rm 4 j+4}^\dagger b_{\rm 4 j+4}^\dagger ) + G_- ( a_{\rm 4 j+2}^\dagger b_{\rm 4 j+2} \nonumber \\ && \hspace{0.5cm}
      + a_{\rm 4 j+3}^\dagger b_{\rm 4 j+3}) \big]
    - J \sum^{4N-1}_{\rm j=1} b^\dagger_{\rm j} b_{\rm j+1}   +  {\rm h.c.},
\end{eqnarray}
where the coupling constant $G_+$ ($G_-$) models the parametric down conversion (coherent exchange)
process induced by a coherent blue-sideband (red-sideband) drive, and $J$ describes the hopping amplitude between the mechanical
modes $b_{\rm i}$ and $a_{\rm i}$ are the lowering operator associated with the electromagnetic modes (see~\cite{suppl} for more details).

In addition to the unitary dynamics described by Eq.~\eqref{eq:1}, we model the
coupling of each site ${\rm i}$ to the environment introducing the (noise) input
operators $a_{\rm in,i}$ and $b_{\rm in,i}$ with expectation values $\langle a_{\rm in,i}(t)
a^\dagger_{\rm in,i}(t)\rangle = \left(n_{\rm c}+1 \right) \delta (t-t')$ and
$\langle b_{\rm in,i}(t) b^\dagger_{\rm in,i}(t)\rangle = \left(n_{\rm m}+1
\right) \delta (t-t')$, respectively, where the system-environment coupling
rates for optical and mechanical modes are given by $\kappa$ and $\gamma$
respectively. We note that, while the thermal occupation for the optical bath
$n_{\rm c}$ can be set to zero, especially for the infrared and the
visible-light case, it is not possible to neglect the thermal population of the
mechanical bath $n_{\rm m}$.
The symmetry between the mechanical and optical excitations in
Eq.~(\ref{eq:1}) would only allow to replace the hopping of phonons with hopping
of photons~\cite{Eichenfield.2009} at zero temperature.
We find that a finite value of $n_{\rm m}$ has detrimental effects on the
entanglement properties of the photonic implementation, and therefore, along the
lines of Ref.~\cite{Ludwig.2013}, we focus here on the phononic hopping, which,
as discussed below, allows for an effective cooling of the mechanical end modes.
Nevertheless, while it is necessary to assume that the hopping term is
predominant for the ``hot'' (mechanical) degrees of freedom, we have checked
that our results are robust upon inclusion of a small hopping-disorder term for
the ``cold'' (optical) degrees of freedom.

\paragraph{Results.} Through a standard approach~\cite{suppl}, the dynamics of the system's degrees
of freedom can be solved to give
\begin{equation}
  \label{eq:2}
  \mathbf{a} = \exp \left[ -i \mathcal{H}_{\rm NH} t \right] \mathbf{a}_0 + \int_0 ^t
               \exp\left[ -i \mathcal{H}_{\rm NH} (t-\tau) \right]
               \sqrt{\bar{\eta}} \mathbf{a}_{\rm in}d \tau
\end{equation}
where $\mathbf{a}=\left[a_1,b_1,\dots,a_{4N},b_{4N};;a^\dagger_1,b^\dagger_1,\dots
  a^\dagger_{4N},b^\dagger_{4N} \right]^T$ and an analogous definition holds for
$\mathbf{a}_{\rm in}$.

In Eq.~\eqref{eq:2}, $-i \mathcal{H}_{\rm NH} = -i(\bar{\sigma}_z \mathcal{H} -
i \bar{\eta}/2) $ --where $\mathcal{H}$ is the BdG Hamiltonian associated with
the Hamiltonian $H$ given in Eq.~\eqref{eq:1}, $\bar{\sigma}_z ={\rm
  diag}\left(1,\dots,1;-1,\dots,-1\right)$ and $\bar{\eta}={\rm diag}
\left(\kappa,\gamma,\dots;\kappa,\gamma,\dots\kappa,\gamma\right)$-- can be
interpreted as the Jacobian matrix for the quantum dynamical system whose
solution is given by Eq.~\eqref{eq:2}. As a consequence, from the dynamical
systems perspective, the spectrum of $\mathcal{H}_{\rm NH}$ characterizes the
stability of the system. If $E_i={\rm eig}( \mathcal{H}_{\rm NH})$, the system
is stable for ${\rm Im}\, E_i <0$ only.  This is confirmed by the third
quantization  analysis~\cite{prosen_quantization_2010} of the dynamics of our system~\cite{suppl}.

Crucial insight into the properties of the system, can be obtained
exploring the topological properties of the Hamiltonian $\mathcal{H}_{\rm NH}$.
To this end we can define the effective Hermitian Hamiltonian
$
{\mathcal H}_{{\rm eff}}(\eta,k)=\eta S-iS{\mathcal H}_{{\rm NH}}(k),
$
where ${\mathcal H}_{{\rm NH}}(k)$ is the k-space BdG Hamiltonian and
$S=\frac{1}{2}\sigma_{0}\otimes\left[\sigma_{z}\otimes\sigma_{0}\otimes\left(\sigma_{z}+\sigma_{0}\right)+\sigma_{0}\otimes\sigma_{z}\otimes\left(\sigma_{0}-\sigma_{z}\right)\right]$
encodes the symmetry properties of ${\mathcal H}_{{\rm NH}}(k)$ (see
Ref.~\cite{Brzezicki.2023,suppl}).
\begin{figure}[ht]
  \includegraphics[width=0.95\columnwidth]{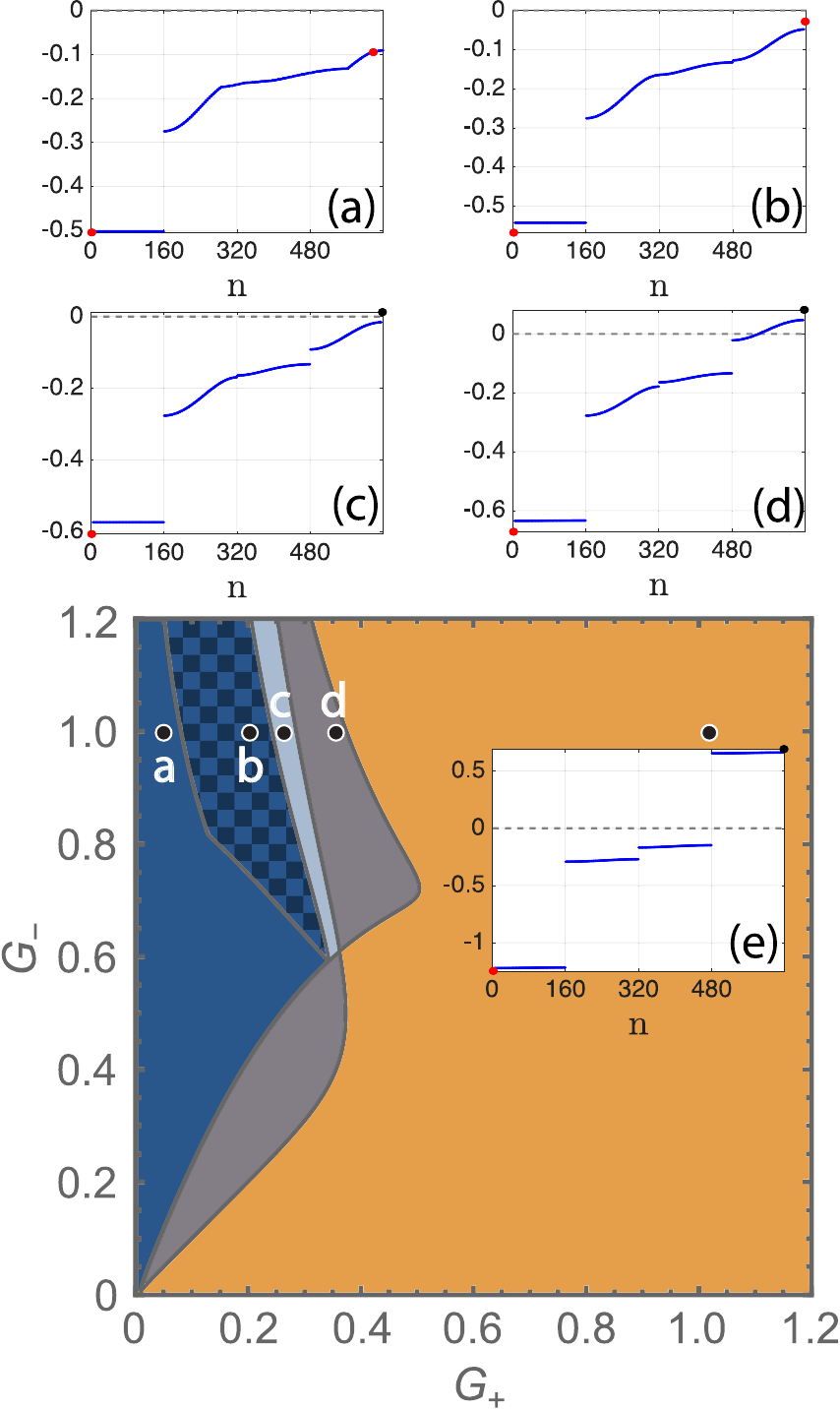}
  \caption{ Phase diagram of $\mathcal{H}_{\rm NH}$ in the
    $\left(G_-,G_+\right)$ plane. (a-e). Imaginary part of the spectrum (${\rm
      Im} E$) for the different values of $(G_+,G_-)$ indicated in the main
    panel. All other parameters as in Fig.~\ref{fig:1}. Blue lines: ${\rm Im}\,
    E_{\rm  b}$, red dots: ${\rm Im}\, E_{\rm  e}$ ($<0$), black dots: ${\rm Im}\, E_{\rm  e}$ ($>0$)}
  \label{fig:2}
\end{figure}
In general, it is possible to prove~\cite{Brzezicki.2019} that a nontrivial
Chern number associated with $\mathcal{H}_{\rm eff}$, determines the nontrivial
topology for the corresponding non-hermitian Hamiltonian $\mathcal{H}_{\rm NH}$
and therefore the appearance of topologically protected end states. For the
optomechanical chain under consideration, we have  $C=4$ indicating that there
are $4$ topologically protected end states with ${\rm Re} \, E=0$ at each end of
the chain~\cite{suppl}.

\begin{figure}[ht]
  \includegraphics[width=0.95\columnwidth]{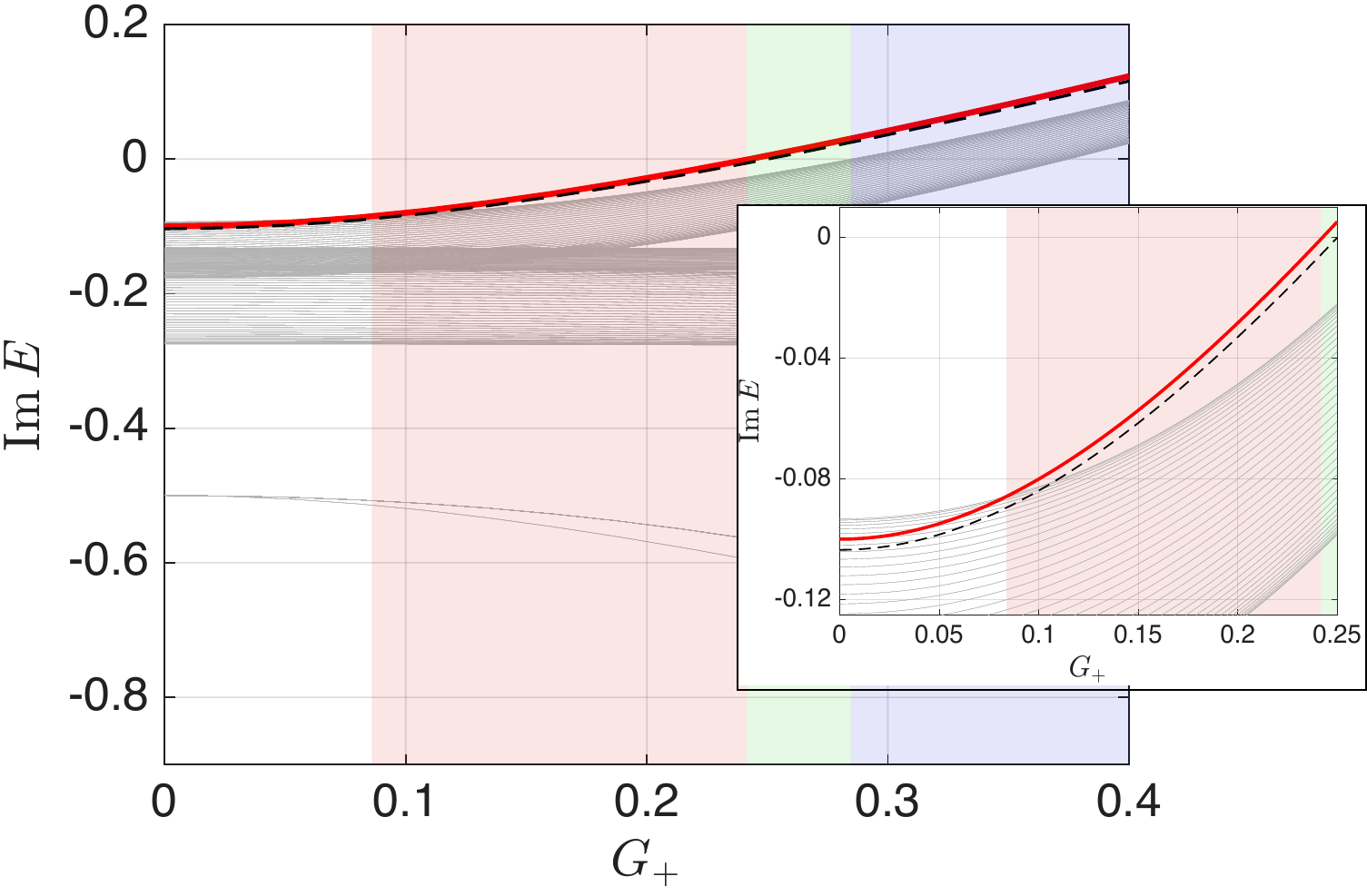}
  \caption{Imaginary part of the spectrum for the OBC chain, same parameters as
    in Fig.~\ref{fig:1}. The red line denotes ${\rm Im}\,E_{\rm e}$, the pink
    stripe the region for which ${\rm Im}\,E_{\rm b}< {\rm Im}\,E_{\rm e} <0$
    (stable end and bulk states), the green one the region ${\rm Im}\,E_{\rm
      b}< 0< {\rm Im}\,E_{\rm e} $ (unstable end, stable bulk), and the blue
    one $0<{\rm Im}\,E_{\rm b}< {\rm Im}\,E_{\rm e} $ (unstable end and bulk).
    The black line denotes the largest value of ${\rm Im}\,E_{\rm e}$ for the
    two-site problem~\cite{suppl}.}
  \label{fig:3}
\end{figure}

In Fig.~\ref{fig:2} we have depicted a phase diagram for  $\mathcal{H}_{\rm NH}$ as a function of $G_-$ and $G_+$, characterizing the properties of
the imaginary parts of the end and bulk state energies ${\rm Im}\, E_{\rm e}$ and ${\rm Im}\, E_{\rm b}$.
On the $G_-=1$ line, we have that, for small $G_+$, ${\rm Im}\, E_{\rm e}$ lies
within the ${\rm Im}\, E_{\rm b}$ band (panel (a)). Increasing $G_+$ leads to
the appearance of a finite gap between ${\rm   Im}\,E_{\rm e}$ and the
$\rm{Im}\, E_{\rm b}$ band (panel (b)). Upon further increase of $G_+$ we
observe a transition to a situation where, at first, ${\rm Im}\, E_{\rm   e}>0$
and ${\rm Im}\, E_{\rm b}<0$  (panel (c)) and eventually, ${\rm Im}\, E_{\rm
  e,b}>0$  (panels (d,e)). While the situation depicted in panel (d) is
characterized by a band crossing the ${\rm Im}\, E=0$ line, in panel (e)
a gap in the bulk spectrum appears.
A cross section of the phase diagram across the points (a-d) in Fig. \ref{fig:2}, depicting the evolution of the spectrum as
a function of $G_+$ is depicted in Fig.~\ref{fig:3}.
The case ${\rm Im}\, E_{\rm e,b}<0$, ${\rm Im}\, E_{\rm
  e} \to 0^-$, leads to an increase in the population of the end states in
stationary conditions, and to its divergence when ${\rm Im}\, E_{\rm e} > 0$
(while ${\rm Im}\, E_{\rm b}<0$).

As anticipated, the nature of the end states is best captured considering the
equilibrium properties of the chain when coupled to a thermal bath. Given that
we are considering Gaussian states and a linear dynamics, the properties of the
system are entirely captured by its covariance matrix
\begin{equation}
  \label{eq:7}
  V_{\bar{a},\bar{a}^\dagger}=
  \begin{bmatrix}
    \langle a_1 a_1^\dagger \rangle & \langle a_1 b_1^\dagger \rangle & \langle a_1 a_1\rangle & \langle a_1 b_1 \rangle & \cdots \\
    \langle b_1 a_1^\dagger \rangle & \ddots &  & \\
    \vdots & & & \\
  \end{bmatrix},
\end{equation}
which can be evaluated in equilibrium conditions, through the solution of the
stationary-state Lyapunov equation for $ V_{\bar{a},\bar{a}^\dagger}$
\begin{equation}
  \label{eq:8}
   \mathcal{H}_{\rm NH}  V_{\bar{a},\bar{a}^\dagger} -  V_{\bar{a},\bar{a}^\dagger} \mathcal{H}^\dagger_{\rm NH} = -i D
\end{equation}
with $D= \left[\kappa \left(n_a+1\right),\gamma \left(n_b+1 \right),\kappa
  n_a,\gamma n_b,\dots \right]$.
As discussed in~\cite{suppl}, the expression for $D$ can be derived
both from a dynamical systems and a third quantization perspective. As expected,
keeping all other parameters fixed, for $G_+$ approaching the instability transition ${\rm Im}\, E_{\rm e} \to 0^-$ (but ${\rm
  Im}\, E_{\rm b} < 0$), the end populations of mechanical and optical modes
diverge.
This is reminiscent of what happens for a single optomechanical system driven on
the blue sideband, and therefore suggests the possibility of optomechanical
lasing for the topologically protected end states. Although the full  description of the lasing
(\cite{Bahari.2017,Harari.2018}) falls beyond the linearized model
employed here, the dynamics of the end states is well
captured by a localized description of the end modes in terms of a two-sites
linearized effective model, so that the physics of our system for ${\rm
  Im}\, E_{\rm e}>0$ and ${\rm Im}\, E_{\rm b}<0$ will be equivalent to the one
controlling optomechanical cavities in the optomechanical lasing
regime~\cite{Metzger.2008,Marquardt.2009,Safavi-Naeini.2011}.

In addition to the local population, through a standard
analysis~\cite{Adesso.2007}, we can explore the entanglement properties of the
chain, evaluating the negativity $E_\mathcal{N}$ between two modes~\cite{suppl}.
Of particular interest is that, at zero temperature, irrespective of the strength
of the drives, optical and mechanical modes pertaining to the first site are
entangled (see \cite{suppl} for a full caracterization of the entanglement properties within a superlattice cell).

\begin{figure}[ht]
  \includegraphics[width=0.95\columnwidth]{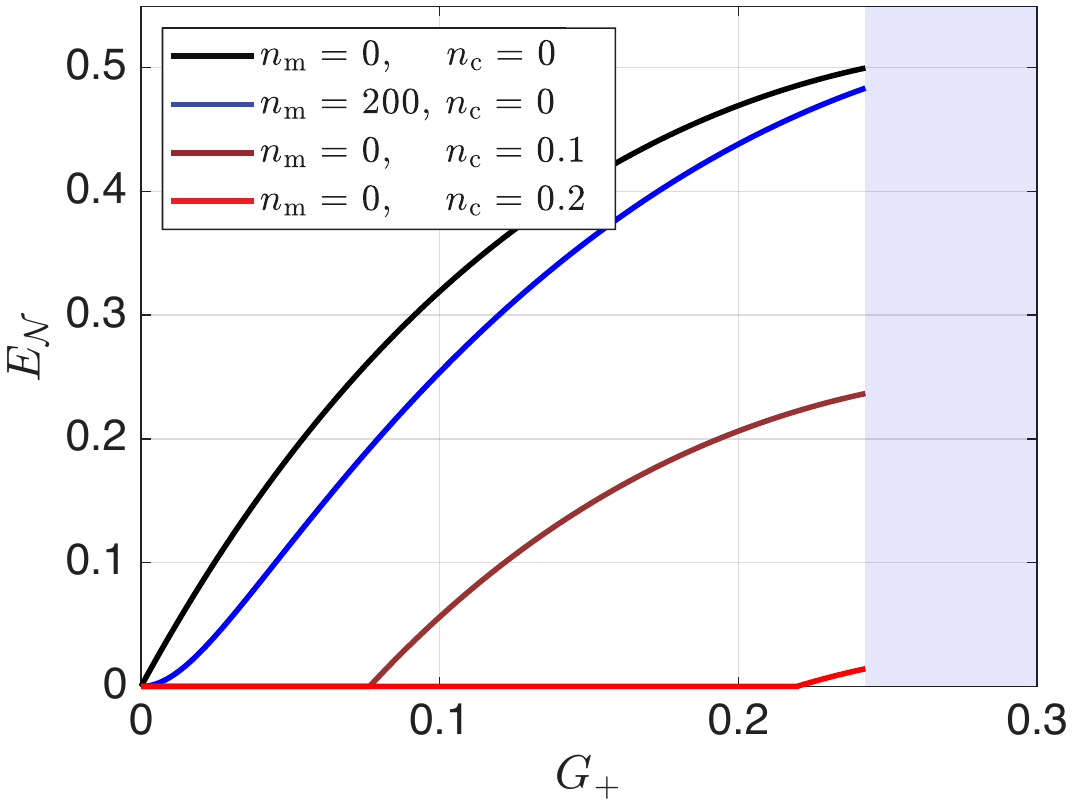}
  \caption{
    Negativity $E_\mathcal{N}$ measuring the entanglement between
    optical and mechanical modes within the first site ($a_1$ and $b_1$) as a function of $G_+$ for different values of the
     mechanical bath population. The entanglement is not strongly affected by a  thermal population of  the mechanical bath.
  }
  \label{fig:5}
\end{figure}

For the single cavity in the presence of a blue-sideband drive, while, in principle, it is possible to have stationary
entanglement~\cite{Genes.2008wrq}, its value is severely limited by stability
constraints, which, in the rotating-wave approximation can be expressed by the
condition $G_+ < \sqrt{\gamma \kappa}/2$. As a consequence, optomechanical
entanglement for a single optomechanical cavity has been predicted either far
from the sideband-resolved regime or for a pulsed drive~\cite{Genes.2008wrq,
  Abdi.2011, Hofer.2011} only. Owing to the demanding parameter regime required
in the former case, experimental verification has been provided in the latter case
only~\cite{Palomaki.2013}.

For the setup described here, we can show that significant entanglement between
optical and mechanical degrees of freedom survives within the topologically
protected end state, even beyond the single-site instability threshold (see
Fig.~\ref{fig:5}). The stability and entanglement properties of the end state
can be understood in terms of an effective bath induced by the chain: mechanical
excitations can hop from the end sites of the chain (driven on the blue
sideband) towards the interior of the chain. The red-detuned drive $G_-$ on the
sites next to the end leads to the cooling of the mechanical excitations
to the (near-zero) effective temperature of the optical bath. This process is
well captured by a two-site model, describing the $G_+$-dependence of the
end-states dissipation properties~\cite{suppl}. More specifically, in the limit
of vanishing blue-sideband drive $G_+ \to 0 $, we have
\begin{equation}
  \label{eq:9}
  {\rm Im} E_{\rm 2-site} = -\frac{\gamma}{2} - \left(\kappa-\gamma\right) \frac{J^2}{G_-^2}\left(1-\frac{J^2}{G_-^2}\right) + O(J/G_-)^5
\end{equation}
while for $G_+ \to \infty$ we have
\begin{equation}
  \label{eq:10}
   {\rm Im}\, E_{\rm 2-site} \simeq -\frac{\kappa+\gamma}{2}
  +\sqrt{\frac{(\kappa-\gamma)^2}{4}+ 4 G_+^2}.
\end{equation}
Importantly, for typical experimental parameters, entanglement is robust with
respect to a finite (even large) thermal population of the mechanical bath
(see Fig.~\ref{fig:5}). This is  in agreement with what is predicted for
single optomechanical systems driven on the red sideband away from the
sideband-resolved regime~\cite{suppl}.

The topological character of the end states  is apparent when
considering experimental imperfections that might lead to random offsets in the
on-site frequencies for mechanical and optical modes and to the appearance of a
random hopping for the optical degrees of freedom. We have verified that adding
disorder for the hopping (for mechanical and optical modes) and on-site
potential, does not alter the picture outlined here~\cite{suppl}.
This aspect is particularly relevant when discussing the experimental
observability of optomechanical entanglement. Ananlogously to what was suggested
in the case of a single optomechanical system~\cite{vitali_optomechanical_2007}, we
envisage the measurement of the mechanical spectrum through the coupling to a
probe cavity driven on the red sideband, whose spectrum reproduces the
mechanical one, which, however, owing to backaction, alters the values of $\gamma$ and $n_\mathrm{m}$~\cite{suppl}.

\paragraph{Conclusions.} We have discussed here the properties of a chain of
optomechanical systems in the presence of site-dependent optical drives. We have
demonstrated that the chain possesses nontrivial topological properties leading
to the existence of topologically protected states both for mechanical and
optical modes. Furthermore, we have demonstrated that mechanical and optical
end states are entangled and that such entanglement is robust with respect to
external perturbations, paving the way for entanglement in topological lasers.
Our findings provide evidence of a strong quantum property (entanglement) within
a non-hermitian topological system.

\paragraph{Acknowledgements.}
WB acknowledge support by Narodowe Centrum Nauki (NCN, National Science Centre,
Poland) Project No. 2019/34/E/ST3/00404 and partial support by the Foundation for Polish Science through the IRA Programme co-financed by EU within SG OP.
FM thanks Kjetil Børkje and Mika Sillanpää for fruitful discussion. FM
acknowledges financial support from the Research Council of Norway (Grant No. 333937) through participation in the QuantERA ERA-NET Cofund in Quantum
Technologies (project MQSens) implemented within the European Union’s Horizon 2020 Programme.

\bibliography{OMtopo}

\begin{thebibliography}{63}%
\makeatletter
\providecommand \@ifxundefined [1]{%
 \@ifx{#1\undefined}
}%
\providecommand \@ifnum [1]{%
 \ifnum #1\expandafter \@firstoftwo
 \else \expandafter \@secondoftwo
 \fi
}%
\providecommand \@ifx [1]{%
 \ifx #1\expandafter \@firstoftwo
 \else \expandafter \@secondoftwo
 \fi
}%
\providecommand \natexlab [1]{#1}%
\providecommand \enquote  [1]{``#1''}%
\providecommand \bibnamefont  [1]{#1}%
\providecommand \bibfnamefont [1]{#1}%
\providecommand \citenamefont [1]{#1}%
\providecommand \href@noop [0]{\@secondoftwo}%
\providecommand \href [0]{\begingroup \@sanitize@url \@href}%
\providecommand \@href[1]{\@@startlink{#1}\@@href}%
\providecommand \@@href[1]{\endgroup#1\@@endlink}%
\providecommand \@sanitize@url [0]{\catcode `\\12\catcode `\$12\catcode
  `\&12\catcode `\#12\catcode `\^12\catcode `\_12\catcode `\%12\relax}%
\providecommand \@@startlink[1]{}%
\providecommand \@@endlink[0]{}%
\providecommand \url  [0]{\begingroup\@sanitize@url \@url }%
\providecommand \@url [1]{\endgroup\@href {#1}{\urlprefix }}%
\providecommand \urlprefix  [0]{URL }%
\providecommand \Eprint [0]{\href }%
\providecommand \doibase [0]{https://doi.org/}%
\providecommand \selectlanguage [0]{\@gobble}%
\providecommand \bibinfo  [0]{\@secondoftwo}%
\providecommand \bibfield  [0]{\@secondoftwo}%
\providecommand \translation [1]{[#1]}%
\providecommand \BibitemOpen [0]{}%
\providecommand \bibitemStop [0]{}%
\providecommand \bibitemNoStop [0]{.\EOS\space}%
\providecommand \EOS [0]{\spacefactor3000\relax}%
\providecommand \BibitemShut  [1]{\csname bibitem#1\endcsname}%
\let\auto@bib@innerbib\@empty
\bibitem [{\citenamefont {Thouless}\ \emph {et~al.}(1982)\citenamefont
  {Thouless}, \citenamefont {Kohmoto}, \citenamefont {Nightingale},\ and\
  \citenamefont {Nijs}}]{Thouless.1982}%
  \BibitemOpen
  \bibfield  {author} {\bibinfo {author} {\bibfnamefont {D.}~\bibnamefont
  {Thouless}}, \bibinfo {author} {\bibfnamefont {M.}~\bibnamefont {Kohmoto}},
  \bibinfo {author} {\bibfnamefont {M.}~\bibnamefont {Nightingale}},\ and\
  \bibinfo {author} {\bibfnamefont {M.~d.}\ \bibnamefont {Nijs}},\ }\bibfield
  {title} {{\selectlanguage {English}\bibinfo {title} {{Quantized Hall
  Conductance in a Two-Dimensional Periodic Potential}}},\ }\href
  {https://doi.org/10.1103/physrevlett.49.405} {\bibfield  {journal} {\bibinfo
  {journal} {Physical Review Letters}\ }\textbf {\bibinfo {volume} {49}},\
  \bibinfo {pages} {405 } (\bibinfo {year} {1982})}\BibitemShut {NoStop}%
\bibitem [{\citenamefont {Altland}\ and\ \citenamefont
  {Zirnbauer}(1997)}]{Altland.1997}%
  \BibitemOpen
  \bibfield  {author} {\bibinfo {author} {\bibfnamefont {A.}~\bibnamefont
  {Altland}}\ and\ \bibinfo {author} {\bibfnamefont {M.~R.}\ \bibnamefont
  {Zirnbauer}},\ }\bibfield  {title} {\bibinfo {title} {{Nonstandard symmetry
  classes in mesoscopic normal-superconducting hybrid structures}},\ }\href
  {https://doi.org/10.1103/physrevb.55.1142} {\bibfield  {journal} {\bibinfo
  {journal} {Physical Review B}\ }\textbf {\bibinfo {volume} {55}},\ \bibinfo
  {pages} {1142} (\bibinfo {year} {1997})}\BibitemShut {NoStop}%
\bibitem [{\citenamefont {Schnyder}\ \emph {et~al.}(2008)\citenamefont
  {Schnyder}, \citenamefont {Ryu}, \citenamefont {Furusaki},\ and\
  \citenamefont {Ludwig}}]{Schnyder.2008}%
  \BibitemOpen
  \bibfield  {author} {\bibinfo {author} {\bibfnamefont {A.~P.}\ \bibnamefont
  {Schnyder}}, \bibinfo {author} {\bibfnamefont {S.}~\bibnamefont {Ryu}},
  \bibinfo {author} {\bibfnamefont {A.}~\bibnamefont {Furusaki}},\ and\
  \bibinfo {author} {\bibfnamefont {A.~W.~W.}\ \bibnamefont {Ludwig}},\
  }\bibfield  {title} {\bibinfo {title} {Classification of topological
  insulators and superconductors in three spatial dimensions},\ }\href
  {https://doi.org/10.1103/PhysRevB.78.195125} {\bibfield  {journal} {\bibinfo
  {journal} {Phys. Rev. B}\ }\textbf {\bibinfo {volume} {78}},\ \bibinfo
  {pages} {195125} (\bibinfo {year} {2008})}\BibitemShut {NoStop}%
\bibitem [{\citenamefont {Kitaev}(2009)}]{Kitaev09}%
  \BibitemOpen
  \bibfield  {author} {\bibinfo {author} {\bibfnamefont {A.}~\bibnamefont
  {Kitaev}},\ }\bibfield  {title} {\bibinfo {title} {{Periodic table for
  topological insulators and superconductors}},\ }\href
  {https://doi.org/10.1063/1.3149495} {\bibfield  {journal} {\bibinfo
  {journal} {AIP Conference Proceedings}\ }\textbf {\bibinfo {volume} {1134}},\
  \bibinfo {pages} {22} (\bibinfo {year} {2009})}\BibitemShut {NoStop}%
\bibitem [{\citenamefont {Ryu}\ \emph {et~al.}(2010)\citenamefont {Ryu},
  \citenamefont {Schnyder}, \citenamefont {Furusaki},\ and\ \citenamefont
  {Ludwig}}]{Ryu_2010}%
  \BibitemOpen
  \bibfield  {author} {\bibinfo {author} {\bibfnamefont {S.}~\bibnamefont
  {Ryu}}, \bibinfo {author} {\bibfnamefont {A.~P.}\ \bibnamefont {Schnyder}},
  \bibinfo {author} {\bibfnamefont {A.}~\bibnamefont {Furusaki}},\ and\
  \bibinfo {author} {\bibfnamefont {A.~W.~W.}\ \bibnamefont {Ludwig}},\
  }\bibfield  {title} {\bibinfo {title} {Topological insulators and
  superconductors: tenfold way and dimensional hierarchy},\ }\href
  {https://doi.org/10.1088/1367-2630/12/6/065010} {\bibfield  {journal}
  {\bibinfo  {journal} {New Journal of Physics}\ }\textbf {\bibinfo {volume}
  {12}},\ \bibinfo {pages} {065010} (\bibinfo {year} {2010})}\BibitemShut
  {NoStop}%
\bibitem [{\citenamefont {Haldane}\ and\ \citenamefont
  {Raghu}(2008)}]{Haldane.2008}%
  \BibitemOpen
  \bibfield  {author} {\bibinfo {author} {\bibfnamefont {F.~D.~M.}\
  \bibnamefont {Haldane}}\ and\ \bibinfo {author} {\bibfnamefont
  {S.}~\bibnamefont {Raghu}},\ }\bibfield  {title} {\bibinfo {title} {{Possible
  Realization of Directional Optical Waveguides in Photonic Crystals with
  Broken Time-Reversal Symmetry}},\ }\href
  {https://doi.org/10.1103/physrevlett.100.013904} {\bibfield  {journal}
  {\bibinfo  {journal} {Physical Review Letters}\ }\textbf {\bibinfo {volume}
  {100}},\ \bibinfo {pages} {013904} (\bibinfo {year} {2008})}\BibitemShut
  {NoStop}%
\bibitem [{\citenamefont {Raghu}\ and\ \citenamefont
  {Haldane}(2008)}]{Raghu.2008czi}%
  \BibitemOpen
  \bibfield  {author} {\bibinfo {author} {\bibfnamefont {S.}~\bibnamefont
  {Raghu}}\ and\ \bibinfo {author} {\bibfnamefont {F.~D.~M.}\ \bibnamefont
  {Haldane}},\ }\bibfield  {title} {\bibinfo {title} {{Analogs of
  quantum-Hall-effect edge states in photonic crystals}},\ }\href
  {https://doi.org/10.1103/physreva.78.033834} {\bibfield  {journal} {\bibinfo
  {journal} {Physical Review A}\ }\textbf {\bibinfo {volume} {78}},\ \bibinfo
  {pages} {033834} (\bibinfo {year} {2008})}\BibitemShut {NoStop}%
\bibitem [{\citenamefont {Xu}\ \emph {et~al.}(2016)\citenamefont {Xu},
  \citenamefont {Mason}, \citenamefont {Jiang},\ and\ \citenamefont
  {Harris}}]{Xu.201614}%
  \BibitemOpen
  \bibfield  {author} {\bibinfo {author} {\bibfnamefont {H.}~\bibnamefont
  {Xu}}, \bibinfo {author} {\bibfnamefont {D.}~\bibnamefont {Mason}}, \bibinfo
  {author} {\bibfnamefont {L.}~\bibnamefont {Jiang}},\ and\ \bibinfo {author}
  {\bibfnamefont {J.~G.~E.}\ \bibnamefont {Harris}},\ }\bibfield  {title}
  {{\selectlanguage {English}\bibinfo {title} {{Topological energy transfer in
  an optomechanical system with exceptional points.}}},\ }\href
  {https://doi.org/10.1038/nature18604} {\bibfield  {journal} {\bibinfo
  {journal} {Nature}\ }\textbf {\bibinfo {volume} {537}},\ \bibinfo {pages} {80
  } (\bibinfo {year} {2016})}\BibitemShut {NoStop}%
\bibitem [{\citenamefont {El-Ganainy}\ \emph {et~al.}(2018)\citenamefont
  {El-Ganainy}, \citenamefont {Makris}, \citenamefont {Khajavikhan},
  \citenamefont {Musslimani}, \citenamefont {Rotter},\ and\ \citenamefont
  {Christodoulides}}]{El-Ganainy.2018}%
  \BibitemOpen
  \bibfield  {author} {\bibinfo {author} {\bibfnamefont {R.}~\bibnamefont
  {El-Ganainy}}, \bibinfo {author} {\bibfnamefont {K.~G.}\ \bibnamefont
  {Makris}}, \bibinfo {author} {\bibfnamefont {M.}~\bibnamefont {Khajavikhan}},
  \bibinfo {author} {\bibfnamefont {Z.~H.}\ \bibnamefont {Musslimani}},
  \bibinfo {author} {\bibfnamefont {S.}~\bibnamefont {Rotter}},\ and\ \bibinfo
  {author} {\bibfnamefont {D.~N.}\ \bibnamefont {Christodoulides}},\ }\bibfield
   {title} {\bibinfo {title} {{Non-Hermitian physics and PT symmetry}},\ }\href
  {https://doi.org/10.1038/nphys4323} {\bibfield  {journal} {\bibinfo
  {journal} {Nature Physics}\ }\textbf {\bibinfo {volume} {14}},\ \bibinfo
  {pages} {11} (\bibinfo {year} {2018})}\BibitemShut {NoStop}%
\bibitem [{\citenamefont {Bandres}\ \emph {et~al.}(2018)\citenamefont
  {Bandres}, \citenamefont {Wittek}, \citenamefont {Harari}, \citenamefont
  {Parto}, \citenamefont {Ren}, \citenamefont {Segev}, \citenamefont
  {Christodoulides},\ and\ \citenamefont {Khajavikhan}}]{Bandres.2018lml}%
  \BibitemOpen
  \bibfield  {author} {\bibinfo {author} {\bibfnamefont {M.~A.}\ \bibnamefont
  {Bandres}}, \bibinfo {author} {\bibfnamefont {S.}~\bibnamefont {Wittek}},
  \bibinfo {author} {\bibfnamefont {G.}~\bibnamefont {Harari}}, \bibinfo
  {author} {\bibfnamefont {M.}~\bibnamefont {Parto}}, \bibinfo {author}
  {\bibfnamefont {J.}~\bibnamefont {Ren}}, \bibinfo {author} {\bibfnamefont
  {M.}~\bibnamefont {Segev}}, \bibinfo {author} {\bibfnamefont {D.~N.}\
  \bibnamefont {Christodoulides}},\ and\ \bibinfo {author} {\bibfnamefont
  {M.}~\bibnamefont {Khajavikhan}},\ }\bibfield  {title} {\bibinfo {title}
  {{Topological insulator laser: Experiments}},\ }\bibfield  {journal}
  {\bibinfo  {journal} {Science}\ }\textbf {\bibinfo {volume} {359}},\ \href
  {https://doi.org/10.1126/science.aar4005} {10.1126/science.aar4005} (\bibinfo
  {year} {2018})\BibitemShut {NoStop}%
\bibitem [{\citenamefont {Mittal}\ \emph {et~al.}(2018)\citenamefont {Mittal},
  \citenamefont {Goldschmidt},\ and\ \citenamefont {Hafezi}}]{Mittal.2018}%
  \BibitemOpen
  \bibfield  {author} {\bibinfo {author} {\bibfnamefont {S.}~\bibnamefont
  {Mittal}}, \bibinfo {author} {\bibfnamefont {E.~A.}\ \bibnamefont
  {Goldschmidt}},\ and\ \bibinfo {author} {\bibfnamefont {M.}~\bibnamefont
  {Hafezi}},\ }\bibfield  {title} {\bibinfo {title} {{A topological source of
  quantum light}},\ }\href {https://doi.org/10.1038/s41586-018-0478-3}
  {\bibfield  {journal} {\bibinfo  {journal} {Nature}\ }\textbf {\bibinfo
  {volume} {561}},\ \bibinfo {pages} {502} (\bibinfo {year}
  {2018})}\BibitemShut {NoStop}%
\bibitem [{\citenamefont {Blanco-Redondo}\ \emph {et~al.}(2018)\citenamefont
  {Blanco-Redondo}, \citenamefont {Bell}, \citenamefont {Oren}, \citenamefont
  {Eggleton},\ and\ \citenamefont {Segev}}]{Blanco-Redondo.2018}%
  \BibitemOpen
  \bibfield  {author} {\bibinfo {author} {\bibfnamefont {A.}~\bibnamefont
  {Blanco-Redondo}}, \bibinfo {author} {\bibfnamefont {B.}~\bibnamefont
  {Bell}}, \bibinfo {author} {\bibfnamefont {D.}~\bibnamefont {Oren}}, \bibinfo
  {author} {\bibfnamefont {B.~J.}\ \bibnamefont {Eggleton}},\ and\ \bibinfo
  {author} {\bibfnamefont {M.}~\bibnamefont {Segev}},\ }\bibfield  {title}
  {\bibinfo {title} {{Topological protection of biphoton states}},\ }\href
  {https://doi.org/10.1126/science.aau4296} {\bibfield  {journal} {\bibinfo
  {journal} {Science}\ }\textbf {\bibinfo {volume} {362}},\ \bibinfo {pages}
  {568} (\bibinfo {year} {2018})}\BibitemShut {NoStop}%
\bibitem [{\citenamefont {Miri}\ and\ \citenamefont {Alù}(2019)}]{Miri.2019}%
  \BibitemOpen
  \bibfield  {author} {\bibinfo {author} {\bibfnamefont {M.-A.}\ \bibnamefont
  {Miri}}\ and\ \bibinfo {author} {\bibfnamefont {A.}~\bibnamefont {Alù}},\
  }\bibfield  {title} {\bibinfo {title} {{Exceptional points in optics and
  photonics}},\ }\bibfield  {journal} {\bibinfo  {journal} {Science}\ }\textbf
  {\bibinfo {volume} {363}},\ \href {https://doi.org/10.1126/science.aar7709}
  {10.1126/science.aar7709} (\bibinfo {year} {2019})\BibitemShut {NoStop}%
\bibitem [{\citenamefont {Ozawa}\ \emph {et~al.}(2019)\citenamefont {Ozawa},
  \citenamefont {Price}, \citenamefont {Amo}, \citenamefont {Goldman},
  \citenamefont {Hafezi}, \citenamefont {Lu}, \citenamefont {Rechtsman},
  \citenamefont {Schuster}, \citenamefont {Simon}, \citenamefont {Zilberberg},\
  and\ \citenamefont {Carusotto}}]{Ozawa.2019}%
  \BibitemOpen
  \bibfield  {author} {\bibinfo {author} {\bibfnamefont {T.}~\bibnamefont
  {Ozawa}}, \bibinfo {author} {\bibfnamefont {H.~M.}\ \bibnamefont {Price}},
  \bibinfo {author} {\bibfnamefont {A.}~\bibnamefont {Amo}}, \bibinfo {author}
  {\bibfnamefont {N.}~\bibnamefont {Goldman}}, \bibinfo {author} {\bibfnamefont
  {M.}~\bibnamefont {Hafezi}}, \bibinfo {author} {\bibfnamefont
  {L.}~\bibnamefont {Lu}}, \bibinfo {author} {\bibfnamefont {M.~C.}\
  \bibnamefont {Rechtsman}}, \bibinfo {author} {\bibfnamefont {D.}~\bibnamefont
  {Schuster}}, \bibinfo {author} {\bibfnamefont {J.}~\bibnamefont {Simon}},
  \bibinfo {author} {\bibfnamefont {O.}~\bibnamefont {Zilberberg}},\ and\
  \bibinfo {author} {\bibfnamefont {I.}~\bibnamefont {Carusotto}},\ }\bibfield
  {title} {\bibinfo {title} {{Topological photonics}},\ }\href
  {https://doi.org/10.1103/revmodphys.91.015006} {\bibfield  {journal}
  {\bibinfo  {journal} {Reviews of Modern Physics}\ }\textbf {\bibinfo {volume}
  {91}},\ \bibinfo {pages} {015006} (\bibinfo {year} {2019})}\BibitemShut
  {NoStop}%
\bibitem [{\citenamefont {Pino}\ \emph {et~al.}(2022)\citenamefont {Pino},
  \citenamefont {Slim},\ and\ \citenamefont {Verhagen}}]{Pino.2022}%
  \BibitemOpen
  \bibfield  {author} {\bibinfo {author} {\bibfnamefont {J.~d.}\ \bibnamefont
  {Pino}}, \bibinfo {author} {\bibfnamefont {J.~J.}\ \bibnamefont {Slim}},\
  and\ \bibinfo {author} {\bibfnamefont {E.}~\bibnamefont {Verhagen}},\
  }\bibfield  {title} {\bibinfo {title} {{Non-Hermitian chiral phononics
  through optomechanically induced squeezing}},\ }\href
  {https://doi.org/10.1038/s41586-022-04609-0} {\bibfield  {journal} {\bibinfo
  {journal} {Nature}\ }\textbf {\bibinfo {volume} {606}},\ \bibinfo {pages}
  {82} (\bibinfo {year} {2022})}\BibitemShut {NoStop}%
\bibitem [{\citenamefont {Patil}\ \emph {et~al.}(2022)\citenamefont {Patil},
  \citenamefont {Höller}, \citenamefont {Henry}, \citenamefont {Guria},
  \citenamefont {Zhang}, \citenamefont {Jiang}, \citenamefont {Kralj},
  \citenamefont {Read},\ and\ \citenamefont {Harris}}]{Patil.2022}%
  \BibitemOpen
  \bibfield  {author} {\bibinfo {author} {\bibfnamefont {Y.~S.~S.}\
  \bibnamefont {Patil}}, \bibinfo {author} {\bibfnamefont {J.}~\bibnamefont
  {Höller}}, \bibinfo {author} {\bibfnamefont {P.~A.}\ \bibnamefont {Henry}},
  \bibinfo {author} {\bibfnamefont {C.}~\bibnamefont {Guria}}, \bibinfo
  {author} {\bibfnamefont {Y.}~\bibnamefont {Zhang}}, \bibinfo {author}
  {\bibfnamefont {L.}~\bibnamefont {Jiang}}, \bibinfo {author} {\bibfnamefont
  {N.}~\bibnamefont {Kralj}}, \bibinfo {author} {\bibfnamefont
  {N.}~\bibnamefont {Read}},\ and\ \bibinfo {author} {\bibfnamefont {J.~G.~E.}\
  \bibnamefont {Harris}},\ }\bibfield  {title} {\bibinfo {title} {{Measuring
  the knot of non-Hermitian degeneracies and non-commuting braids}},\ }\href
  {https://doi.org/10.1038/s41586-022-04796-w} {\bibfield  {journal} {\bibinfo
  {journal} {Nature}\ }\textbf {\bibinfo {volume} {607}},\ \bibinfo {pages}
  {271} (\bibinfo {year} {2022})}\BibitemShut {NoStop}%
\bibitem [{\citenamefont {Dai}\ \emph {et~al.}(2024)\citenamefont {Dai},
  \citenamefont {Ao}, \citenamefont {Mao}, \citenamefont {Yang}, \citenamefont
  {Zheng}, \citenamefont {Zhai}, \citenamefont {Li}, \citenamefont {Yuan},
  \citenamefont {Tang}, \citenamefont {Li}, \citenamefont {Luo}, \citenamefont
  {Wang}, \citenamefont {Hu}, \citenamefont {Gong},\ and\ \citenamefont
  {Wang}}]{Dai.2024}%
  \BibitemOpen
  \bibfield  {author} {\bibinfo {author} {\bibfnamefont {T.}~\bibnamefont
  {Dai}}, \bibinfo {author} {\bibfnamefont {Y.}~\bibnamefont {Ao}}, \bibinfo
  {author} {\bibfnamefont {J.}~\bibnamefont {Mao}}, \bibinfo {author}
  {\bibfnamefont {Y.}~\bibnamefont {Yang}}, \bibinfo {author} {\bibfnamefont
  {Y.}~\bibnamefont {Zheng}}, \bibinfo {author} {\bibfnamefont
  {C.}~\bibnamefont {Zhai}}, \bibinfo {author} {\bibfnamefont {Y.}~\bibnamefont
  {Li}}, \bibinfo {author} {\bibfnamefont {J.}~\bibnamefont {Yuan}}, \bibinfo
  {author} {\bibfnamefont {B.}~\bibnamefont {Tang}}, \bibinfo {author}
  {\bibfnamefont {Z.}~\bibnamefont {Li}}, \bibinfo {author} {\bibfnamefont
  {J.}~\bibnamefont {Luo}}, \bibinfo {author} {\bibfnamefont {W.}~\bibnamefont
  {Wang}}, \bibinfo {author} {\bibfnamefont {X.}~\bibnamefont {Hu}}, \bibinfo
  {author} {\bibfnamefont {Q.}~\bibnamefont {Gong}},\ and\ \bibinfo {author}
  {\bibfnamefont {J.}~\bibnamefont {Wang}},\ }\bibfield  {title} {\bibinfo
  {title} {{Non-Hermitian topological phase transitions controlled by
  nonlinearity}},\ }\href {https://doi.org/10.1038/s41567-023-02244-8}
  {\bibfield  {journal} {\bibinfo  {journal} {Nature Physics}\ }\textbf
  {\bibinfo {volume} {20}},\ \bibinfo {pages} {101} (\bibinfo {year}
  {2024})}\BibitemShut {NoStop}%
\bibitem [{\citenamefont {Guria}\ \emph {et~al.}(2024)\citenamefont {Guria},
  \citenamefont {Zhong}, \citenamefont {Ozdemir}, \citenamefont {Patil},
  \citenamefont {El-Ganainy},\ and\ \citenamefont {Harris}}]{Guria.2024}%
  \BibitemOpen
  \bibfield  {author} {\bibinfo {author} {\bibfnamefont {C.}~\bibnamefont
  {Guria}}, \bibinfo {author} {\bibfnamefont {Q.}~\bibnamefont {Zhong}},
  \bibinfo {author} {\bibfnamefont {S.~K.}\ \bibnamefont {Ozdemir}}, \bibinfo
  {author} {\bibfnamefont {Y.~S.~S.}\ \bibnamefont {Patil}}, \bibinfo {author}
  {\bibfnamefont {R.}~\bibnamefont {El-Ganainy}},\ and\ \bibinfo {author}
  {\bibfnamefont {J.~G.~E.}\ \bibnamefont {Harris}},\ }\bibfield  {title}
  {\bibinfo {title} {{Resolving the topology of encircling multiple exceptional
  points}},\ }\href {https://doi.org/10.1038/s41467-024-45530-6} {\bibfield
  {journal} {\bibinfo  {journal} {Nature Communications}\ }\textbf {\bibinfo
  {volume} {15}},\ \bibinfo {pages} {1369} (\bibinfo {year}
  {2024})}\BibitemShut {NoStop}%
\bibitem [{\citenamefont {Metelmann}\ and\ \citenamefont
  {Clerk}(2015)}]{Metelmann.2015}%
  \BibitemOpen
  \bibfield  {author} {\bibinfo {author} {\bibfnamefont {A.}~\bibnamefont
  {Metelmann}}\ and\ \bibinfo {author} {\bibfnamefont {A.~A.}\ \bibnamefont
  {Clerk}},\ }\bibfield  {title} {{\selectlanguage {English}\bibinfo {title}
  {{Nonreciprocal Photon Transmission and Amplification via Reservoir
  Engineering}}},\ }\href {https://doi.org/10.1103/physrevx.5.021025}
  {\bibfield  {journal} {\bibinfo  {journal} {Physical Review X}\ }\textbf
  {\bibinfo {volume} {5}},\ \bibinfo {pages} {021025} (\bibinfo {year}
  {2015})}\BibitemShut {NoStop}%
\bibitem [{\citenamefont {Bernier}\ \emph {et~al.}(2017)\citenamefont
  {Bernier}, \citenamefont {Tóth}, \citenamefont {Koottandavida},
  \citenamefont {Ioannou}, \citenamefont {Malz}, \citenamefont {Nunnenkamp},
  \citenamefont {Feofanov},\ and\ \citenamefont {Kippenberg}}]{Bernier.2017}%
  \BibitemOpen
  \bibfield  {author} {\bibinfo {author} {\bibfnamefont {N.~R.}\ \bibnamefont
  {Bernier}}, \bibinfo {author} {\bibfnamefont {L.~D.}\ \bibnamefont {Tóth}},
  \bibinfo {author} {\bibfnamefont {A.}~\bibnamefont {Koottandavida}}, \bibinfo
  {author} {\bibfnamefont {M.~A.}\ \bibnamefont {Ioannou}}, \bibinfo {author}
  {\bibfnamefont {D.}~\bibnamefont {Malz}}, \bibinfo {author} {\bibfnamefont
  {A.}~\bibnamefont {Nunnenkamp}}, \bibinfo {author} {\bibfnamefont {A.~K.}\
  \bibnamefont {Feofanov}},\ and\ \bibinfo {author} {\bibfnamefont {T.~J.}\
  \bibnamefont {Kippenberg}},\ }\bibfield  {title} {\bibinfo {title}
  {{Nonreciprocal reconfigurable microwave optomechanical circuit}},\ }\href
  {https://doi.org/10.1038/s41467-017-00447-1} {\bibfield  {journal} {\bibinfo
  {journal} {Nature Communications}\ }\textbf {\bibinfo {volume} {8}},\
  \bibinfo {pages} {604} (\bibinfo {year} {2017})}\BibitemShut {NoStop}%
\bibitem [{\citenamefont {Verhagen}\ and\ \citenamefont
  {Alù}(2017)}]{Verhagen.2017}%
  \BibitemOpen
  \bibfield  {author} {\bibinfo {author} {\bibfnamefont {E.}~\bibnamefont
  {Verhagen}}\ and\ \bibinfo {author} {\bibfnamefont {A.}~\bibnamefont
  {Alù}},\ }\bibfield  {title} {\bibinfo {title} {{Optomechanical
  nonreciprocity}},\ }\href {https://doi.org/10.1038/nphys4283} {\bibfield
  {journal} {\bibinfo  {journal} {Nature Physics}\ }\textbf {\bibinfo {volume}
  {13}},\ \bibinfo {pages} {922} (\bibinfo {year} {2017})}\BibitemShut
  {NoStop}%
\bibitem [{\citenamefont {Malz}\ \emph {et~al.}(2018)\citenamefont {Malz},
  \citenamefont {Tóth}, \citenamefont {Bernier}, \citenamefont {Feofanov},
  \citenamefont {Kippenberg},\ and\ \citenamefont {Nunnenkamp}}]{Malz.2018}%
  \BibitemOpen
  \bibfield  {author} {\bibinfo {author} {\bibfnamefont {D.}~\bibnamefont
  {Malz}}, \bibinfo {author} {\bibfnamefont {L.~D.}\ \bibnamefont {Tóth}},
  \bibinfo {author} {\bibfnamefont {N.~R.}\ \bibnamefont {Bernier}}, \bibinfo
  {author} {\bibfnamefont {A.~K.}\ \bibnamefont {Feofanov}}, \bibinfo {author}
  {\bibfnamefont {T.~J.}\ \bibnamefont {Kippenberg}},\ and\ \bibinfo {author}
  {\bibfnamefont {A.}~\bibnamefont {Nunnenkamp}},\ }\bibfield  {title}
  {\bibinfo {title} {{Quantum-Limited Directional Amplifiers with
  Optomechanics}},\ }\href {https://doi.org/10.1103/physrevlett.120.023601}
  {\bibfield  {journal} {\bibinfo  {journal} {Physical Review Letters}\
  }\textbf {\bibinfo {volume} {120}},\ \bibinfo {pages} {023601} (\bibinfo
  {year} {2018})}\BibitemShut {NoStop}%
\bibitem [{\citenamefont {Youssefi}\ \emph {et~al.}(2022)\citenamefont
  {Youssefi}, \citenamefont {Kono}, \citenamefont {Bancora}, \citenamefont
  {Chegnizadeh}, \citenamefont {Pan}, \citenamefont {Vovk},\ and\ \citenamefont
  {Kippenberg}}]{Youssefi.2022}%
  \BibitemOpen
  \bibfield  {author} {\bibinfo {author} {\bibfnamefont {A.}~\bibnamefont
  {Youssefi}}, \bibinfo {author} {\bibfnamefont {S.}~\bibnamefont {Kono}},
  \bibinfo {author} {\bibfnamefont {A.}~\bibnamefont {Bancora}}, \bibinfo
  {author} {\bibfnamefont {M.}~\bibnamefont {Chegnizadeh}}, \bibinfo {author}
  {\bibfnamefont {J.}~\bibnamefont {Pan}}, \bibinfo {author} {\bibfnamefont
  {T.}~\bibnamefont {Vovk}},\ and\ \bibinfo {author} {\bibfnamefont {T.~J.}\
  \bibnamefont {Kippenberg}},\ }\bibfield  {title} {\bibinfo {title}
  {{Topological lattices realized in superconducting circuit optomechanics}},\
  }\href {https://doi.org/10.1038/s41586-022-05367-9} {\bibfield  {journal}
  {\bibinfo  {journal} {Nature}\ }\textbf {\bibinfo {volume} {612}},\ \bibinfo
  {pages} {666} (\bibinfo {year} {2022})}\BibitemShut {NoStop}%
\bibitem [{\citenamefont {Peano}\ \emph {et~al.}(2015)\citenamefont {Peano},
  \citenamefont {Brendel}, \citenamefont {Schmidt},\ and\ \citenamefont
  {Marquardt}}]{Peano.2015}%
  \BibitemOpen
  \bibfield  {author} {\bibinfo {author} {\bibfnamefont {V.}~\bibnamefont
  {Peano}}, \bibinfo {author} {\bibfnamefont {C.}~\bibnamefont {Brendel}},
  \bibinfo {author} {\bibfnamefont {M.}~\bibnamefont {Schmidt}},\ and\ \bibinfo
  {author} {\bibfnamefont {F.}~\bibnamefont {Marquardt}},\ }\bibfield  {title}
  {{\selectlanguage {English}\bibinfo {title} {{Topological Phases of Sound and
  Light}}},\ }\href {https://doi.org/10.1103/physrevx.5.031011} {\bibfield
  {journal} {\bibinfo  {journal} {Physical Review X}\ }\textbf {\bibinfo
  {volume} {5}},\ \bibinfo {pages} {031011} (\bibinfo {year}
  {2015})}\BibitemShut {NoStop}%
\bibitem [{\citenamefont {Sanavio}\ \emph {et~al.}(2020)\citenamefont
  {Sanavio}, \citenamefont {Peano},\ and\ \citenamefont
  {Xuereb}}]{Sanavio.2020}%
  \BibitemOpen
  \bibfield  {author} {\bibinfo {author} {\bibfnamefont {C.}~\bibnamefont
  {Sanavio}}, \bibinfo {author} {\bibfnamefont {V.}~\bibnamefont {Peano}},\
  and\ \bibinfo {author} {\bibfnamefont {A.}~\bibnamefont {Xuereb}},\
  }\bibfield  {title} {\bibinfo {title} {{Nonreciprocal topological phononics
  in optomechanical arrays}},\ }\href
  {https://doi.org/10.1103/physrevb.101.085108} {\bibfield  {journal} {\bibinfo
   {journal} {Physical Review B}\ }\textbf {\bibinfo {volume} {101}},\ \bibinfo
  {pages} {085108} (\bibinfo {year} {2020})}\BibitemShut {NoStop}%
\bibitem [{\citenamefont {Ren}\ \emph {et~al.}(2022)\citenamefont {Ren},
  \citenamefont {Shah}, \citenamefont {Pfeifer}, \citenamefont {Brendel},
  \citenamefont {Peano}, \citenamefont {Marquardt},\ and\ \citenamefont
  {Painter}}]{Ren.2022}%
  \BibitemOpen
  \bibfield  {author} {\bibinfo {author} {\bibfnamefont {H.}~\bibnamefont
  {Ren}}, \bibinfo {author} {\bibfnamefont {T.}~\bibnamefont {Shah}}, \bibinfo
  {author} {\bibfnamefont {H.}~\bibnamefont {Pfeifer}}, \bibinfo {author}
  {\bibfnamefont {C.}~\bibnamefont {Brendel}}, \bibinfo {author} {\bibfnamefont
  {V.}~\bibnamefont {Peano}}, \bibinfo {author} {\bibfnamefont
  {F.}~\bibnamefont {Marquardt}},\ and\ \bibinfo {author} {\bibfnamefont
  {O.}~\bibnamefont {Painter}},\ }\bibfield  {title} {\bibinfo {title}
  {{Topological phonon transport in an optomechanical system}},\ }\href
  {https://doi.org/10.1038/s41467-022-30941-0} {\bibfield  {journal} {\bibinfo
  {journal} {Nature Communications}\ }\textbf {\bibinfo {volume} {13}},\
  \bibinfo {pages} {3476} (\bibinfo {year} {2022})}\BibitemShut {NoStop}%
\bibitem [{\citenamefont {Budich}\ and\ \citenamefont
  {Bergholtz}(2020)}]{Budich20}%
  \BibitemOpen
  \bibfield  {author} {\bibinfo {author} {\bibfnamefont {J.~C.}\ \bibnamefont
  {Budich}}\ and\ \bibinfo {author} {\bibfnamefont {E.~J.}\ \bibnamefont
  {Bergholtz}},\ }\bibfield  {title} {\bibinfo {title} {{Non-Hermitian
  Topological Sensors}},\ }\href
  {https://doi.org/10.1103/PhysRevLett.125.180403} {\bibfield  {journal}
  {\bibinfo  {journal} {Phys. Rev. Lett.}\ }\textbf {\bibinfo {volume} {125}},\
  \bibinfo {pages} {180403} (\bibinfo {year} {2020})}\BibitemShut {NoStop}%
\bibitem [{\citenamefont {Koch}\ and\ \citenamefont {Budich}(2022)}]{Koch22}%
  \BibitemOpen
  \bibfield  {author} {\bibinfo {author} {\bibfnamefont {F.}~\bibnamefont
  {Koch}}\ and\ \bibinfo {author} {\bibfnamefont {J.~C.}\ \bibnamefont
  {Budich}},\ }\bibfield  {title} {\bibinfo {title} {{Quantum non-Hermitian
  topological sensors}},\ }\href
  {https://doi.org/10.1103/PhysRevResearch.4.013113} {\bibfield  {journal}
  {\bibinfo  {journal} {Phys. Rev. Res.}\ }\textbf {\bibinfo {volume} {4}},\
  \bibinfo {pages} {013113} (\bibinfo {year} {2022})}\BibitemShut {NoStop}%
\bibitem [{\citenamefont {Wu}\ \emph {et~al.}(2024)\citenamefont {Wu},
  \citenamefont {Xu}, \citenamefont {Jing}, \citenamefont {Kuang},
  \citenamefont {Nori},\ and\ \citenamefont {Liao}}]{wu_manipulating_2024}%
  \BibitemOpen
  \bibfield  {author} {\bibinfo {author} {\bibfnamefont {J.-K.}\ \bibnamefont
  {Wu}}, \bibinfo {author} {\bibfnamefont {X.-W.}\ \bibnamefont {Xu}}, \bibinfo
  {author} {\bibfnamefont {H.}~\bibnamefont {Jing}}, \bibinfo {author}
  {\bibfnamefont {L.-M.}\ \bibnamefont {Kuang}}, \bibinfo {author}
  {\bibfnamefont {F.}~\bibnamefont {Nori}},\ and\ \bibinfo {author}
  {\bibfnamefont {J.-Q.}\ \bibnamefont {Liao}},\ }\href
  {http://arxiv.org/abs/2405.05753} {\bibinfo {title} {Manipulating
  {Topological} {Polaritons} in {Optomechanical} {Ladders}}} (\bibinfo {year}
  {2024}),\ \bibinfo {note} {arXiv:2405.05753 [quant-ph]}\BibitemShut {NoStop}%
\bibitem [{\citenamefont {Bergholtz}\ \emph {et~al.}(2021)\citenamefont
  {Bergholtz}, \citenamefont {Budich},\ and\ \citenamefont
  {Kunst}}]{Bergholtz21}%
  \BibitemOpen
  \bibfield  {author} {\bibinfo {author} {\bibfnamefont {E.~J.}\ \bibnamefont
  {Bergholtz}}, \bibinfo {author} {\bibfnamefont {J.~C.}\ \bibnamefont
  {Budich}},\ and\ \bibinfo {author} {\bibfnamefont {F.~K.}\ \bibnamefont
  {Kunst}},\ }\bibfield  {title} {\bibinfo {title} {{Exceptional topology of
  non-Hermitian systems}},\ }\href
  {https://doi.org/10.1103/RevModPhys.93.015005} {\bibfield  {journal}
  {\bibinfo  {journal} {Rev. Mod. Phys.}\ }\textbf {\bibinfo {volume} {93}},\
  \bibinfo {pages} {015005} (\bibinfo {year} {2021})}\BibitemShut {NoStop}%
\bibitem [{\citenamefont {Kawabata}\ \emph {et~al.}(2019)\citenamefont
  {Kawabata}, \citenamefont {Shiozaki}, \citenamefont {Ueda},\ and\
  \citenamefont {Sato}}]{Kawabata.2019}%
  \BibitemOpen
  \bibfield  {author} {\bibinfo {author} {\bibfnamefont {K.}~\bibnamefont
  {Kawabata}}, \bibinfo {author} {\bibfnamefont {K.}~\bibnamefont {Shiozaki}},
  \bibinfo {author} {\bibfnamefont {M.}~\bibnamefont {Ueda}},\ and\ \bibinfo
  {author} {\bibfnamefont {M.}~\bibnamefont {Sato}},\ }\bibfield  {title}
  {\bibinfo {title} {{Symmetry and Topology in Non-Hermitian Physics}},\ }\href
  {https://doi.org/10.1103/physrevx.9.041015} {\bibfield  {journal} {\bibinfo
  {journal} {Physical Review X}\ }\textbf {\bibinfo {volume} {9}},\ \bibinfo
  {pages} {041015} (\bibinfo {year} {2019})}\BibitemShut {NoStop}%
\bibitem [{\citenamefont {Palomaki}\ \emph {et~al.}(2013)\citenamefont
  {Palomaki}, \citenamefont {Teufel}, \citenamefont {Simmonds},\ and\
  \citenamefont {Lehnert}}]{Palomaki.2013}%
  \BibitemOpen
  \bibfield  {author} {\bibinfo {author} {\bibfnamefont {T.~A.}\ \bibnamefont
  {Palomaki}}, \bibinfo {author} {\bibfnamefont {J.~D.}\ \bibnamefont
  {Teufel}}, \bibinfo {author} {\bibfnamefont {R.~W.}\ \bibnamefont
  {Simmonds}},\ and\ \bibinfo {author} {\bibfnamefont {K.~W.}\ \bibnamefont
  {Lehnert}},\ }\bibfield  {title} {{\selectlanguage {English}\bibinfo {title}
  {{Entangling mechanical motion with microwave fields.}}},\ }\href
  {https://doi.org/10.1126/science.1244563} {\bibfield  {journal} {\bibinfo
  {journal} {Science}\ }\textbf {\bibinfo {volume} {342}},\ \bibinfo {pages}
  {710 } (\bibinfo {year} {2013})}\BibitemShut {NoStop}%
\bibitem [{\citenamefont {Genes}\ \emph {et~al.}(2008)\citenamefont {Genes},
  \citenamefont {Mari}, \citenamefont {Tombesi},\ and\ \citenamefont
  {Vitali}}]{Genes.2008wrq}%
  \BibitemOpen
  \bibfield  {author} {\bibinfo {author} {\bibfnamefont {C.}~\bibnamefont
  {Genes}}, \bibinfo {author} {\bibfnamefont {A.}~\bibnamefont {Mari}},
  \bibinfo {author} {\bibfnamefont {P.}~\bibnamefont {Tombesi}},\ and\ \bibinfo
  {author} {\bibfnamefont {D.}~\bibnamefont {Vitali}},\ }\bibfield  {title}
  {\bibinfo {title} {{Robust entanglement of a micromechanical resonator with
  output optical fields}},\ }\href {https://doi.org/10.1103/physreva.78.032316}
  {\bibfield  {journal} {\bibinfo  {journal} {Physical Review A}\ }\textbf
  {\bibinfo {volume} {78}},\ \bibinfo {pages} {032316} (\bibinfo {year}
  {2008})}\BibitemShut {NoStop}%
\bibitem [{\citenamefont {Peterson}\ \emph {et~al.}(2019)\citenamefont
  {Peterson}, \citenamefont {Kotler}, \citenamefont {Lecocq}, \citenamefont
  {Cicak}, \citenamefont {Jin}, \citenamefont {Simmonds}, \citenamefont
  {Aumentado},\ and\ \citenamefont {Teufel}}]{Peterson.2019}%
  \BibitemOpen
  \bibfield  {author} {\bibinfo {author} {\bibfnamefont {G.~A.}\ \bibnamefont
  {Peterson}}, \bibinfo {author} {\bibfnamefont {S.}~\bibnamefont {Kotler}},
  \bibinfo {author} {\bibfnamefont {F.}~\bibnamefont {Lecocq}}, \bibinfo
  {author} {\bibfnamefont {K.}~\bibnamefont {Cicak}}, \bibinfo {author}
  {\bibfnamefont {X.~Y.}\ \bibnamefont {Jin}}, \bibinfo {author} {\bibfnamefont
  {R.~W.}\ \bibnamefont {Simmonds}}, \bibinfo {author} {\bibfnamefont
  {J.}~\bibnamefont {Aumentado}},\ and\ \bibinfo {author} {\bibfnamefont
  {J.~D.}\ \bibnamefont {Teufel}},\ }\bibfield  {title} {\bibinfo {title}
  {{Ultrastrong Parametric Coupling between a Superconducting Cavity and a
  Mechanical Resonator}},\ }\href
  {https://doi.org/10.1103/physrevlett.123.247701} {\bibfield  {journal}
  {\bibinfo  {journal} {Physical Review Letters}\ }\textbf {\bibinfo {volume}
  {123}},\ \bibinfo {pages} {247701} (\bibinfo {year} {2019})}\BibitemShut
  {NoStop}%
\bibitem [{\citenamefont {Brzezicki}\ and\ \citenamefont
  {Hyart}(2019)}]{Brzezicki.2019}%
  \BibitemOpen
  \bibfield  {author} {\bibinfo {author} {\bibfnamefont {W.}~\bibnamefont
  {Brzezicki}}\ and\ \bibinfo {author} {\bibfnamefont {T.}~\bibnamefont
  {Hyart}},\ }\bibfield  {title} {\bibinfo {title} {{Hidden Chern number in
  one-dimensional non-Hermitian chiral-symmetric systems}},\ }\href
  {https://doi.org/10.1103/physrevb.100.161105} {\bibfield  {journal} {\bibinfo
   {journal} {Physical Review B}\ }\textbf {\bibinfo {volume} {100}},\ \bibinfo
  {pages} {161105} (\bibinfo {year} {2019})}\BibitemShut {NoStop}%
\bibitem [{\citenamefont {Brzezicki}\ \emph {et~al.}(2023)\citenamefont
  {Brzezicki}, \citenamefont {Silveri}, \citenamefont {Płodzień},
  \citenamefont {Massel},\ and\ \citenamefont {Hyart}}]{Brzezicki.2023}%
  \BibitemOpen
  \bibfield  {author} {\bibinfo {author} {\bibfnamefont {W.}~\bibnamefont
  {Brzezicki}}, \bibinfo {author} {\bibfnamefont {M.}~\bibnamefont {Silveri}},
  \bibinfo {author} {\bibfnamefont {M.}~\bibnamefont {Płodzień}}, \bibinfo
  {author} {\bibfnamefont {F.}~\bibnamefont {Massel}},\ and\ \bibinfo {author}
  {\bibfnamefont {T.}~\bibnamefont {Hyart}},\ }\bibfield  {title} {\bibinfo
  {title} {{Non-Hermitian topological quantum states in a reservoir-engineered
  transmon chain}},\ }\href {https://doi.org/10.1103/physrevb.107.115146}
  {\bibfield  {journal} {\bibinfo  {journal} {Physical Review B}\ }\textbf
  {\bibinfo {volume} {107}},\ \bibinfo {pages} {115146} (\bibinfo {year}
  {2023})}\BibitemShut {NoStop}%
\bibitem [{\citenamefont {Aspelmeyer}\ \emph {et~al.}(2014)\citenamefont
  {Aspelmeyer}, \citenamefont {Kippenberg},\ and\ \citenamefont
  {Marquardt}}]{Aspelmeyer.2014}%
  \BibitemOpen
  \bibfield  {author} {\bibinfo {author} {\bibfnamefont {M.}~\bibnamefont
  {Aspelmeyer}}, \bibinfo {author} {\bibfnamefont {T.~J.}\ \bibnamefont
  {Kippenberg}},\ and\ \bibinfo {author} {\bibfnamefont {F.}~\bibnamefont
  {Marquardt}},\ }\bibfield  {title} {{\selectlanguage {English}\bibinfo
  {title} {{Cavity optomechanics}}},\ }\href
  {https://doi.org/10.1103/revmodphys.86.1391} {\bibfield  {journal} {\bibinfo
  {journal} {Reviews of Modern Physics}\ }\textbf {\bibinfo {volume} {86}},\
  \bibinfo {pages} {1391 } (\bibinfo {year} {2014})}\BibitemShut {NoStop}%
\bibitem [{\citenamefont {Bowen}\ and\ \citenamefont
  {Milburn}(2015)}]{Bowen.2015}%
  \BibitemOpen
  \bibfield  {author} {\bibinfo {author} {\bibfnamefont {W.~P.}\ \bibnamefont
  {Bowen}}\ and\ \bibinfo {author} {\bibfnamefont {G.~J.}\ \bibnamefont
  {Milburn}},\ }\href {https://doi.org/10.1103/physreva.86.043803}
  {{\selectlanguage {English}\emph {\bibinfo {title} {{Quantum
  Optomechanics}}}}},\ CRC Press\ (\bibinfo  {publisher} {CRC Press},\ \bibinfo
  {year} {2015})\BibitemShut {NoStop}%
\bibitem [{\citenamefont {Marquardt}\ \emph {et~al.}(2007)\citenamefont
  {Marquardt}, \citenamefont {Chen}, \citenamefont {Clerk},\ and\ \citenamefont
  {Girvin}}]{Marquardt.2007}%
  \BibitemOpen
  \bibfield  {author} {\bibinfo {author} {\bibfnamefont {F.}~\bibnamefont
  {Marquardt}}, \bibinfo {author} {\bibfnamefont {J.~P.}\ \bibnamefont {Chen}},
  \bibinfo {author} {\bibfnamefont {A.~A.}\ \bibnamefont {Clerk}},\ and\
  \bibinfo {author} {\bibfnamefont {S.~M.}\ \bibnamefont {Girvin}},\ }\bibfield
   {title} {{\selectlanguage {English}\bibinfo {title} {{Quantum theory of
  cavity-assisted sideband cooling of mechanical motion}}},\ }\href
  {https://doi.org/10.1103/physrevlett.99.093902} {\bibfield  {journal}
  {\bibinfo  {journal} {Physical Review Letters}\ }\textbf {\bibinfo {volume}
  {99}},\ \bibinfo {pages} {093902} (\bibinfo {year} {2007})}\BibitemShut
  {NoStop}%
\bibitem [{\citenamefont {Teufel}\ \emph {et~al.}(2011)\citenamefont {Teufel},
  \citenamefont {Donner}, \citenamefont {Li}, \citenamefont {Harlow},
  \citenamefont {Allman}, \citenamefont {Cicak}, \citenamefont {Sirois},
  \citenamefont {Whittaker}, \citenamefont {Lehnert},\ and\ \citenamefont
  {Simmonds}}]{Teufel.2011ju}%
  \BibitemOpen
  \bibfield  {author} {\bibinfo {author} {\bibfnamefont {J.~D.}\ \bibnamefont
  {Teufel}}, \bibinfo {author} {\bibfnamefont {T.}~\bibnamefont {Donner}},
  \bibinfo {author} {\bibfnamefont {D.}~\bibnamefont {Li}}, \bibinfo {author}
  {\bibfnamefont {J.~W.}\ \bibnamefont {Harlow}}, \bibinfo {author}
  {\bibfnamefont {M.~S.}\ \bibnamefont {Allman}}, \bibinfo {author}
  {\bibfnamefont {K.}~\bibnamefont {Cicak}}, \bibinfo {author} {\bibfnamefont
  {A.~J.}\ \bibnamefont {Sirois}}, \bibinfo {author} {\bibfnamefont {J.~D.}\
  \bibnamefont {Whittaker}}, \bibinfo {author} {\bibfnamefont {K.~W.}\
  \bibnamefont {Lehnert}},\ and\ \bibinfo {author} {\bibfnamefont {R.~W.}\
  \bibnamefont {Simmonds}},\ }\bibfield  {title} {{\selectlanguage
  {English}\bibinfo {title} {{Sideband cooling of micromechanical motion to the
  quantum ground state}}},\ }\href {https://doi.org/10.1038/nature10261}
  {\bibfield  {journal} {\bibinfo  {journal} {Nature}\ }\textbf {\bibinfo
  {volume} {475}},\ \bibinfo {pages} {359 } (\bibinfo {year}
  {2011})}\BibitemShut {NoStop}%
\bibitem [{\citenamefont {Ludwig}\ \emph {et~al.}(2008)\citenamefont {Ludwig},
  \citenamefont {Kubala},\ and\ \citenamefont {Marquardt}}]{Ludwig.2008}%
  \BibitemOpen
  \bibfield  {author} {\bibinfo {author} {\bibfnamefont {M.}~\bibnamefont
  {Ludwig}}, \bibinfo {author} {\bibfnamefont {B.}~\bibnamefont {Kubala}},\
  and\ \bibinfo {author} {\bibfnamefont {F.}~\bibnamefont {Marquardt}},\
  }\bibfield  {title} {{\selectlanguage {English}\bibinfo {title} {{The
  optomechanical instability in the quantum regime}}},\ }\href
  {https://doi.org/10.1088/1367-2630/10/9/095013} {\bibfield  {journal}
  {\bibinfo  {journal} {New Journal Of Physics}\ }\textbf {\bibinfo {volume}
  {10}},\ \bibinfo {pages} {095013} (\bibinfo {year} {2008})}\BibitemShut
  {NoStop}%
\bibitem [{\citenamefont {Massel}\ \emph {et~al.}(2011)\citenamefont {Massel},
  \citenamefont {Heikkilä}, \citenamefont {Pirkkalainen}, \citenamefont {Cho},
  \citenamefont {Saloniemi}, \citenamefont {Hakonen},\ and\ \citenamefont
  {Sillanpää}}]{Massel.2011}%
  \BibitemOpen
  \bibfield  {author} {\bibinfo {author} {\bibfnamefont {F.}~\bibnamefont
  {Massel}}, \bibinfo {author} {\bibfnamefont {T.~T.}\ \bibnamefont
  {Heikkilä}}, \bibinfo {author} {\bibfnamefont {J.~M.}\ \bibnamefont
  {Pirkkalainen}}, \bibinfo {author} {\bibfnamefont {S.~U.}\ \bibnamefont
  {Cho}}, \bibinfo {author} {\bibfnamefont {H.}~\bibnamefont {Saloniemi}},
  \bibinfo {author} {\bibfnamefont {P.~J.}\ \bibnamefont {Hakonen}},\ and\
  \bibinfo {author} {\bibfnamefont {M.~A.}\ \bibnamefont {Sillanpää}},\
  }\bibfield  {title} {{\selectlanguage {English}\bibinfo {title} {{Microwave
  amplification with nanomechanical resonators}}},\ }\href
  {https://doi.org/10.1038/nature10628} {\bibfield  {journal} {\bibinfo
  {journal} {Nature}\ }\textbf {\bibinfo {volume} {480}},\ \bibinfo {pages}
  {351 } (\bibinfo {year} {2011})}\BibitemShut {NoStop}%
\bibitem [{\citenamefont {Safavi-Naeini}\ and\ \citenamefont
  {Painter}(2010)}]{Safavi-Naeini.2010}%
  \BibitemOpen
  \bibfield  {author} {\bibinfo {author} {\bibfnamefont {A.~H.}\ \bibnamefont
  {Safavi-Naeini}}\ and\ \bibinfo {author} {\bibfnamefont {O.}~\bibnamefont
  {Painter}},\ }\bibfield  {title} {\bibinfo {title} {{Design of optomechanical
  cavities and waveguides on a simultaneous bandgap phononic-photonic crystal
  slab}},\ }\href {https://doi.org/10.1364/oe.18.014926} {\bibfield  {journal}
  {\bibinfo  {journal} {Optics Express}\ }\textbf {\bibinfo {volume} {18}},\
  \bibinfo {pages} {14926} (\bibinfo {year} {2010})}\BibitemShut {NoStop}%
\bibitem [{\citenamefont {Painter}\ \emph {et~al.}(1999)\citenamefont
  {Painter}, \citenamefont {Lee}, \citenamefont {Scherer}, \citenamefont
  {Yariv}, \citenamefont {O'Brien}, \citenamefont {Dapkus},\ and\ \citenamefont
  {Kim}}]{Painter.1999}%
  \BibitemOpen
  \bibfield  {author} {\bibinfo {author} {\bibfnamefont {O.}~\bibnamefont
  {Painter}}, \bibinfo {author} {\bibfnamefont {R.~K.}\ \bibnamefont {Lee}},
  \bibinfo {author} {\bibfnamefont {A.}~\bibnamefont {Scherer}}, \bibinfo
  {author} {\bibfnamefont {A.}~\bibnamefont {Yariv}}, \bibinfo {author}
  {\bibfnamefont {J.~D.}\ \bibnamefont {O'Brien}}, \bibinfo {author}
  {\bibfnamefont {P.~D.}\ \bibnamefont {Dapkus}},\ and\ \bibinfo {author}
  {\bibfnamefont {I.}~\bibnamefont {Kim}},\ }\bibfield  {title} {\bibinfo
  {title} {{Two-Dimensional Photonic Band-Gap Defect Mode Laser}},\ }\href
  {https://doi.org/10.1126/science.284.5421.1819} {\bibfield  {journal}
  {\bibinfo  {journal} {Science}\ }\textbf {\bibinfo {volume} {284}},\ \bibinfo
  {pages} {1819} (\bibinfo {year} {1999})}\BibitemShut {NoStop}%
\bibitem [{\citenamefont {Vučković}\ \emph {et~al.}(2002)\citenamefont
  {Vučković}, \citenamefont {Lončar}, \citenamefont {Mabuchi},\ and\
  \citenamefont {Scherer}}]{Vučković.2002}%
  \BibitemOpen
  \bibfield  {author} {\bibinfo {author} {\bibfnamefont {J.}~\bibnamefont
  {Vučković}}, \bibinfo {author} {\bibfnamefont {M.}~\bibnamefont {Lončar}},
  \bibinfo {author} {\bibfnamefont {H.}~\bibnamefont {Mabuchi}},\ and\ \bibinfo
  {author} {\bibfnamefont {A.}~\bibnamefont {Scherer}},\ }\bibfield  {title}
  {\bibinfo {title} {{Design of photonic crystal microcavities for cavity
  QED}},\ }\href {https://doi.org/10.1103/physreve.65.016608} {\bibfield
  {journal} {\bibinfo  {journal} {Physical Review E}\ }\textbf {\bibinfo
  {volume} {65}},\ \bibinfo {pages} {016608} (\bibinfo {year}
  {2002})}\BibitemShut {NoStop}%
\bibitem [{\citenamefont {Vasseur}\ \emph {et~al.}(2007)\citenamefont
  {Vasseur}, \citenamefont {Hladky-Hennion}, \citenamefont {Djafari-Rouhani},
  \citenamefont {Duval}, \citenamefont {Dubus}, \citenamefont {Pennec},\ and\
  \citenamefont {Deymier}}]{Vasseur.2007}%
  \BibitemOpen
  \bibfield  {author} {\bibinfo {author} {\bibfnamefont {J.~O.}\ \bibnamefont
  {Vasseur}}, \bibinfo {author} {\bibfnamefont {A.-C.}\ \bibnamefont
  {Hladky-Hennion}}, \bibinfo {author} {\bibfnamefont {B.}~\bibnamefont
  {Djafari-Rouhani}}, \bibinfo {author} {\bibfnamefont {F.}~\bibnamefont
  {Duval}}, \bibinfo {author} {\bibfnamefont {B.}~\bibnamefont {Dubus}},
  \bibinfo {author} {\bibfnamefont {Y.}~\bibnamefont {Pennec}},\ and\ \bibinfo
  {author} {\bibfnamefont {P.~A.}\ \bibnamefont {Deymier}},\ }\bibfield
  {title} {\bibinfo {title} {Waveguiding in two-dimensional piezoelectric
  phononic crystal plates},\ }\href {https://doi.org/10.1063/1.2740352}
  {\bibfield  {journal} {\bibinfo  {journal} {Journal of Applied Physics}\
  }\textbf {\bibinfo {volume} {101}},\ \bibinfo {pages} {114904} (\bibinfo
  {year} {2007})}\BibitemShut {NoStop}%
\bibitem [{\citenamefont {III}\ and\ \citenamefont {El-Kady}(2009)}]{III.2009}%
  \BibitemOpen
  \bibfield  {author} {\bibinfo {author} {\bibfnamefont {R.~H.~O.}\
  \bibnamefont {III}}\ and\ \bibinfo {author} {\bibfnamefont {I.}~\bibnamefont
  {El-Kady}},\ }\bibfield  {title} {\bibinfo {title} {{Microfabricated phononic
  crystal devices and applications}},\ }\href
  {https://doi.org/10.1088/0957-0233/20/1/012002} {\bibfield  {journal}
  {\bibinfo  {journal} {Measurement Science and Technology}\ }\textbf {\bibinfo
  {volume} {20}},\ \bibinfo {pages} {012002} (\bibinfo {year}
  {2009})}\BibitemShut {NoStop}%
\bibitem [{sup(2024)}]{suppl}%
  \BibitemOpen
  \bibfield  {title} {\bibinfo {title} {{Supplemental Material. Additional
  references appear in the supplemental material
  (Refs.~\cite{Gardiner.2004,gardiner_handbook_2004,Walls.2008})}},\
  }\href@noop {} {\  (\bibinfo {year} {2024})}\BibitemShut {NoStop}%
\bibitem [{\citenamefont {Eichenfield}\ \emph {et~al.}(2009)\citenamefont
  {Eichenfield}, \citenamefont {Chan}, \citenamefont {Camacho}, \citenamefont
  {Vahala},\ and\ \citenamefont {Painter}}]{Eichenfield.2009}%
  \BibitemOpen
  \bibfield  {author} {\bibinfo {author} {\bibfnamefont {M.}~\bibnamefont
  {Eichenfield}}, \bibinfo {author} {\bibfnamefont {J.}~\bibnamefont {Chan}},
  \bibinfo {author} {\bibfnamefont {R.~M.}\ \bibnamefont {Camacho}}, \bibinfo
  {author} {\bibfnamefont {K.~J.}\ \bibnamefont {Vahala}},\ and\ \bibinfo
  {author} {\bibfnamefont {O.}~\bibnamefont {Painter}},\ }\bibfield  {title}
  {{\selectlanguage {English}\bibinfo {title} {{Optomechanical crystals}}},\
  }\href {https://doi.org/10.1038/nature08524} {\bibfield  {journal} {\bibinfo
  {journal} {Nature}\ }\textbf {\bibinfo {volume} {462}},\ \bibinfo {pages} {78
  } (\bibinfo {year} {2009})}\BibitemShut {NoStop}%
\bibitem [{\citenamefont {Ludwig}\ and\ \citenamefont
  {Marquardt}(2013)}]{Ludwig.2013}%
  \BibitemOpen
  \bibfield  {author} {\bibinfo {author} {\bibfnamefont {M.}~\bibnamefont
  {Ludwig}}\ and\ \bibinfo {author} {\bibfnamefont {F.}~\bibnamefont
  {Marquardt}},\ }\bibfield  {title} {{\selectlanguage {English}\bibinfo
  {title} {{Quantum Many-Body Dynamics in Optomechanical Arrays}}},\ }\href
  {https://doi.org/10.1103/physrevlett.111.073603} {\bibfield  {journal}
  {\bibinfo  {journal} {Physical Review Letters}\ }\textbf {\bibinfo {volume}
  {111}},\ \bibinfo {pages} {073603} (\bibinfo {year} {2013})}\BibitemShut
  {NoStop}%
\bibitem [{\citenamefont {Prosen}\ and\ \citenamefont
  {Seligman}(2010)}]{prosen_quantization_2010}%
  \BibitemOpen
  \bibfield  {author} {\bibinfo {author} {\bibfnamefont {T.}~\bibnamefont
  {Prosen}}\ and\ \bibinfo {author} {\bibfnamefont {T.~H.}\ \bibnamefont
  {Seligman}},\ }\bibfield  {title} {\bibinfo {title} {Quantization over boson
  operator spaces},\ }\href {https://doi.org/10.1088/1751-8113/43/39/392004}
  {\bibfield  {journal} {\bibinfo  {journal} {Journal of Physics A:
  Mathematical and Theoretical}\ }\textbf {\bibinfo {volume} {43}},\ \bibinfo
  {pages} {392004} (\bibinfo {year} {2010})}\BibitemShut {NoStop}%
\bibitem [{\citenamefont {Bahari}\ \emph {et~al.}(2017)\citenamefont {Bahari},
  \citenamefont {Ndao}, \citenamefont {Vallini}, \citenamefont {Amili},
  \citenamefont {Fainman},\ and\ \citenamefont {Kanté}}]{Bahari.2017}%
  \BibitemOpen
  \bibfield  {author} {\bibinfo {author} {\bibfnamefont {B.}~\bibnamefont
  {Bahari}}, \bibinfo {author} {\bibfnamefont {A.}~\bibnamefont {Ndao}},
  \bibinfo {author} {\bibfnamefont {F.}~\bibnamefont {Vallini}}, \bibinfo
  {author} {\bibfnamefont {A.~E.}\ \bibnamefont {Amili}}, \bibinfo {author}
  {\bibfnamefont {Y.}~\bibnamefont {Fainman}},\ and\ \bibinfo {author}
  {\bibfnamefont {B.}~\bibnamefont {Kanté}},\ }\bibfield  {title} {\bibinfo
  {title} {{Nonreciprocal lasing in topological cavities of arbitrary
  geometries}},\ }\href {https://doi.org/10.1126/science.aao4551} {\bibfield
  {journal} {\bibinfo  {journal} {Science}\ }\textbf {\bibinfo {volume}
  {358}},\ \bibinfo {pages} {636} (\bibinfo {year} {2017})}\BibitemShut
  {NoStop}%
\bibitem [{\citenamefont {Harari}\ \emph {et~al.}(2018)\citenamefont {Harari},
  \citenamefont {Bandres}, \citenamefont {Lumer}, \citenamefont {Rechtsman},
  \citenamefont {Chong}, \citenamefont {Khajavikhan}, \citenamefont
  {Christodoulides},\ and\ \citenamefont {Segev}}]{Harari.2018}%
  \BibitemOpen
  \bibfield  {author} {\bibinfo {author} {\bibfnamefont {G.}~\bibnamefont
  {Harari}}, \bibinfo {author} {\bibfnamefont {M.~A.}\ \bibnamefont {Bandres}},
  \bibinfo {author} {\bibfnamefont {Y.}~\bibnamefont {Lumer}}, \bibinfo
  {author} {\bibfnamefont {M.~C.}\ \bibnamefont {Rechtsman}}, \bibinfo {author}
  {\bibfnamefont {Y.~D.}\ \bibnamefont {Chong}}, \bibinfo {author}
  {\bibfnamefont {M.}~\bibnamefont {Khajavikhan}}, \bibinfo {author}
  {\bibfnamefont {D.~N.}\ \bibnamefont {Christodoulides}},\ and\ \bibinfo
  {author} {\bibfnamefont {M.}~\bibnamefont {Segev}},\ }\bibfield  {title}
  {\bibinfo {title} {{Topological insulator laser: Theory}},\ }\bibfield
  {journal} {\bibinfo  {journal} {Science}\ }\textbf {\bibinfo {volume}
  {359}},\ \href {https://doi.org/10.1126/science.aar4003}
  {10.1126/science.aar4003} (\bibinfo {year} {2018})\BibitemShut {NoStop}%
\bibitem [{\citenamefont {Metzger}\ \emph {et~al.}(2008)\citenamefont
  {Metzger}, \citenamefont {Ludwig}, \citenamefont {Neuenhahn}, \citenamefont
  {Ortlieb}, \citenamefont {Favero}, \citenamefont {Karrai},\ and\
  \citenamefont {Marquardt}}]{Metzger.2008}%
  \BibitemOpen
  \bibfield  {author} {\bibinfo {author} {\bibfnamefont {C.}~\bibnamefont
  {Metzger}}, \bibinfo {author} {\bibfnamefont {M.}~\bibnamefont {Ludwig}},
  \bibinfo {author} {\bibfnamefont {C.}~\bibnamefont {Neuenhahn}}, \bibinfo
  {author} {\bibfnamefont {A.}~\bibnamefont {Ortlieb}}, \bibinfo {author}
  {\bibfnamefont {I.}~\bibnamefont {Favero}}, \bibinfo {author} {\bibfnamefont
  {K.}~\bibnamefont {Karrai}},\ and\ \bibinfo {author} {\bibfnamefont
  {F.}~\bibnamefont {Marquardt}},\ }\bibfield  {title} {\bibinfo {title}
  {{Self-Induced Oscillations in an Optomechanical System Driven by Bolometric
  Backaction}},\ }\href {https://doi.org/10.1103/physrevlett.101.133903}
  {\bibfield  {journal} {\bibinfo  {journal} {Physical Review Letters}\
  }\textbf {\bibinfo {volume} {101}},\ \bibinfo {pages} {133903} (\bibinfo
  {year} {2008})}\BibitemShut {NoStop}%
\bibitem [{\citenamefont {Marquardt}\ and\ \citenamefont
  {Girvin}(2009)}]{Marquardt.2009}%
  \BibitemOpen
  \bibfield  {author} {\bibinfo {author} {\bibfnamefont {F.}~\bibnamefont
  {Marquardt}}\ and\ \bibinfo {author} {\bibfnamefont {S.}~\bibnamefont
  {Girvin}},\ }\bibfield  {title} {{\selectlanguage {English}\bibinfo {title}
  {{Optomechanics}}},\ }\href {https://doi.org/10.1103/physics.2.40} {\bibfield
   {journal} {\bibinfo  {journal} {Physics}\ }\textbf {\bibinfo {volume} {2}},\
  \bibinfo {pages} {40} (\bibinfo {year} {2009})}\BibitemShut {NoStop}%
\bibitem [{\citenamefont {Safavi-Naeini}\ \emph {et~al.}(2011)\citenamefont
  {Safavi-Naeini}, \citenamefont {Alegre}, \citenamefont {Chan}, \citenamefont
  {Eichenfield}, \citenamefont {Winger}, \citenamefont {Lin}, \citenamefont
  {Hill}, \citenamefont {Chang},\ and\ \citenamefont
  {Painter}}]{Safavi-Naeini.2011}%
  \BibitemOpen
  \bibfield  {author} {\bibinfo {author} {\bibfnamefont {A.~H.}\ \bibnamefont
  {Safavi-Naeini}}, \bibinfo {author} {\bibfnamefont {T.~P.~M.}\ \bibnamefont
  {Alegre}}, \bibinfo {author} {\bibfnamefont {J.}~\bibnamefont {Chan}},
  \bibinfo {author} {\bibfnamefont {M.}~\bibnamefont {Eichenfield}}, \bibinfo
  {author} {\bibfnamefont {M.}~\bibnamefont {Winger}}, \bibinfo {author}
  {\bibfnamefont {Q.}~\bibnamefont {Lin}}, \bibinfo {author} {\bibfnamefont
  {J.~T.}\ \bibnamefont {Hill}}, \bibinfo {author} {\bibfnamefont {D.~E.}\
  \bibnamefont {Chang}},\ and\ \bibinfo {author} {\bibfnamefont
  {O.}~\bibnamefont {Painter}},\ }\bibfield  {title} {{\selectlanguage
  {English}\bibinfo {title} {{Electromagnetically induced transparency and slow
  light with optomechanics}}},\ }\href {https://doi.org/10.1038/nature09933}
  {\bibfield  {journal} {\bibinfo  {journal} {Nature}\ }\textbf {\bibinfo
  {volume} {472}},\ \bibinfo {pages} {69 } (\bibinfo {year}
  {2011})}\BibitemShut {NoStop}%
\bibitem [{\citenamefont {Adesso}\ and\ \citenamefont
  {Illuminati}(2007)}]{Adesso.2007}%
  \BibitemOpen
  \bibfield  {author} {\bibinfo {author} {\bibfnamefont {G.}~\bibnamefont
  {Adesso}}\ and\ \bibinfo {author} {\bibfnamefont {F.}~\bibnamefont
  {Illuminati}},\ }\bibfield  {title} {{\selectlanguage {English}\bibinfo
  {title} {{Entanglement in continuous-variable systems: recent advances and
  current perspectives}}},\ }\href
  {https://doi.org/10.1088/1751-8113/40/28/s01} {\bibfield  {journal} {\bibinfo
   {journal} {Journal of Physics A: Mathematical and Theoretical}\ }\textbf
  {\bibinfo {volume} {40}},\ \bibinfo {pages} {7821 } (\bibinfo {year}
  {2007})}\BibitemShut {NoStop}%
\bibitem [{\citenamefont {Abdi}\ \emph {et~al.}(2011)\citenamefont {Abdi},
  \citenamefont {Barzanjeh}, \citenamefont {Tombesi},\ and\ \citenamefont
  {Vitali}}]{Abdi.2011}%
  \BibitemOpen
  \bibfield  {author} {\bibinfo {author} {\bibfnamefont {M.}~\bibnamefont
  {Abdi}}, \bibinfo {author} {\bibfnamefont {S.}~\bibnamefont {Barzanjeh}},
  \bibinfo {author} {\bibfnamefont {P.}~\bibnamefont {Tombesi}},\ and\ \bibinfo
  {author} {\bibfnamefont {D.}~\bibnamefont {Vitali}},\ }\bibfield  {title}
  {\bibinfo {title} {{Effect of phase noise on the generation of stationary
  entanglement in cavity optomechanics}},\ }\href
  {https://doi.org/10.1103/physreva.84.032325} {\bibfield  {journal} {\bibinfo
  {journal} {Physical Review A}\ }\textbf {\bibinfo {volume} {84}},\ \bibinfo
  {pages} {032325} (\bibinfo {year} {2011})}\BibitemShut {NoStop}%
\bibitem [{\citenamefont {Hofer}\ \emph {et~al.}(2011)\citenamefont {Hofer},
  \citenamefont {Wieczorek}, \citenamefont {Aspelmeyer},\ and\ \citenamefont
  {Hammerer}}]{Hofer.2011}%
  \BibitemOpen
  \bibfield  {author} {\bibinfo {author} {\bibfnamefont {S.~G.}\ \bibnamefont
  {Hofer}}, \bibinfo {author} {\bibfnamefont {W.}~\bibnamefont {Wieczorek}},
  \bibinfo {author} {\bibfnamefont {M.}~\bibnamefont {Aspelmeyer}},\ and\
  \bibinfo {author} {\bibfnamefont {K.}~\bibnamefont {Hammerer}},\ }\bibfield
  {title} {{\selectlanguage {English}\bibinfo {title} {{Quantum entanglement
  and teleportation in pulsed cavity optomechanics}}},\ }\href
  {https://doi.org/10.1103/physreva.84.052327} {\bibfield  {journal} {\bibinfo
  {journal} {Physical Review A}\ }\textbf {\bibinfo {volume} {84}},\ \bibinfo
  {pages} {052327} (\bibinfo {year} {2011})}\BibitemShut {NoStop}%
\bibitem [{\citenamefont {Vitali}\ \emph {et~al.}(2007)\citenamefont {Vitali},
  \citenamefont {Gigan}, \citenamefont {Ferreira}, \citenamefont {Böhm},
  \citenamefont {Tombesi}, \citenamefont {Guerreiro}, \citenamefont {Vedral},
  \citenamefont {Zeilinger},\ and\ \citenamefont
  {Aspelmeyer}}]{vitali_optomechanical_2007}%
  \BibitemOpen
  \bibfield  {author} {\bibinfo {author} {\bibfnamefont {D.}~\bibnamefont
  {Vitali}}, \bibinfo {author} {\bibfnamefont {S.}~\bibnamefont {Gigan}},
  \bibinfo {author} {\bibfnamefont {A.}~\bibnamefont {Ferreira}}, \bibinfo
  {author} {\bibfnamefont {H.~R.}\ \bibnamefont {Böhm}}, \bibinfo {author}
  {\bibfnamefont {P.}~\bibnamefont {Tombesi}}, \bibinfo {author} {\bibfnamefont
  {A.}~\bibnamefont {Guerreiro}}, \bibinfo {author} {\bibfnamefont
  {V.}~\bibnamefont {Vedral}}, \bibinfo {author} {\bibfnamefont
  {A.}~\bibnamefont {Zeilinger}},\ and\ \bibinfo {author} {\bibfnamefont
  {M.}~\bibnamefont {Aspelmeyer}},\ }\bibfield  {title} {\bibinfo {title}
  {Optomechanical {Entanglement} between a {Movable} {Mirror} and a {Cavity}
  {Field}},\ }\href {https://doi.org/10.1103/physrevlett.98.030405} {\bibfield
  {journal} {\bibinfo  {journal} {Physical Review Letters}\ }\textbf {\bibinfo
  {volume} {98}},\ \bibinfo {pages} {030405 } (\bibinfo {year}
  {2007})}\BibitemShut {NoStop}%
\bibitem [{\citenamefont {Gardiner}\ and\ \citenamefont
  {Zoller}(2004)}]{Gardiner.2004}%
  \BibitemOpen
  \bibfield  {author} {\bibinfo {author} {\bibfnamefont {C.~W.}\ \bibnamefont
  {Gardiner}}\ and\ \bibinfo {author} {\bibfnamefont {P.}~\bibnamefont
  {Zoller}},\ }\href@noop {} {{\selectlanguage {English}\emph {\bibinfo {title}
  {Quantum noise}}}},\ Springer\ (\bibinfo  {publisher} {Springer},\ \bibinfo
  {year} {2004})\BibitemShut {NoStop}%
\bibitem [{\citenamefont {Gardiner}(2004)}]{gardiner_handbook_2004}%
  \BibitemOpen
  \bibfield  {author} {\bibinfo {author} {\bibfnamefont {C.~W.}\ \bibnamefont
  {Gardiner}},\ }\href@noop {} {\emph {\bibinfo {title} {{Handbook of
  Stochastic Methods for Physics, Chemistry and the Natural Sciences}}}},\
  Springer Series in Synergetics\ (\bibinfo  {publisher} {Springer series in
  synergetics},\ \bibinfo {year} {2004})\BibitemShut {NoStop}%
\bibitem [{\citenamefont {Walls}\ and\ \citenamefont
  {Milburn}(2008)}]{Walls.2008}%
  \BibitemOpen
  \bibfield  {author} {\bibinfo {author} {\bibfnamefont {D.~F.}\ \bibnamefont
  {Walls}}\ and\ \bibinfo {author} {\bibfnamefont {G.~J.}\ \bibnamefont
  {Milburn}},\ }\href {https://doi.org/10.1007/978-3-540-28574-8} {\emph
  {\bibinfo {title} {{Quantum Optics}}}},\ Quantum Optics\ (\bibinfo
  {publisher} {Springer Berlin Heidelberg},\ \bibinfo {year}
  {2008})\BibitemShut {NoStop}%
\end{thebibliography}%


\begin{thebibliography}{16}%
\makeatletter
\providecommand \@ifxundefined [1]{%
 \@ifx{#1\undefined}
}%
\providecommand \@ifnum [1]{%
 \ifnum #1\expandafter \@firstoftwo
 \else \expandafter \@secondoftwo
 \fi
}%
\providecommand \@ifx [1]{%
 \ifx #1\expandafter \@firstoftwo
 \else \expandafter \@secondoftwo
 \fi
}%
\providecommand \natexlab [1]{#1}%
\providecommand \enquote  [1]{``#1''}%
\providecommand \bibnamefont  [1]{#1}%
\providecommand \bibfnamefont [1]{#1}%
\providecommand \citenamefont [1]{#1}%
\providecommand \href@noop [0]{\@secondoftwo}%
\providecommand \href [0]{\begingroup \@sanitize@url \@href}%
\providecommand \@href[1]{\@@startlink{#1}\@@href}%
\providecommand \@@href[1]{\endgroup#1\@@endlink}%
\providecommand \@sanitize@url [0]{\catcode `\\12\catcode `\$12\catcode
  `\&12\catcode `\#12\catcode `\^12\catcode `\_12\catcode `\%12\relax}%
\providecommand \@@startlink[1]{}%
\providecommand \@@endlink[0]{}%
\providecommand \url  [0]{\begingroup\@sanitize@url \@url }%
\providecommand \@url [1]{\endgroup\@href {#1}{\urlprefix }}%
\providecommand \urlprefix  [0]{URL }%
\providecommand \Eprint [0]{\href }%
\providecommand \doibase [0]{https://doi.org/}%
\providecommand \selectlanguage [0]{\@gobble}%
\providecommand \bibinfo  [0]{\@secondoftwo}%
\providecommand \bibfield  [0]{\@secondoftwo}%
\providecommand \translation [1]{[#1]}%
\providecommand \BibitemOpen [0]{}%
\providecommand \bibitemStop [0]{}%
\providecommand \bibitemNoStop [0]{.\EOS\space}%
\providecommand \EOS [0]{\spacefactor3000\relax}%
\providecommand \BibitemShut  [1]{\csname bibitem#1\endcsname}%
\let\auto@bib@innerbib\@empty
\bibitem [{\citenamefont {Safavi-Naeini}\ and\ \citenamefont
  {Painter}(2010)}]{Safavi-Naeini.2010}%
  \BibitemOpen
  \bibfield  {author} {\bibinfo {author} {\bibfnamefont {A.~H.}\ \bibnamefont
  {Safavi-Naeini}}\ and\ \bibinfo {author} {\bibfnamefont {O.}~\bibnamefont
  {Painter}},\ }\bibfield  {title} {\bibinfo {title} {{Design of optomechanical
  cavities and waveguides on a simultaneous bandgap phononic-photonic crystal
  slab}},\ }\href {https://doi.org/10.1364/oe.18.014926} {\bibfield  {journal}
  {\bibinfo  {journal} {Optics Express}\ }\textbf {\bibinfo {volume} {18}},\
  \bibinfo {pages} {14926} (\bibinfo {year} {2010})}\BibitemShut {NoStop}%
\bibitem [{\citenamefont {Painter}\ \emph {et~al.}(1999)\citenamefont
  {Painter}, \citenamefont {Lee}, \citenamefont {Scherer}, \citenamefont
  {Yariv}, \citenamefont {O'Brien}, \citenamefont {Dapkus},\ and\ \citenamefont
  {Kim}}]{Painter.1999}%
  \BibitemOpen
  \bibfield  {author} {\bibinfo {author} {\bibfnamefont {O.}~\bibnamefont
  {Painter}}, \bibinfo {author} {\bibfnamefont {R.~K.}\ \bibnamefont {Lee}},
  \bibinfo {author} {\bibfnamefont {A.}~\bibnamefont {Scherer}}, \bibinfo
  {author} {\bibfnamefont {A.}~\bibnamefont {Yariv}}, \bibinfo {author}
  {\bibfnamefont {J.~D.}\ \bibnamefont {O'Brien}}, \bibinfo {author}
  {\bibfnamefont {P.~D.}\ \bibnamefont {Dapkus}},\ and\ \bibinfo {author}
  {\bibfnamefont {I.}~\bibnamefont {Kim}},\ }\bibfield  {title} {\bibinfo
  {title} {{Two-Dimensional Photonic Band-Gap Defect Mode Laser}},\ }\href
  {https://doi.org/10.1126/science.284.5421.1819} {\bibfield  {journal}
  {\bibinfo  {journal} {Science}\ }\textbf {\bibinfo {volume} {284}},\ \bibinfo
  {pages} {1819} (\bibinfo {year} {1999})}\BibitemShut {NoStop}%
\bibitem [{\citenamefont {Vučković}\ \emph {et~al.}(2002)\citenamefont
  {Vučković}, \citenamefont {Lončar}, \citenamefont {Mabuchi},\ and\
  \citenamefont {Scherer}}]{Vučković.2002}%
  \BibitemOpen
  \bibfield  {author} {\bibinfo {author} {\bibfnamefont {J.}~\bibnamefont
  {Vučković}}, \bibinfo {author} {\bibfnamefont {M.}~\bibnamefont {Lončar}},
  \bibinfo {author} {\bibfnamefont {H.}~\bibnamefont {Mabuchi}},\ and\ \bibinfo
  {author} {\bibfnamefont {A.}~\bibnamefont {Scherer}},\ }\bibfield  {title}
  {\bibinfo {title} {{Design of photonic crystal microcavities for cavity
  QED}},\ }\href {https://doi.org/10.1103/physreve.65.016608} {\bibfield
  {journal} {\bibinfo  {journal} {Physical Review E}\ }\textbf {\bibinfo
  {volume} {65}},\ \bibinfo {pages} {016608} (\bibinfo {year}
  {2002})}\BibitemShut {NoStop}%
\bibitem [{\citenamefont {Vasseur}\ \emph {et~al.}(2007)\citenamefont
  {Vasseur}, \citenamefont {Hladky-Hennion}, \citenamefont {Djafari-Rouhani},
  \citenamefont {Duval}, \citenamefont {Dubus}, \citenamefont {Pennec},\ and\
  \citenamefont {Deymier}}]{Vasseur.2007}%
  \BibitemOpen
  \bibfield  {author} {\bibinfo {author} {\bibfnamefont {J.~O.}\ \bibnamefont
  {Vasseur}}, \bibinfo {author} {\bibfnamefont {A.-C.}\ \bibnamefont
  {Hladky-Hennion}}, \bibinfo {author} {\bibfnamefont {B.}~\bibnamefont
  {Djafari-Rouhani}}, \bibinfo {author} {\bibfnamefont {F.}~\bibnamefont
  {Duval}}, \bibinfo {author} {\bibfnamefont {B.}~\bibnamefont {Dubus}},
  \bibinfo {author} {\bibfnamefont {Y.}~\bibnamefont {Pennec}},\ and\ \bibinfo
  {author} {\bibfnamefont {P.~A.}\ \bibnamefont {Deymier}},\ }\bibfield
  {title} {\bibinfo {title} {Waveguiding in two-dimensional piezoelectric
  phononic crystal plates},\ }\href {https://doi.org/10.1063/1.2740352}
  {\bibfield  {journal} {\bibinfo  {journal} {Journal of Applied Physics}\
  }\textbf {\bibinfo {volume} {101}},\ \bibinfo {pages} {114904} (\bibinfo
  {year} {2007})}\BibitemShut {NoStop}%
\bibitem [{\citenamefont {III}\ and\ \citenamefont {El-Kady}(2009)}]{III.2009}%
  \BibitemOpen
  \bibfield  {author} {\bibinfo {author} {\bibfnamefont {R.~H.~O.}\
  \bibnamefont {III}}\ and\ \bibinfo {author} {\bibfnamefont {I.}~\bibnamefont
  {El-Kady}},\ }\bibfield  {title} {\bibinfo {title} {{Microfabricated phononic
  crystal devices and applications}},\ }\href
  {https://doi.org/10.1088/0957-0233/20/1/012002} {\bibfield  {journal}
  {\bibinfo  {journal} {Measurement Science and Technology}\ }\textbf {\bibinfo
  {volume} {20}},\ \bibinfo {pages} {012002} (\bibinfo {year}
  {2009})}\BibitemShut {NoStop}%
\bibitem [{\citenamefont {Ludwig}\ and\ \citenamefont
  {Marquardt}(2013)}]{Ludwig.2013}%
  \BibitemOpen
  \bibfield  {author} {\bibinfo {author} {\bibfnamefont {M.}~\bibnamefont
  {Ludwig}}\ and\ \bibinfo {author} {\bibfnamefont {F.}~\bibnamefont
  {Marquardt}},\ }\bibfield  {title} {{\selectlanguage {English}\bibinfo
  {title} {{Quantum Many-Body Dynamics in Optomechanical Arrays}}},\ }\href
  {https://doi.org/10.1103/physrevlett.111.073603} {\bibfield  {journal}
  {\bibinfo  {journal} {Physical Review Letters}\ }\textbf {\bibinfo {volume}
  {111}},\ \bibinfo {pages} {073603} (\bibinfo {year} {2013})}\BibitemShut
  {NoStop}%
\bibitem [{\citenamefont {Brzezicki}\ \emph {et~al.}(2023)\citenamefont
  {Brzezicki}, \citenamefont {Silveri}, \citenamefont {Płodzień},
  \citenamefont {Massel},\ and\ \citenamefont {Hyart}}]{Brzezicki.2023}%
  \BibitemOpen
  \bibfield  {author} {\bibinfo {author} {\bibfnamefont {W.}~\bibnamefont
  {Brzezicki}}, \bibinfo {author} {\bibfnamefont {M.}~\bibnamefont {Silveri}},
  \bibinfo {author} {\bibfnamefont {M.}~\bibnamefont {Płodzień}}, \bibinfo
  {author} {\bibfnamefont {F.}~\bibnamefont {Massel}},\ and\ \bibinfo {author}
  {\bibfnamefont {T.}~\bibnamefont {Hyart}},\ }\bibfield  {title} {\bibinfo
  {title} {{Non-Hermitian topological quantum states in a reservoir-engineered
  transmon chain}},\ }\href {https://doi.org/10.1103/physrevb.107.115146}
  {\bibfield  {journal} {\bibinfo  {journal} {Physical Review B}\ }\textbf
  {\bibinfo {volume} {107}},\ \bibinfo {pages} {115146} (\bibinfo {year}
  {2023})}\BibitemShut {NoStop}%
\bibitem [{\citenamefont {Aspelmeyer}\ \emph {et~al.}(2014)\citenamefont
  {Aspelmeyer}, \citenamefont {Kippenberg},\ and\ \citenamefont
  {Marquardt}}]{Aspelmeyer.2014}%
  \BibitemOpen
  \bibfield  {author} {\bibinfo {author} {\bibfnamefont {M.}~\bibnamefont
  {Aspelmeyer}}, \bibinfo {author} {\bibfnamefont {T.~J.}\ \bibnamefont
  {Kippenberg}},\ and\ \bibinfo {author} {\bibfnamefont {F.}~\bibnamefont
  {Marquardt}},\ }\bibfield  {title} {{\selectlanguage {English}\bibinfo
  {title} {{Cavity optomechanics}}},\ }\href
  {https://doi.org/10.1103/revmodphys.86.1391} {\bibfield  {journal} {\bibinfo
  {journal} {Reviews of Modern Physics}\ }\textbf {\bibinfo {volume} {86}},\
  \bibinfo {pages} {1391 } (\bibinfo {year} {2014})}\BibitemShut {NoStop}%
\bibitem [{\citenamefont {Gardiner}\ and\ \citenamefont
  {Zoller}(2004)}]{Gardiner.2004}%
  \BibitemOpen
  \bibfield  {author} {\bibinfo {author} {\bibfnamefont {C.~W.}\ \bibnamefont
  {Gardiner}}\ and\ \bibinfo {author} {\bibfnamefont {P.}~\bibnamefont
  {Zoller}},\ }\href@noop {} {{\selectlanguage {English}\emph {\bibinfo {title}
  {Quantum noise}}}},\ Springer\ (\bibinfo  {publisher} {Springer},\ \bibinfo
  {year} {2004})\BibitemShut {NoStop}%
\bibitem [{\citenamefont {Gardiner}(2004)}]{gardiner_handbook_2004}%
  \BibitemOpen
  \bibfield  {author} {\bibinfo {author} {\bibfnamefont {C.~W.}\ \bibnamefont
  {Gardiner}},\ }\href@noop {} {\emph {\bibinfo {title} {{Handbook of
  Stochastic Methods for Physics, Chemistry and the Natural Sciences}}}},\
  Springer Series in Synergetics\ (\bibinfo  {publisher} {Springer series in
  synergetics},\ \bibinfo {year} {2004})\BibitemShut {NoStop}%
\bibitem [{\citenamefont {Adesso}\ and\ \citenamefont
  {Illuminati}(2007)}]{Adesso.2007}%
  \BibitemOpen
  \bibfield  {author} {\bibinfo {author} {\bibfnamefont {G.}~\bibnamefont
  {Adesso}}\ and\ \bibinfo {author} {\bibfnamefont {F.}~\bibnamefont
  {Illuminati}},\ }\bibfield  {title} {{\selectlanguage {English}\bibinfo
  {title} {{Entanglement in continuous-variable systems: recent advances and
  current perspectives}}},\ }\href
  {https://doi.org/10.1088/1751-8113/40/28/s01} {\bibfield  {journal} {\bibinfo
   {journal} {Journal of Physics A: Mathematical and Theoretical}\ }\textbf
  {\bibinfo {volume} {40}},\ \bibinfo {pages} {7821 } (\bibinfo {year}
  {2007})}\BibitemShut {NoStop}%
\bibitem [{\citenamefont {Brzezicki}\ and\ \citenamefont
  {Hyart}(2019)}]{Brzezicki.2019}%
  \BibitemOpen
  \bibfield  {author} {\bibinfo {author} {\bibfnamefont {W.}~\bibnamefont
  {Brzezicki}}\ and\ \bibinfo {author} {\bibfnamefont {T.}~\bibnamefont
  {Hyart}},\ }\bibfield  {title} {\bibinfo {title} {{Hidden Chern number in
  one-dimensional non-Hermitian chiral-symmetric systems}},\ }\href
  {https://doi.org/10.1103/physrevb.100.161105} {\bibfield  {journal} {\bibinfo
   {journal} {Physical Review B}\ }\textbf {\bibinfo {volume} {100}},\ \bibinfo
  {pages} {161105} (\bibinfo {year} {2019})}\BibitemShut {NoStop}%
\bibitem [{\citenamefont {Youssefi}\ \emph {et~al.}(2022)\citenamefont
  {Youssefi}, \citenamefont {Kono}, \citenamefont {Bancora}, \citenamefont
  {Chegnizadeh}, \citenamefont {Pan}, \citenamefont {Vovk},\ and\ \citenamefont
  {Kippenberg}}]{Youssefi.2022}%
  \BibitemOpen
  \bibfield  {author} {\bibinfo {author} {\bibfnamefont {A.}~\bibnamefont
  {Youssefi}}, \bibinfo {author} {\bibfnamefont {S.}~\bibnamefont {Kono}},
  \bibinfo {author} {\bibfnamefont {A.}~\bibnamefont {Bancora}}, \bibinfo
  {author} {\bibfnamefont {M.}~\bibnamefont {Chegnizadeh}}, \bibinfo {author}
  {\bibfnamefont {J.}~\bibnamefont {Pan}}, \bibinfo {author} {\bibfnamefont
  {T.}~\bibnamefont {Vovk}},\ and\ \bibinfo {author} {\bibfnamefont {T.~J.}\
  \bibnamefont {Kippenberg}},\ }\bibfield  {title} {\bibinfo {title}
  {{Topological lattices realized in superconducting circuit optomechanics}},\
  }\href {https://doi.org/10.1038/s41586-022-05367-9} {\bibfield  {journal}
  {\bibinfo  {journal} {Nature}\ }\textbf {\bibinfo {volume} {612}},\ \bibinfo
  {pages} {666} (\bibinfo {year} {2022})}\BibitemShut {NoStop}%
\bibitem [{\citenamefont {Prosen}\ and\ \citenamefont
  {Seligman}(2010)}]{prosen_quantization_2010}%
  \BibitemOpen
  \bibfield  {author} {\bibinfo {author} {\bibfnamefont {T.}~\bibnamefont
  {Prosen}}\ and\ \bibinfo {author} {\bibfnamefont {T.~H.}\ \bibnamefont
  {Seligman}},\ }\bibfield  {title} {\bibinfo {title} {Quantization over boson
  operator spaces},\ }\href {https://doi.org/10.1088/1751-8113/43/39/392004}
  {\bibfield  {journal} {\bibinfo  {journal} {Journal of Physics A:
  Mathematical and Theoretical}\ }\textbf {\bibinfo {volume} {43}},\ \bibinfo
  {pages} {392004} (\bibinfo {year} {2010})}\BibitemShut {NoStop}%
\bibitem [{\citenamefont {Vitali}\ \emph {et~al.}(2007)\citenamefont {Vitali},
  \citenamefont {Gigan}, \citenamefont {Ferreira}, \citenamefont {Böhm},
  \citenamefont {Tombesi}, \citenamefont {Guerreiro}, \citenamefont {Vedral},
  \citenamefont {Zeilinger},\ and\ \citenamefont
  {Aspelmeyer}}]{vitali_optomechanical_2007}%
  \BibitemOpen
  \bibfield  {author} {\bibinfo {author} {\bibfnamefont {D.}~\bibnamefont
  {Vitali}}, \bibinfo {author} {\bibfnamefont {S.}~\bibnamefont {Gigan}},
  \bibinfo {author} {\bibfnamefont {A.}~\bibnamefont {Ferreira}}, \bibinfo
  {author} {\bibfnamefont {H.~R.}\ \bibnamefont {Böhm}}, \bibinfo {author}
  {\bibfnamefont {P.}~\bibnamefont {Tombesi}}, \bibinfo {author} {\bibfnamefont
  {A.}~\bibnamefont {Guerreiro}}, \bibinfo {author} {\bibfnamefont
  {V.}~\bibnamefont {Vedral}}, \bibinfo {author} {\bibfnamefont
  {A.}~\bibnamefont {Zeilinger}},\ and\ \bibinfo {author} {\bibfnamefont
  {M.}~\bibnamefont {Aspelmeyer}},\ }\bibfield  {title} {\bibinfo {title}
  {Optomechanical {Entanglement} between a {Movable} {Mirror} and a {Cavity}
  {Field}},\ }\href {https://doi.org/10.1103/physrevlett.98.030405} {\bibfield
  {journal} {\bibinfo  {journal} {Physical Review Letters}\ }\textbf {\bibinfo
  {volume} {98}},\ \bibinfo {pages} {030405 } (\bibinfo {year}
  {2007})}\BibitemShut {NoStop}%
\bibitem [{\citenamefont {Walls}\ and\ \citenamefont
  {Milburn}(2008)}]{Walls.2008}%
  \BibitemOpen
  \bibfield  {author} {\bibinfo {author} {\bibfnamefont {D.~F.}\ \bibnamefont
  {Walls}}\ and\ \bibinfo {author} {\bibfnamefont {G.~J.}\ \bibnamefont
  {Milburn}},\ }\href {https://doi.org/10.1007/978-3-540-28574-8} {\emph
  {\bibinfo {title} {{Quantum Optics}}}},\ Quantum Optics\ (\bibinfo
  {publisher} {Springer Berlin Heidelberg},\ \bibinfo {year}
  {2008})\BibitemShut {NoStop}%
\end{thebibliography}%

\end{document}


\title{Non-hermitian topology in optomechanics -- Supplemental material}

\author{Wojciech Brzezicki}
\affiliation{Institute of Theoretical Physics, Jagiellonian University, ulica S. \L{}ojasiewicza 11, PL-30348 Krak\'ow, Poland}
\affiliation{International Research Centre MagTop, Institute of
  Physics, Polish Academy of Sciences, Aleja Lotnik\'ow 32/46, PL-02668 Warsaw,
  Poland}

\author{Timo Hyart}
\affiliation{Computational Physics Laboratory, Physics Unit, Faculty of
  Engineering and Natural Sciences,
  Tampere University, FI-33014 Tampere, Finland}
 \affiliation{Department of Applied Physics, Aalto University, 00076 Aalto, Espoo, Finland}

\author{Francesco Massel}
\affiliation{Department of Science and Industry Systems, University of
  South-Eastern Norway, PO Box 235, Kongsberg, Norway}

\maketitle
\tableofcontents
\section{Linearized optomechanical Hamiltonian: derivation}
\label{sec:LinH}

We consider here a 1D lattice of optomechanical systems whose phononic degrees
of freedom are coupled through a hopping Hamiltonian. Namely
\begin{equation}
  \label{eq:1}
  H_{\rm opt}=  \sum_{\rm i=0}^{N} H_{\rm i} +H_J
\end{equation}
with
\begin{equation}
  \begin{aligned}
    \label{eq:2}
    H_{\rm i}&= \omega_{\rm c, i} a_{\rm i}^\dagger a_{\rm i} + \omega_{\rm m, i} b_{\rm i}^\dagger b_{\rm i} +    g_{\rm i} a_{\rm i}^\dagger a_{\rm i} \left(b_{\rm i}^\dagger +b_{\rm i} \right) \\
    H_J&=-J \sum_{\rm i=0}^{N} \left(b_{\rm i}^\dagger b_{\rm i+1} +\, {\rm h.c.}\right)
  \end{aligned}
\end{equation}
where, for each site $\rm i$ $a_{\rm i}$, $b_{\rm i}$ are the photonic and mechanical degrees of freedom, with resonant frequencies given by $\omega_{\rm c, i} $ and $\omega_{\rm m, i} $, respectively, and $J$ the hopping energy.
This setup could be realized, for instance, in an optomechanical crystal. The
experimental realization of optomechanical crystals was first discussed in
Ref.~\cite{Safavi-Naeini.2010}, building upon previous work in the context of
photonic~\cite{Painter.1999,Vučković.2002} and
phononic~\cite{Vasseur.2007,III.2009} crystals. While the concomitant presence of
hopping both for optical and mechanical degrees of freedom can be included in
our analysis, analogously to what was discussed in Ref.~\cite{Ludwig.2013}, we
focus here on the case of purely mechanical hopping.

In our analysis, we assume that all sites are identical ($\omega_{\rm c,
  i}=\omega_{\rm c}$, $\omega_{\rm m , i}=\omega_{\rm m}$, $g_{i}=g$). Along the
lines of \cite{Brzezicki.2023}, where the coupling to the environment of a 1D
transmon chain was modified to obtain a periodic modulation of the dissipation
properties, we consider here a spatial-dependent drive of the optomechanical
chain in a BRRB pattern. Each site of the chain is either driven on the blue
(``B'') or the red (``R'') sideband~\cite{Aspelmeyer.2014}.

The linearization of the equations of motion (EOMs) generated by $H_{\rm
  i}+H_{J}$, in the presence of an external drive modeled by $H_{\rm
  d}=\alpha_{\rm d} e^{-i \omega_{\rm d}t}$ can be performed with the same
strategy used to linearize the Hamiltonian of a single optomechanical system.
Focusing on site $\rm i$ the EOMs for $a_{\rm i}$ and $b_{\rm i}$ can be written as
\begin{subequations}
  \begin{align}
  \begin{split}
    \label{eq:3}
      \dot{a}_{\rm i}= - i \omega_{\rm c} a_{\rm i}  - i g_0 a_{\rm i}\left(b_{\rm i}^\dagger
                    - \frac{\kappa}{2} a_{\rm i} - \alpha_\mathrm{d} e^{-i \omega_{\rm d}t}
                    + b_{\rm i}\right)
                    +\sqrt{\kappa}\, a_{\rm in, i}
   \end{split}\\
   \begin{split}
     \label{eq:4}
      \dot{b}_{\rm i}= - i \omega_{\rm m} b_{\rm i} -i g_0  a^\dagger_{\rm i}a_{\rm i}
                      + i J (b_{\rm i+1}+ b_{\rm i-1})
                    - \frac{\gamma}{2} b_{\rm i} +\sqrt{\gamma}\, b_{\rm in, i}
   \end{split}
   \end{align}
\end{subequations}
Writing the optical and mechanical operators as a combination of a strong
classical tone imposed by the drive $\alpha_{\rm d}$ and of a quantum
fluctuation. Namely,
$$
a_{\rm i} \to \alpha_{\rm i} + a_{\rm i}, \ \ \ \\
b_{\rm i} \to \beta_{\rm i} + b_{\rm i}.
$$

We can solve Eqs.~(\ref{eq:3},\ref{eq:4}) order-by-order.
The zeroth order is given by
\begin{subequations}
  \begin{align}
  \begin{split}
  \label{eq:5}
 \dot{\alpha}_{\rm i}= i \Delta \alpha_{\rm i} -\frac{\kappa}{2} \alpha_{\rm i} - \alpha_{\rm d}
                       - i g_0 \alpha_{\rm d}\left(\beta_{\rm i}^* + \beta_{\rm i}\right)
  \end{split}\\
    \begin{split}
      \label{eq:6}
  \dot{\beta}_{\rm i}= -i \omega_{\rm m} \beta_{\rm i} -\frac{\kappa}{2} \beta_{\rm i}
                       - i g_0 \left|\alpha_{\rm d}\right|^2 + i J \left(\beta_{\rm i+1} + \beta_{\rm i-1}\right),
    \end{split}
  \end{align}
\end{subequations}
where we have moved to a frame rotating at the external drive frequency
$\omega_{\rm d}$ and $\Delta=\Omega_{\rm d}-\Omega_{\rm c}$. Since the drive
imposes a constant displacement to the cavity and the mechanical field (in the
frame rotating at $\omega_{\rm d}$), we have $\dot{\alpha}_{\rm i}=0$ and
$\dot{\beta}_{\rm i}=0$, allowing us to solve Eqs.~(\ref{eq:5},\ref{eq:6}) as a
set of coupled algebraic equations. We note here that, if $g_0$ (and the
mechanical displacement) are sufficiently
small, namely $g_0 \left(\beta_{\rm i}^* + \beta_{\rm i}\right) \ll \Delta$, the
equations for the cavity field can be solved independently from the mechanical
degrees of freedom.

The first-order equations can be written as
\begin{subequations}
  \begin{align}
  \begin{split}
    \label{eq:7}
      \dot{a}_{\rm i}=  - \frac{\kappa}{2} a_{\rm i}
                       - i g_0 \alpha_{\rm i} e^{-i \bar{\Delta}t} \left( e^{i \omega_{\rm m}t} b_{\rm i}^\dagger + e^{-i \omega_{\rm m}t} b_{\rm i}\right)
                    +\sqrt{\kappa}\, a_{\rm in, i}
   \end{split}\\
   \begin{split}
     \label{eq:8}
     \dot{b}_{\rm i}= - \frac{\gamma}{2} b_{\rm i}
                     - i g_0 e^{i \omega_{\rm m}t} \left(\alpha_i e^{-i \bar{\Delta}t} a^\dagger + \alpha_{\rm i}^* e^{i \bar{\Delta}t} a_{\rm i}\right)
                      + i J \left(b_{\rm i+1}+ b_{\rm i-1} \right)
                      +\sqrt{\gamma}\, b_{\rm in, i}
   \end{split}
   \end{align}
\end{subequations}
where we have now moved to a frame rotating at the cavity frequency $\omega_{\rm
  c}$ and at the mechanical frequency $\omega_{\rm m}$ for $a_{\rm i}$ and
$b_{\rm i}$, respectively. We now consider two specific values for the driving
frequency $\omega_{\rm d}$. The first one is such as to fulfill the condition $
\bar{\Delta}= -\omega_{\rm m}$ (red-detuned drive). Invoking the rotating wave
approximation, we are allowed to neglect in Eqs.~(\ref{eq:7},\ref{eq:8}) fast
oscillating terms that oscillate at $\bar{\Delta} - \omega_{\rm m}\simeq 2
\omega_{\rm m}$. With this approximation, Eqs.~(\ref{eq:7},\ref{eq:8}) become
\begin{subequations}
  \begin{align}
  \begin{split}
    \label{eq:9}
      \dot{a}_{\rm i}=  - \frac{\kappa}{2} a_{\rm i}
                       - i G_-  b_{\rm i} +\sqrt{\kappa}\, a_{\rm in, i}
   \end{split}\\
   \begin{split}
     \label{eq:10}
     \dot{b}_{\rm i}= - \frac{\gamma}{2} b_{\rm i}
                     - i G_- a_{\rm i}
                      + i J \left(b_{\rm i+1}+ b_{\rm i-1} \right)
                      +\sqrt{\gamma}\, b_{\rm in, i}
   \end{split}
   \end{align}
\end{subequations}
where and we have used the fact that $\bar{\Delta} + \omega_{\rm m}=0$ and
assumed, without loss of generality that $G_-= g_0 \alpha_i$. Conversely,
choosing $\omega_{\rm d}$, in such a way that $\bar{\Delta}= \omega_{\rm m}$
(blue-detuned drive), leads to the following EOMs
\begin{subequations}
  \begin{align}
  \begin{split}
    \label{eq:11}
      \dot{a}_{\rm i}=  - \frac{\kappa}{2} a_{\rm i}
                       - i G_+  b^\dagger_{\rm i} +\sqrt{\kappa}\, a_{\rm in, i}
   \end{split}\\
   \begin{split}
     \label{eq:12}
     \dot{b}_{\rm i}= - \frac{\gamma}{2} b_{\rm i}
                     - i G_+ a^\dagger_{\rm i}
                      + i J \left(b_{\rm i+1}+ b_{\rm i-1} \right)
                      +\sqrt{\gamma}\, b_{\rm in, i}.
   \end{split}
   \end{align}
\end{subequations}
The EOMs given in Eqs.~(\ref{eq:9}-\ref{eq:12}) can be thought of as originating from
the following bilinear Hamlitonians (in the laboratory frame)
\begin{equation}
  \begin{aligned}
    \label{eq:13}
    H_{\rm i, R} = \omega_{\rm c} a_{\rm i}^\dagger a_{\rm i} + \omega_{\rm m} b_{\rm i}^\dagger b_{\rm i} + G_{\rm -} a_{\rm i}^\dagger b_{\rm i} +J (b^\dagger_{i} b_{i+1} +b^\dagger_{i} b_{i-1} ) \, {\rm h.c.}
    \\
    H_{\rm i, B} = \omega_{\rm c} a_{\rm i}^\dagger a_{\rm i} + \omega_{\rm m} b_{\rm i}^\dagger b_{\rm i} + G_{\rm +} a_{\rm i}^\dagger b^\dagger_{\rm i} +J (b^\dagger_{i} b_{i+1} +b^\dagger_{i} b_{i-1} )  + \, {\rm h.c.}
  \end{aligned}
\end{equation}
depending on whether site $\rm i$ is driven on the red or the blue sideband,
respectively. These two Hamiltonians are the building blocks for the Hamiltonian
given in Eq.(1) of the main text.

\section{Dynamical system equations}
\label{sec:DynS}

We consider here the dynamics induced by the Hamiltonian~\eqref{eq:4} in the
presence of a coupling to thermal reservoirs both for the mechanical and the
optical degrees of freedom, introducing dissipation rates $\gamma$ and $\kappa$
and noise operators $b_{\rm in,i}$ and $a_{\rm in,i}$ for the mechanical and
optical degrees of freedom, respectively. The operators $b_{\rm in,i}$ fulfill the
following relations
\begin{subequations}
  \begin{align}
    \label{eq:14}
    & \langle b_{\rm in,i}(t) b^\dagger_{\rm in,i}(t') \rangle = \left(n_{\rm m}+1\right) \delta(t-t') \\
    \label{eq:15}
    & \langle b^\dagger_{\rm in,i}(t) b_{\rm in,i}(t') \rangle = n_{\rm m}\delta(t-t')
  \end{align}
\end{subequations}
 and analogous relations hold for $a_{\rm in,i}$
\begin{subequations}
  \begin{align}
    \begin{split}
      \label{eq:16}
      \langle a_{\rm in,i}(t) a^\dagger_{\rm in,i}(t') \rangle = \left(n_{\rm c}+1\right) \delta(t-t')
    \end{split} \\
    \begin{split}
      \label{eq:17}
      \langle a^\dagger_{\rm in,i}(t) a_{\rm in,i}(t') \rangle = n_{\rm c}\delta(t-t').
    \end{split}
  \end{align}
\end{subequations}
The Hamiltonian given Eq.~(1) of the main text generates the following dynamics,
described by the quantum Langevin equations (QLEs)~\cite{Gardiner.2004}, for
the optical and mechanical degrees of freedom
\begin{subequations}
  \begin{align}
    \begin{split}
      \label{eq:18}
      \dot{a}_{\rm i} &= - i \left[a_{\rm i}, \hat{H} \right] - \frac{\kappa}{2} a_{\rm i} +\sqrt{\kappa} a_{\rm i, in}
    \end{split}\\
    \begin{split}
      \label{eq:19}
      \dot{b}_{\rm i} &= - i \left[b_{\rm i}, \hat{H} \right] - \frac{\gamma}{2} b_{\rm i} +\sqrt{\gamma} b_{\rm i, in}.
    \end{split}
  \end{align}
\end{subequations}
Eqs.(~\ref{eq:18},\ref{eq:19}) can be written more compactly in terms of a
operators vector
$\mathbf{a}=\left[a_1,b_1,\dots,a_{4N},b_{4N};a^\dagger_1,b^\dagger_1,\dots
  a^\dagger_{4N},b^\dagger_{4N} \right]^T$ and an input-operators vector
$\mathbf{a}_{\rm in }=\left[a_{\rm in\, 1},b_{\rm in\, 1},\dots, a_{\rm in\, 4N},b_{\rm in\, 4N};a^\dagger_{\rm in\, 1},b^\dagger_{\rm in\, 1}\dots,\dots,
  a^\dagger_{\rm in\, 4N},b^\dagger_{\rm in\, 4N} \right]^T$
\begin{equation}
  \label{eq:20}
  \dot{\mathbf{a}} = -i \mathcal{H}_{\rm NH} \mathbf{a} + \sqrt{\bar{\eta}} \mathbf{a}_{\rm in}
\end{equation}
with $\bar{\eta}={\rm diag}
\left(\kappa,\gamma,\dots;\kappa,\gamma,\dots\kappa,\gamma\right)$,
and
 \begin{equation}
    \label{eq:21}
     \mathcal{H}_{\rm NH} = \bar{\sigma}_z \mathcal{H} -  \frac{i \bar{\eta}}{2}
  \end{equation}
where $\bar{\sigma}_z ={\rm diag}\left(1,\dots,1;-1,\dots,-1\right)$. The
solution of Eq.~\eqref{eq:20} can be written as
\begin{equation}
  \label{eq:22}
  \mathbf{a} = \exp \left[ -i \mathcal{H}_{\rm NH} t \right] \mathbf{a}_0 + \int_0 ^t
               \exp\left[ -i \mathcal{H}_{\rm NH} (t-\tau) \right]  \sqrt{\bar{\eta}} \mathbf{a}_{\rm in}d \tau,
\end{equation}
which corresponds to Eq.~(2) of the main text.

\section{Stability, covariance matrix, entanglement}
\label{sec:StabCovNeg}

The first aspect in the study of the (quantum) dynamical system described by
Eq.~\eqref{eq:20}, whose solution is given by Eq.~\eqref{eq:22} (Eq.~(2) of the
main text) is its stability. Eq.~\eqref{eq:20} can be construed as a
quantum stochastic differential equation~\cite{Gardiner.2004} (quantum
version of a Ornstein-Uhlenbeck process~\cite{gardiner_handbook_2004}), which can be
written, in general, as
\begin{equation}
  \label{eq:23}
  \dot{\bar{a}} = A \bar{a}  + \bar{\xi}_{\rm in}.
\end{equation}
Its stability is guaranteed by the condition ${\rm Re}\, \lambda_{\rm i}<0$
where $\lambda_{\rm i}$ are the eigenvalues of the matrix associated with the
dynamical system, in our case $A= - i \mathcal{H}_{\rm NH}$. This condition
corresponds to the ${\rm Im}\, E <0$ condition on the eigenenergies of the
non-Hermitian Hamiltonian $\mathcal{H}_{\rm NH}$.

From Eq.~\eqref{eq:23}, it is possible to write the Lyapunov  equation for the
time evolution of the covariance matrix $V$ for the operators
$\left[a_1,b_1,a_1^\dagger, b_1^\dagger,\dots \right]$, namely
\begin{equation}
  \label{eq:24}
  \frac{d V_{\bar{a},\bar{a}^\dagger}}{dt}=A V_{\bar{a},\bar{a}^\dagger} + V_{\bar{a},\bar{a}^\dagger} A^\dagger + D
\end{equation}
where
\begin{equation*}
  V_{\bar{a},\bar{a}^\dagger}=
  \begin{bmatrix}
    \langle a_1 a_1^\dagger \rangle & \langle a_1 b_1^\dagger \rangle & \langle a_1 a_1\rangle & \langle a_1 b_1 \rangle & \cdots \\
    \langle b_1 a_1^\dagger \rangle & \ddots &  & \\
    \vdots & & & \\
  \end{bmatrix}
\end{equation*}
and
$D={\rm diag} \left[ \kappa \left(n_a+1\right),\gamma \left(n_b+1 \right),\kappa n_a,\gamma n_b,\dots \right]$.
In our analysis, we are interested in the stationary value of the covariance
matrix. We therefore seek a numerical solution to the equation
\begin{equation}
  \label{eq:25}
  A V_{\bar{a},\bar{a}^\dagger} + V_{\bar{a},\bar{a}^\dagger} A^\dagger + D = 0.
\end{equation}
From $V_{\bar{a},\bar{a}^\dagger}$ it is possible to obtain the covariance
matrix between the quadratures --either mechanical
$$
\begin{aligned}
& X_{\rm m, i}=1/\sqrt{2}\left( b_{\rm i}^\dagger +b_{\rm i}\right),\\
& P_{\rm m, i}=i/\sqrt{2}\left( b_{\rm i}^\dagger -b_{\rm i}\right)
  \end{aligned}
$$
or optical
$$
\begin{aligned}
&X_{\rm o, i}=1/\sqrt{2}\left( a_{\rm i}^\dagger +a_{\rm i}\right),\\
&P_{\rm o, i}=i/\sqrt{2}\left( a_{\rm i}^\dagger -a_{\rm i}\right).
\end{aligned}
$$
We evaluate the entanglement properties between the quadratures $\xi,\eta = {\rm
  m,c}$ on sites $\rm i, j$, considering the two mode covariance matrix
\begin{equation}
  \label{eq:26}
  V_{\rm \xi,i, \eta, j} =
  \begin{bmatrix}
       \langle X_{\rm \xi,i} X_{\rm \xi, i} \rangle
    &  \langle X_{\rm \xi,i} P_{\rm \xi, i} \rangle
    &  \langle X_{\rm \xi,i} X_{\rm \eta, j} \rangle
    &  \langle X_{\rm \xi,i} P_{\rm \eta, j} \rangle \\
       \langle P_{\rm \xi,i} X_{\rm \xi, i} \rangle
    &  \langle P_{\rm \xi,i} P_{\rm \xi, i} \rangle
    &  \langle P_{\rm \xi,i} X_{\rm \eta, j} \rangle
    &  \langle P_{\rm \xi,i} P_{\rm \eta, j} \rangle \\
       \langle X_{\rm \eta,i} X_{\rm \xi, i} \rangle
    &  \langle X_{\rm \eta,i} P_{\rm \xi, i} \rangle
    &  \langle X_{\rm \eta,i} X_{\rm \eta, j} \rangle
    &  \langle X_{\rm \eta,i} P_{\rm \eta, j} \rangle \\
       \langle P_{\rm \eta,i} X_{\rm \xi, i} \rangle
    &  \langle P_{\rm \eta,i} P_{\rm \xi, i} \rangle
    &  \langle P_{\rm \eta,i} X_{\rm \eta, j} \rangle
    &  \langle P_{\rm \eta,i} P_{\rm \eta, j} \rangle
  \end{bmatrix}.
\end{equation}
Eq.~\eqref{eq:26} can be written in terms of three $2 \times 2$ matrices
$\alpha$, $\beta$, $\gamma$ as
\begin{equation}
  \label{eq:27}
  V_{\rm i, j} =
  \left[
    \begin{array} {@{}c|c@{}}
      \alpha_{\rm i, j}  & \gamma_{\rm i, j}    \\
      \hline
      \gamma_{\rm i, j}^{T}  & \beta_{\rm i, j}
    \end{array}
  \right]
\end{equation}
Following~\cite{Adesso.2007}, we can determine the negativity associated with
the entanglement (or lack thereof).

The positivity of the partial transposed state (PPT criterion) can be recast as
the  condition $\tilde{\nu}_- \geq 1$ on the smallest symplectic eigenvalue  of
the covariance matrix evaluated on the partially transposed state. These can be expressed in
terms of Eq.~\eqref{eq:27} as
\begin{equation}
  \label{eq:28}
  \tilde{\nu}_\pm = \sqrt{
    \frac{\tilde{\Delta}_{\rm i, j}\pm \sqrt{\tilde{\Delta}_{\rm i, j}-4 {\,\rm Det \,}V_{\rm i,j}}
         }{2}
  }
\end{equation}
where
$\tilde{\Delta}_{\rm i, j}={\rm Det\,} \alpha_{\rm i, j} +
{\rm Det\,} \beta_{\rm i, j} - 2 {\,\rm Det\,} \gamma_{\rm i, j}$.

The (logarithmic) negativity
\begin{equation}
  \label{eq:29}
  E_\mathcal{N} = {\rm max}\left[0, -\log_2 \tilde{\nu}_- \right]
\end{equation}
gives a measure of the violation of the condition $\tilde{\nu}_- \geq 1$, and
thus quantifies the entanglement between the modes localized on sites $\rm i$
and $\rm j$. Furthermore, the diagonal elements of $V_{\rm \xi,i,\eta, j}$ allow
us to infer the number of mechanical excitations on sites $\rm i$ and $\rm j$.

\pagebreak
\section{PBC and OBC spectra}
\label{sec:spectra}

\begin{figure}[H]
  \includegraphics[width=0.95\columnwidth]{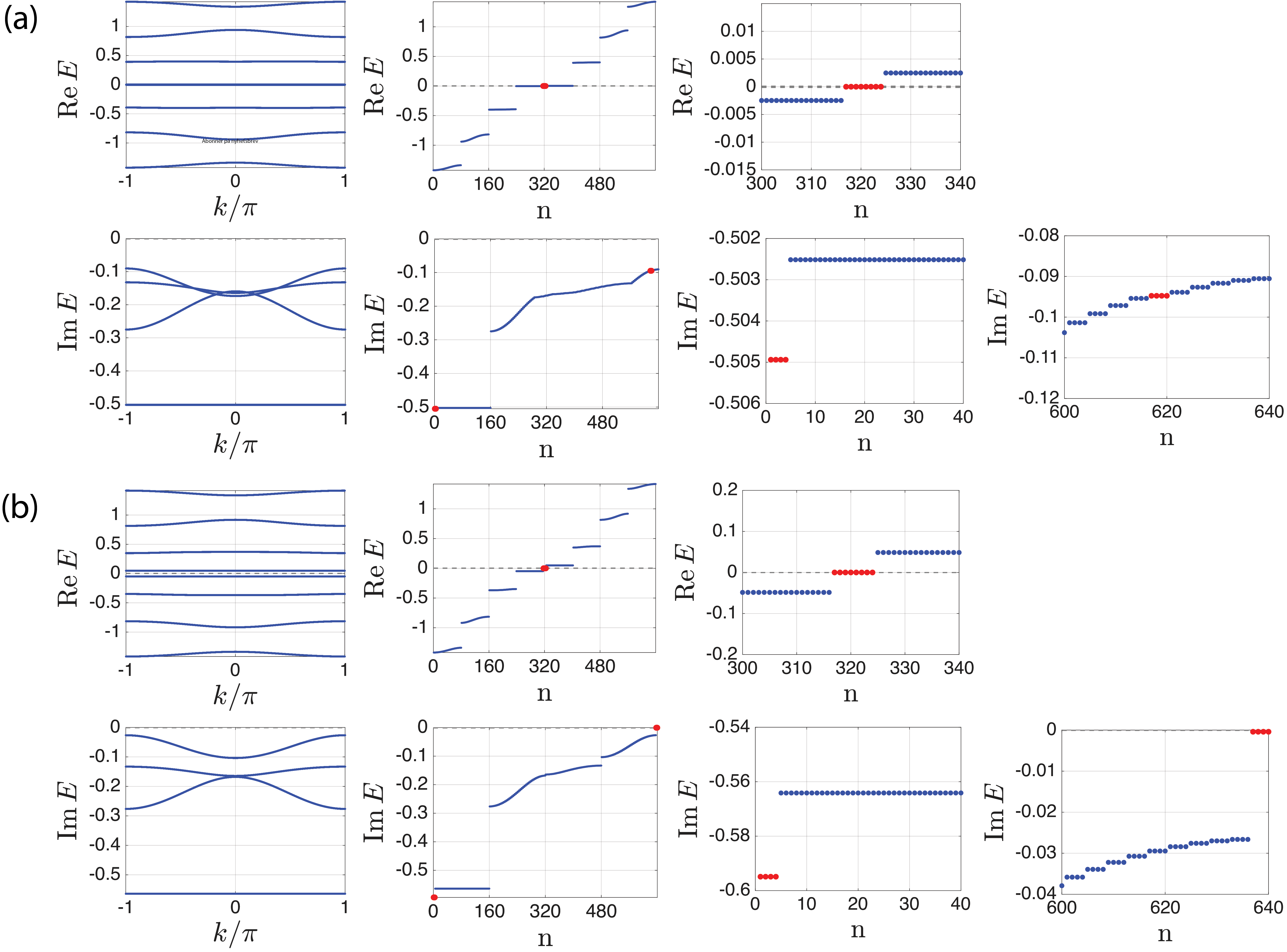}
  \caption{Real and imaginary spectra for periodic (first columm) and open
    (second column) boundary conditions (columns three and four represent a
    zoom-in of column two). Red dots correspond to ${\rm Im}\, E_{\rm e}<0$, i.e.
    (stable) end states, blue dots to ${\rm Im} E_{\rm b}$, i.e. bulk states.
    (a) $G_+=0.05$, (b) $G_+=0.242$. $G_-=1,  J=0.5$, $\gamma = 10^{-4}$ (all
    energies expressed in units of $\kappa$). For $G_+=0.242$, ${\rm Im}\, E_{\rm e}\simeq 4 \cdot  10^{-4}$. }
  \label{fig:1}
\end{figure}

We depict here the spectra for periodic and open boundary conditions for
different values of $G_+$. In Fig.~\ref{fig:1}, we consider the
case where both end states are stable. Panel~\ref{fig:1}(a) illustrates the
case for which  ${\rm Im} E_{\rm e}$ is within the ${\rm Im} E_{\rm b}$ band ($G_+=0.2$),
whereas panel~\ref{fig:1}(b) refers to the case $G_+=0.446$, for which a
gap develops between  ${\rm Im} E_{\rm e} \to 0^-$ and ${\rm Im} E_{\rm b}< 0$ states.

\pagebreak

\begin{figure}[H]
  \includegraphics[width=0.9\columnwidth]{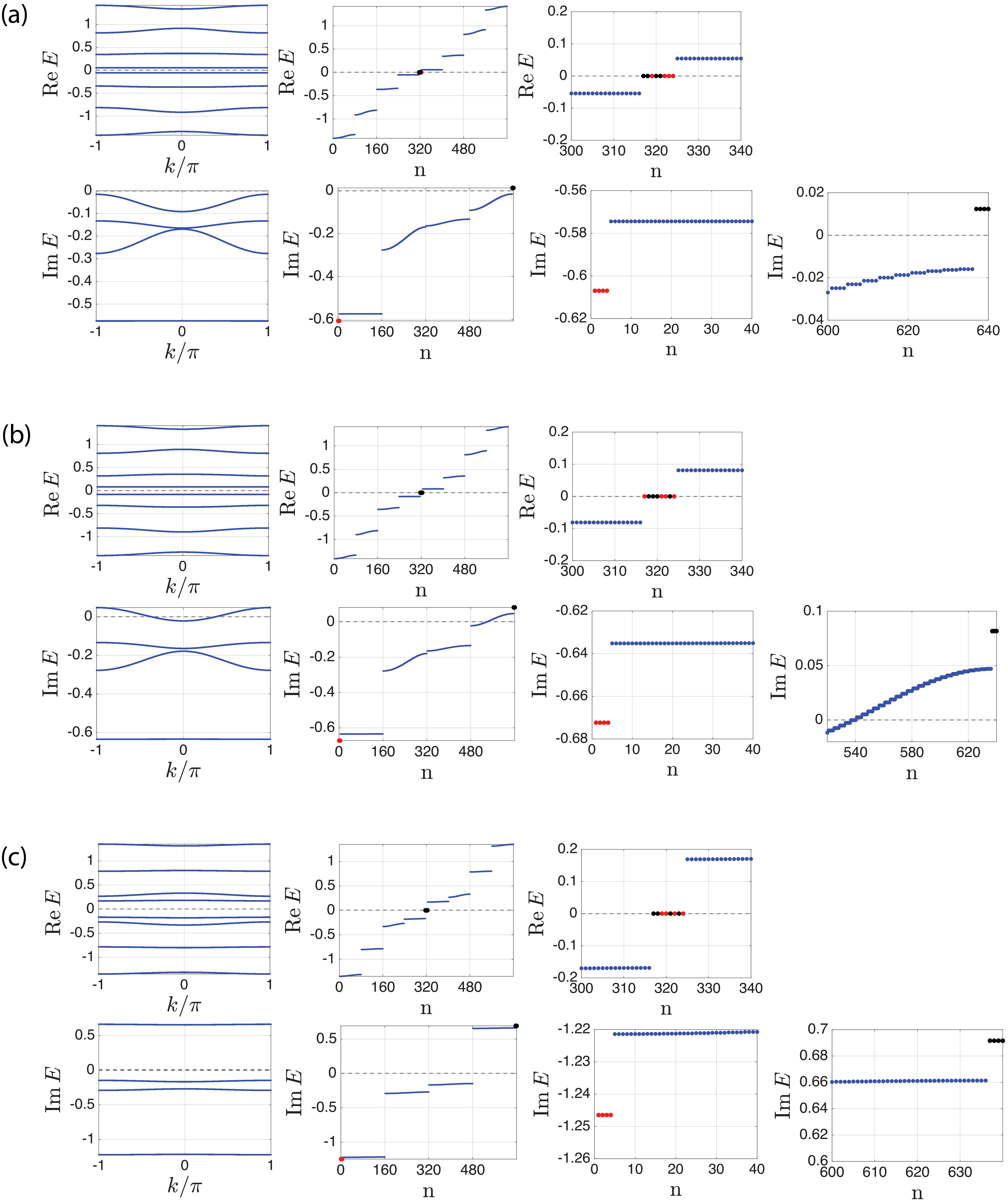}
  \caption{Real and imaginary spectra for periodic (first columm) and open
    (second column) boundary conditions (columns three and four represent a
    zoom-in of column two). Red dots correspond to ${\rm Im} E_{\rm e}<0$, i.e.
    stable end states, black dots correspond to ${\rm Im} E_{\rm e}>0$, i.e.
    unstable end states, blue dots to ${\rm Im} E_{\rm b}$, i.e. bulk states.
    (a) $G_+=0.35$, (b) $G_+=1$. $G_-=1, J=0.5$, $\gamma =
    10^{-4}$ (all energies expressed in units of $\kappa$).}
  \label{fig:2}
\end{figure}

In Fig.~\ref{fig:2}, we further increase $G_+$. At first ($G_+=0.26$, panel~\ref{fig:2}(a)) end states become
unstable, while bulk states are still stable. Further increasing $G_+$
($G_+=0.35$,$G_+=1$) leads the instability of the bulk states as well. The
case $G_+=0.35$ (panel \ref{fig:2}(b)) corresponds to the case for which both the
PBC and the OBC spectra are gapless at ${\rm Im} E=0$, while a gap develops for
$G_+=1$ (anel \ref{fig:2}(c))

\section{Non-Hermitian topology}
\label{sec:nhTopo}

In order to establish the topological character of the (complex) spectrum of the
chain, we rely on the analysis performed in Ref.~\cite{Brzezicki.2019}, where,
as outlined in the main text, the authors establish an exact mapping between the
complex spectrum at ${\rm Re}\, E=0$ of a non-hermitian Hamiltonian --in our
case $\mathcal{H}_{\rm NH}$-- and an effective hermitian Hamiltonian given by
\[
{\mathcal H}_{{\rm eff}}(\eta,k)=\eta S-iS{\mathcal H}_{{\rm NH}}(k).
\]

The definition of $S$ originates from the symmetry properties of
$\mathcal{H}_{\rm NH}$: it is possible to show that $\mathcal{H}_{\rm NH}$ obeys
a non-hermitian chiral symmetry
\begin{equation}
  \label{eq:30}
  S{\cal H}_{{\rm NH}}S^{-1}=-{\cal H}_{{\rm NH}}^{\dagger}
\end{equation}
where $S=\Pi\, \Sigma$,
\[
  \Sigma=\begin{pmatrix}\sigma & 0 & 0 & 0\\
    0 & \sigma & 0 & 0\\
    0 & 0 & \sigma & 0\\
    0 & 0 & 0 & \sigma
  \end{pmatrix},\quad\Pi=\begin{pmatrix}\sigma & 0 & 0 & 0\\
    0 & \tau & 0 & 0\\
    0 & 0 & -\tau & 0\\
    0 & 0 & 0 & -\sigma
  \end{pmatrix},
\]
$\sigma=\sigma_z \otimes \mathbb{I}_2$,$\tau=\mathbb{I}_2 \otimes
\sigma_z$, and $\sigma_z$ is the 2x2  Pauli matrix.

We focus on the eigenstates $\psi$ of $\mathcal{H}_{\rm NH}$ with ${\rm Re} =0$ (for
open boundary conditions). The states $\psi$ must therefore satisfy the
following relation
\begin{equation}
  \label{eq:31}
   \mathcal{H}_{\rm NH} \psi = i \eta \, \psi.
\end{equation}
From the symmetry properties expressed in Eq.~\eqref{eq:30}, it is possible to
show that Eq.~\eqref{eq:31} can be recast as a zero-energy eigenproblem for the
\emph{hermitian} Hamiltonian $\mathcal{H}_{\rm eff}$. Namely
\begin{equation}
  \label{eq:32}
  \mathcal{H}_{\rm eff}(\psi,k) \psi \doteq
  S \left(\eta - i \mathcal{H}_{\rm NH}\right) \psi =0.
\end{equation}
The potentially nontrivial topological properties of $\mathcal{H}_{\rm
  eff}(k,\eta)$ are reflected into the topological protection of the ${\rm
  Re}=0$ (end) modes of  $\mathcal{H}_{\rm NH}$. To ascertain the topological
protection of the end modes, following  \cite{Brzezicki.2023}, we can
evaluate the Berry curvature and the Chern number associated with ${\mathcal
  H}_{{\rm eff}}$ as
\begin{equation}
  \label{eq:3}
  \Omega_{k,\eta} = \sum_{{n\leq 8\atop m>8}} {\rm Im}
             \frac{
               2 \left[\psi_{k,\eta}^{n\,*} \, \left(\partial_{k}\mathcal{H}^{{\rm eff}}\right)\, \psi_{k,\eta}^{m}\right]
                 \left[\psi_{k,\eta}^{m\,*}\,\left(\partial_{\eta}\mathcal{H}^{{\rm eff}}\right)\, \psi_{k,\eta}^{n}\right]}
                    {\left(E_{k,\eta}^{(n)}-E_{k,\eta}^{(m)}\right)^{2}}
\end{equation}
and
\begin{equation}
  \label{eq:4}
C=\frac{1}{2 \pi}\int_{-\infty}^{+\infty}d\eta\int_{0}^{2\pi}dk \ \Omega_{k,\eta},
\end{equation}
where the eigenstates are sorted in ascending order of the eigenenergy.
In order to guarantee the
quantization of the Chern number for $\mathcal{H}_{\rm eff}(k,\eta)$,
one has to
require its periodicity in $k$ and $\eta$. While the periodicity of the former
is guaranteed by the periodicity of  $\mathcal{H}_{\rm NH}(k)$,
$\mathcal{H}_{\rm eff}(k,\eta)$ is not periodic in $\eta$.  This issue can be
solved by defining a compactified version of $\mathcal{H}_{\rm eff}(k,\eta)$,
sharing with it the same spectrum,
through the relation
\begin{equation}
  \label{eq:33}
  \mathcal{H}_{\rm eff}^{cmp}(k,\eta)=
          \mathcal{R}_\eta \mathcal{H}_{\rm eff}(k,\eta)  \mathcal{R}_\eta^\dagger
\end{equation}
with
\begin{equation}
  \label{eq:34}
  \mathcal{R}_\eta = \exp \left[ i \frac{\pi}{4} \left(1+\tanh \eta\right) \mathcal{G}\right],
  \qquad \mathcal{G}=
  \begin{pmatrix}
    & & & \mathbb{I}_4 \\
    & & \mathbb{I}_4 & \\
    & \mathbb{I}_4 & & \\
    \mathbb{I}_4 & & & \\
  \end{pmatrix}
\end{equation}.

\pagebreak

\section{Two-site model}
\label{sec:2sites}

\begin{figure}[H]
  \includegraphics[width=0.9\columnwidth]{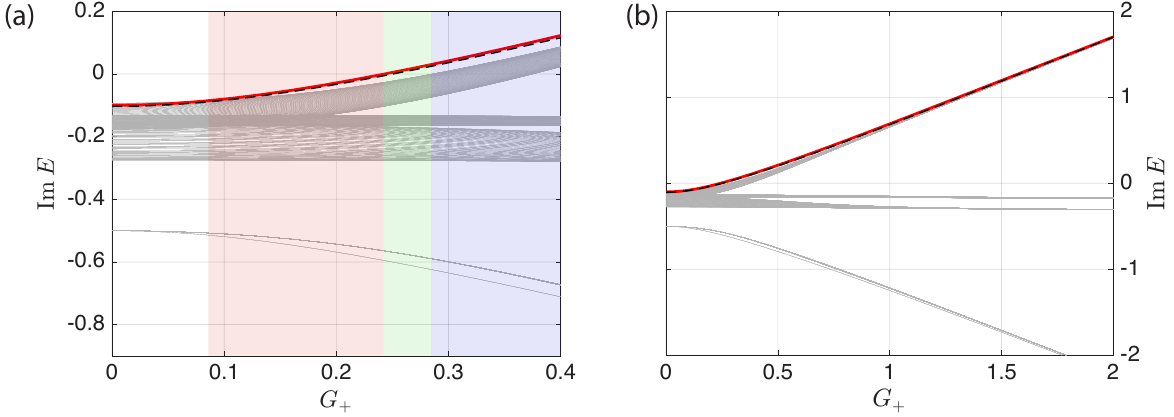}
  \caption{Imaginary part of the OBC spectrum ${\rm Im} E$ as a function of
    $G_+$. Red line: ${\rm Im}\, E_{\rm e}$ ; dashed line: approximate result
    given by the solution of the two-site problem. (a) $G_+ \in \left[0,
      0.4\right]$ corresponding to Fig.~3 of the main text. (b) $G_+ \in
    \left[0, 2 \right]$. $G_-$,$J=0.5$, $\gamma <= 10^{-4}$ (all energies
    expressed in units of $\kappa$). }
  \label{fig:3}
\end{figure}

The stability and dissipation properties of the end states can be approximated
terms of dynamics of a two-site system, with one site driven on the blue
sideband ($G_+$) and the other on the red ($G_-$). Intuitively, this setup
corresponds to $G_-$-driven site acting as an effective bath for the $G_+$
driven one, stabilizing its dynamics. In Fourier space, the dynamics of the
two-site system can be written as
\begin{equation}
  \label{eq:35}
    {\bf a} = A\, \xi_{\rm in}
\end{equation}
where ${\bf a} =\left[a_1, b^\dagger_1,a_2,b_2 \right]$, $\xi_{\rm
  in}=\left[\sqrt{\kappa}\, a_{\rm in, 1},\sqrt{\gamma}\, b^\dagger_{\rm in, 1},
  \sqrt{\kappa}\, a_{\rm in, 2},\sqrt{\gamma}\, b_{\rm in, 2} \right]$ and
\begin{equation}
  \label{eq:36}
  A^{-1}=
  \begin{pmatrix}
    \chi_{\rm c}^{-1} & i G_+ & 0 & 0 \\
    -i G_+ & \chi_{\rm m}^{-1} &  0 & -i J\\
    0  & 0 & \chi_{\rm c}^{-1} & -i G_-  \\
    0  & -i J &  -i G_- & \chi_{\rm m}^{-1} \\
  \end{pmatrix}
\end{equation}
with $\chi_{\rm m} = \left(\gamma/2 - i \omega \right)^{-1}$ and $\chi_{\rm c} =
\left(\kappa/2 - i \omega \right)^{-1}$. With an analysis similar to the one
carried out in Sec.~\ref{sec:DynS}, it is possible to show that the poles of the
matrix $A$ correspond to the (complex) excitation energies for the two-site
system, and, in particular, their imaginary part characterizes its stability
properties. In Fig.~\ref{fig:3} we compare the complex eigenenergies of the
two-site system, with the spectrum of the chain, and we see that the overall
behavior of the end state (in red) is well captured by the two-site
description. While an analytical solution to the two-site problem is possible
the expressions for ${\rm Im} E_{\rm 2-site}$ are not particularly informative. We provide
here the expression for ${\rm Im} E_{\rm 2-site}$ corresponding to the largest
imaginary part of the two-site eigenenergies  in the limit $G_+ \to 0$ and $G_+ \to \infty$. For $G_+ \to 0 $
we have
\begin{equation}
  \label{eq:37}
 {\rm Im}\, E_{\rm 2-site} \simeq -\frac{\Delta^{1/3} + \kappa + 2 \gamma }{6} +\frac {12 (G_-^2+J^2)-\left(\kappa-\gamma\right)^2} {6
     \Delta^{1/3}}
\end{equation}
with
\begin{align*}
  \Delta&=\kappa \left( \kappa^2-18 G_-^2  +36  J^2 \right) +\sqrt{\kappa ^2 \left(\kappa ^2 -18 G_-^2+36    J^2\right)^2-\left(\kappa ^2-12 G_-^2-12 J^2\right)^3} \\
        & - 3 \left( \kappa ^2 -6   G_-^2+12 J^2\right)\gamma
           + O(\gamma^2)
\end{align*}
which, expanding in powers of $J/G_-$, can be written as
\begin{equation}
  \label{eq:39}
  {\rm Im}\, E_{\rm 2-site} \simeq -\frac{\gamma}{2} - \left(\kappa-\gamma\right) \frac{J^2}{G_-^2}\left(1-\frac{J^2}{G_-^2}\right)+ O\left[\left(\frac{J}{G_-}\right)^5\right]
\end{equation}

Conversely, for $G_+ \to \infty$ we get
\begin{eqnarray}
  \label{eq:38}
  {\rm Im} E_{\rm 2-site} \simeq -\frac{\kappa+\gamma}{2}
  +\sqrt{\frac{(\kappa-\gamma)^2}{4}+ 4 G_+^2}.
 \end{eqnarray}

\pagebreak
\section{Topological protection}
\label{sec:topoprot}

\subsection{Hopping disorder}
\label{sec:Jdis}

We plot here the spectra (for periodic- and open-boundary conditions ) in the
presence of disorder for the hopping parameter $J$ ($\delta J$ unformly
distributed in $\left[-0.1 J, 0.1 J \right]$), for different values of $G_+$.
For all values of $G_+$ it is possible to see that the ${\rm Re}\, E=0$ end
states survive in the presence of disorder. The degeneracy of ${\rm Im }\, E$,
however, is partially lifted: the imaginary part of the spectrum shows the
appearance of pairs of states instead of the quadruplet associated with the
corresponding spectrum in the absence of disorder.

\begin{figure}[H]
  \includegraphics[width=0.9\columnwidth]{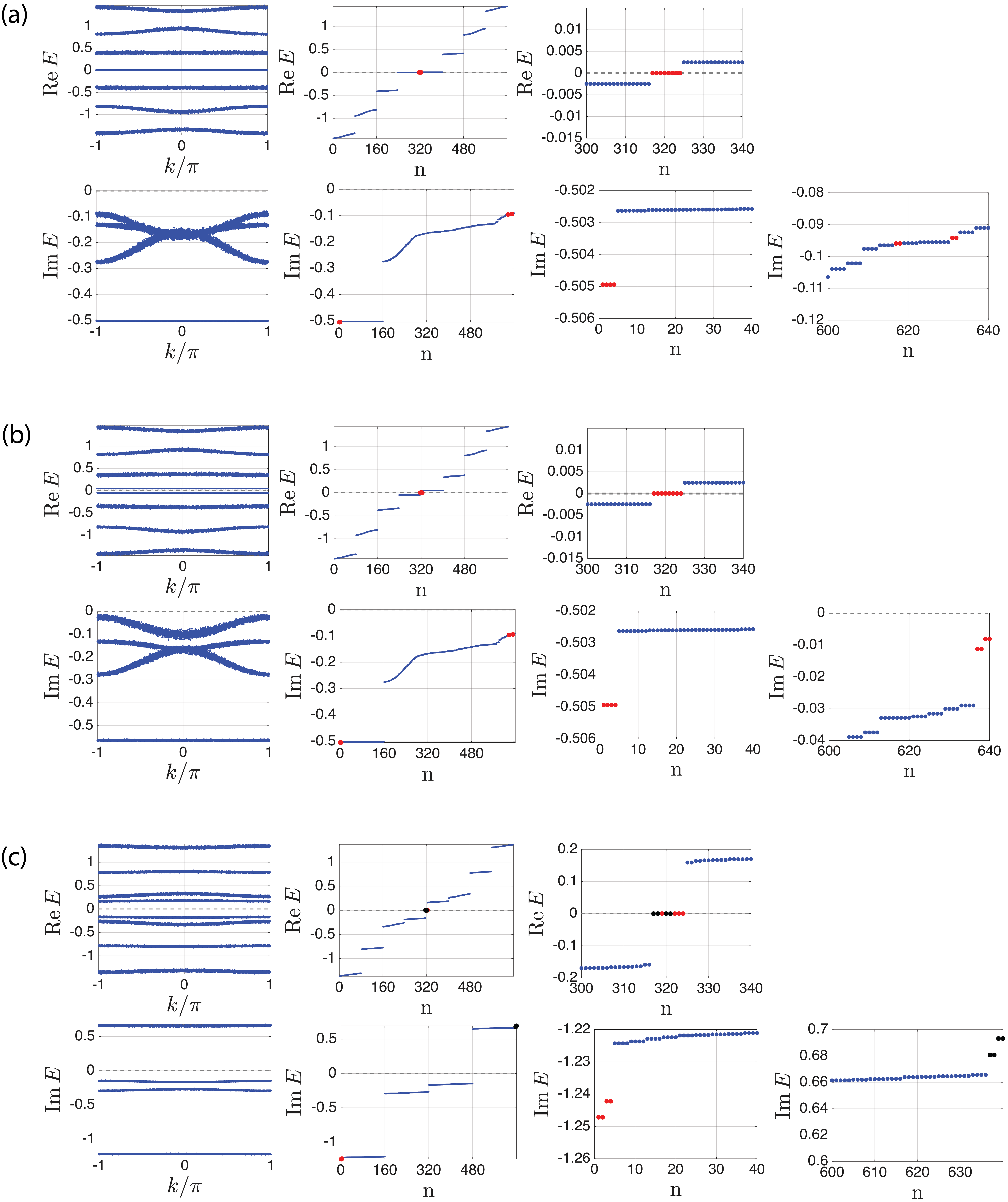}
  \caption{Real and imaginary spectra for periodic (first columm) and open
    (second column) boundary conditions (columns three and four represent a
    zoom-in of column two). Red dots correspond to ${\rm Im} E_{\rm e}<0$, i.e.
    stable end states, black dots correspond to ${\rm Im} E_{\rm e}>0$, i.e.
    unstable end states, blue dots to ${\rm Im} E_{\rm b}$, i.e. bulk states.
    (a) $G_+=0.05$, (b) $G_+=0.242$, (c) $G_+=1$. $G_-=1, J=0.5$, $\gamma =
    10^{-4}$, $\delta J /J =0.1$ (all energies expressed in units of $\kappa$).}
  \label{fig:4}
\end{figure}

\subsection{On-site frequency disorder}
\label{sec:wdis}

\begin{figure}[H]
  \includegraphics[width=0.9\columnwidth]{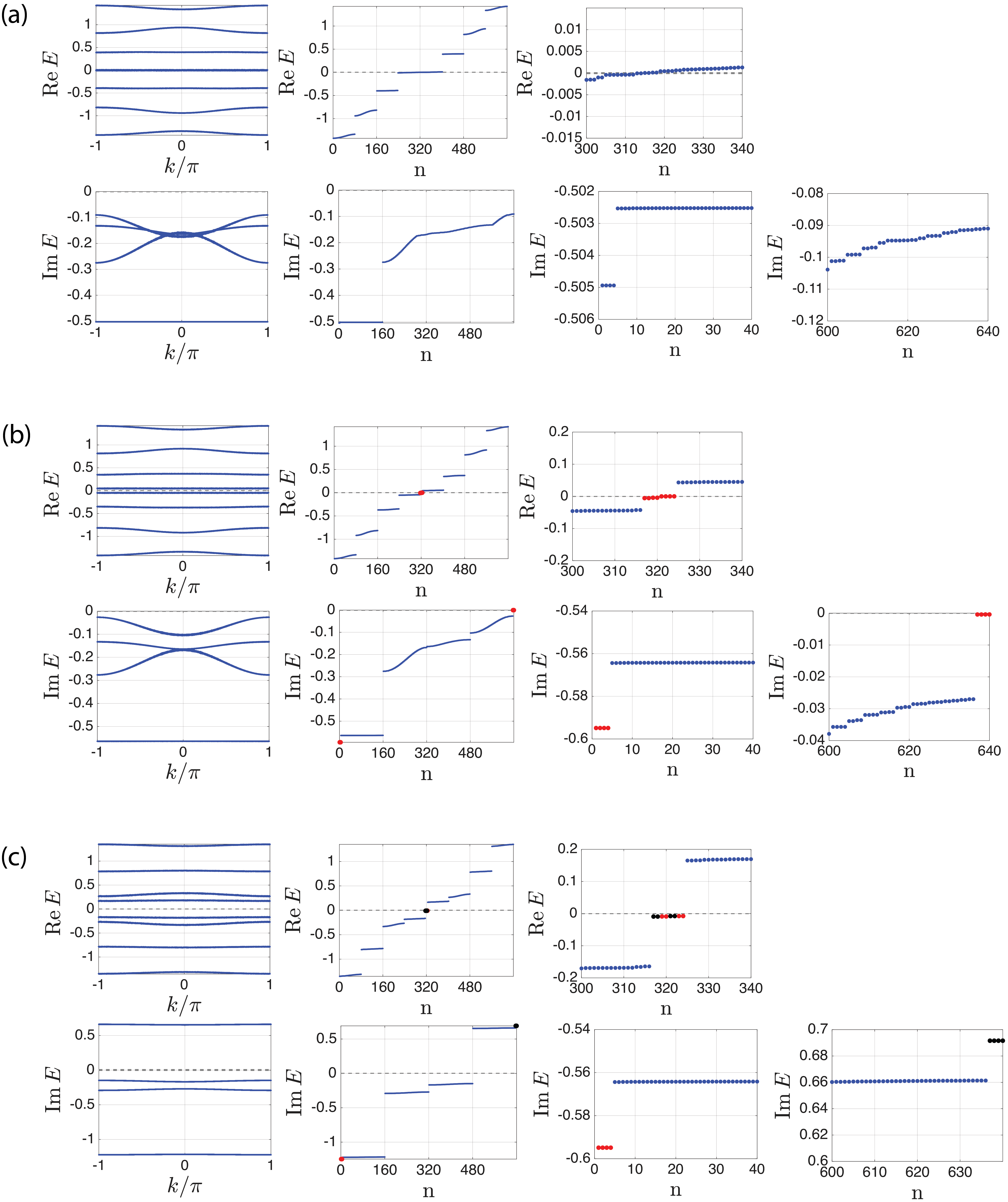}
  \caption{Real and imaginary spectra for periodic (first columm) and open
    (second column) boundary conditions (columns three and four represent a
    zoom-in of column two). Red dots correspond to ${\rm Im} E_{\rm e}<0$, i.e.
    stable end states, black dots correspond to ${\rm Im} E_{\rm e}>0$, i.e.
    unstable end states, blue dots to ${\rm Im} E_{\rm b}$, i.e. bulk states.
    (a) $G_+=0.05$, (b) $G_+=0.242$, (c) $G_+=1$. $G_-=1, J=0.5$, $\gamma =
    10^{-4}$, $\delta \omega_\mathrm{m} /\omega_\mathrm{m} =0.01$ (all energies expressed in units of $\kappa$).}
  \label{fig:5}
\end{figure}

We take here into account the potential difference of on-site frequencies for
either the mechanics or the cavity. Following the derivation in
section~\ref{sec:LinH}  it is easy to show that a site dependence of
$\omega_\mathrm{c}$ can be absorbed into the choice of the driving frequency
$\omega_\mathrm{d}$. Conversely, a potential site dependence of the mechanical
frequency has to be explicitly taken into account. In Fig.~\ref{fig:5}, we have
plotted the spectra for $\omega_m$ uniformly distributed (in the rotating frame)
in $\left[-0.01 \omega_\mathrm{m}, 0.01 \omega_\mathrm{m} \right]$. The disorder
amplitude is compatible with the current experimental capabilities discussed
in~\cite{Youssefi.2022}.

As expected from the breaking of the chiral symmetry, Fig~\ref{fig:5} displays a
deviation from the ${\rm Re}\, E =0$ condition for the end
states. For the parameters considered here, however, the  real part of the
spectrum still exhibits a significant gap for the ${\rm Re}\, E \ne 0$
states.

\subsection{End-sites mechanical dissipation renormalization}
\label{sec:gammaVal}

We depict here the OBC spectra in the presence of measurement-induced
renormalization of the end-states mechanical dissipation. As expected, in these
figures it is possible to see that even for $\gamma_{\rm end} = 100 \gamma$,
the spectrum is essentially unchanged, as expected from a topological protection argument.

\begin{figure}[H]
  \includegraphics[width=0.75\columnwidth]{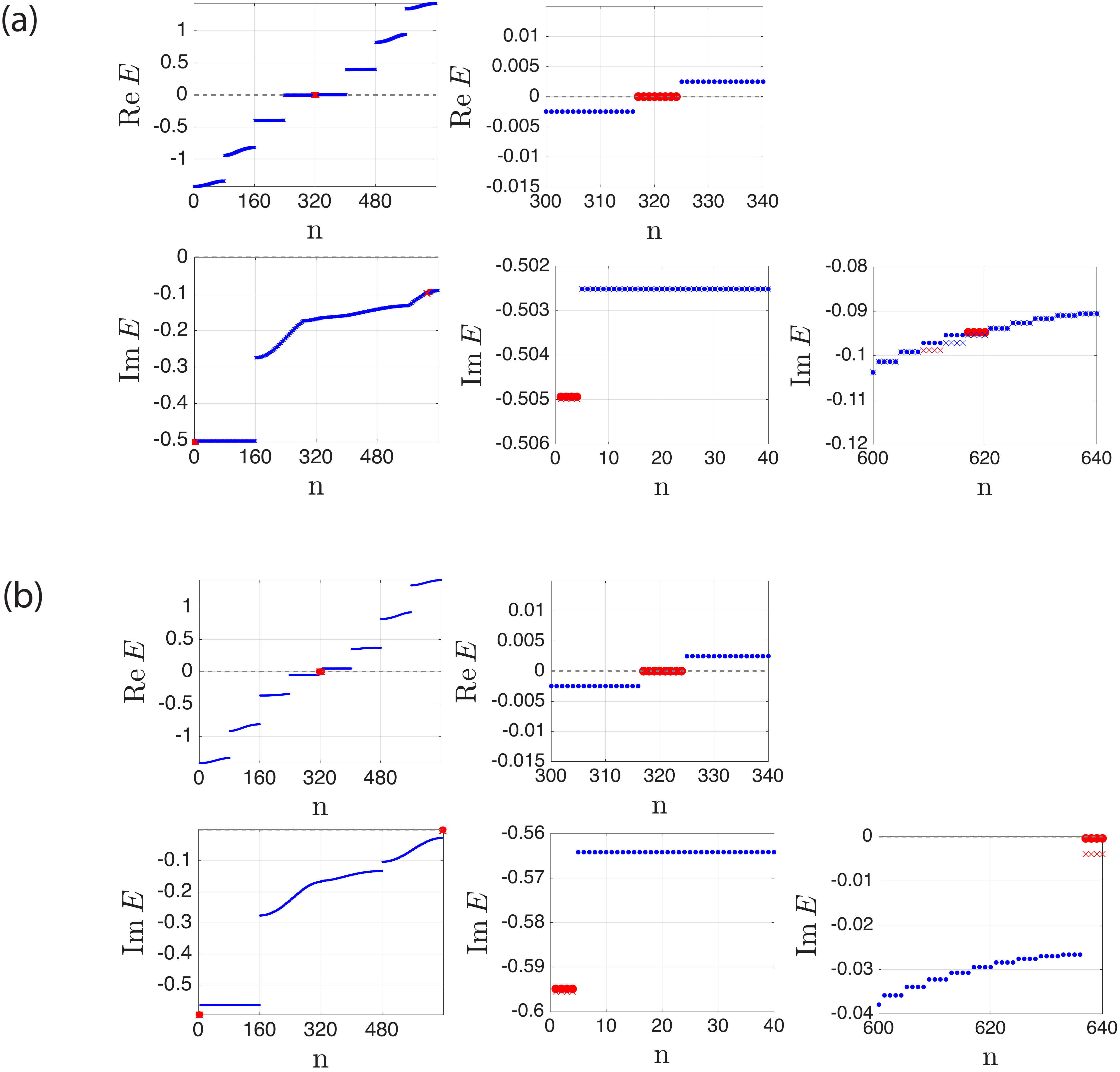}
  \caption{Real and imaginary spectra for periodic (first columm) and open
    (second column) boundary conditions (columns three and four represent a
    zoom-in of column two). Red dots correspond to ${\rm Im} E_{\rm e}<0$, i.e.
    stable end states, black dots correspond to ${\rm Im} E_{\rm e}>0$, i.e.
    unstable end states, blue dots to ${\rm Im} E_{\rm b}$, i.e. bulk states.
    (a) $G_+=0.05$, (b) $G_+=0.242$, $\gamma =
    10^{-4}$, $\gamma_{\rm end}=10^{-2}$ (all energies expressed in units of $\kappa$).}
  \label{fig:6}
\end{figure}

\section{Third quantization approach}
\label{sec:ThirdQuant}

\subsection{Hamiltonian}
The real-space Hamiltonian has a form of:
\begin{eqnarray}
{\cal H}&=&\sum_{i}\left(\vec{d}_{i}^{\,\dagger}\hat{P}\vec{d}_{i}+\vec{d}_{i}\hat{Q}\vec{d}_{i}+\vec{d}_{i}^{\,\dagger}\hat{Q}^{\star}\vec{d}_{i}^{\,\dagger}\right)
+\sum_{i}\left(\vec{d}_{i}^{\,\dagger}\hat{R}\vec{d}_{i+1}+\vec{d}_{i+1}^{\,\dagger}\hat{R}^{\dagger}\vec{d}_{i}\right)
\end{eqnarray}
with $\vec{d}_{i}$ describing bosonic degrees of freedom within unit cell $i$ as a vector:
\begin{eqnarray}
\vec{d}_{i}=\begin{pmatrix}a_{1,i} & b_{1,i} & a_{2,i} & b_{2,i} & a_{3,i} & b_{3,i} & a_{4,i} & b_{4,i}\end{pmatrix}
\end{eqnarray}
with matrices $\hat P$, $\hat Q$ and $\hat R$ describing hopping and pairing of the bosons inside and between unit cells. We have:
\begin{equation}
\hat{P}=\begin{pmatrix}0 & 0 & 0 & 0 & 0 & 0 & 0 & 0\\
0 & 0 & 0 & -1 & 0 & 0 & 0 & 0\\
0 & 0 & 0 & G_{-} & 0 & 0 & 0 & 0\\
0 & -1 & G_{-} & 0 & 0 & -1 & 0 & 0\\
0 & 0 & 0 & 0 & 0 & G_{-} & 0 & 0\\
0 & 0 & 0 & -1 & G_{-} & 0 & 0 & -1\\
0 & 0 & 0 & 0 & 0 & 0 & 0 & 0\\
0 & 0 & 0 & 0 & 0 & -1 & 0 & 0
\end{pmatrix}\!,\,\,
\hat{Q}=\begin{pmatrix}0 & G_{+} & 0 & 0 & 0 & 0 & 0 & 0\\
G_{+} & 0 & 0 & 0 & 0 & 0 & 0 & 0\\
0 & 0 & 0 & 0 & 0 & 0 & 0 & 0\\
0 & 0 & 0 & 0 & 0 & 0 & 0 & 0\\
0 & 0 & 0 & 0 & 0 & 0 & 0 & 0\\
0 & 0 & 0 & 0 & 0 & 0 & 0 & 0\\
0 & 0 & 0 & 0 & 0 & 0 & 0 & G_{+}\\
0 & 0 & 0 & 0 & 0 & 0 & G_{+} & 0
\end{pmatrix}\!,\,\,
\hat{R}=\begin{pmatrix}0 & 0 & 0 & 0 & 0 & 0 & 0 & 0\\
0 & 0 & 0 & 0 & 0 & 0 & 0 & 0\\
0 & 0 & 0 & 0 & 0 & 0 & 0 & 0\\
0 & 0 & 0 & 0 & 0 & 0 & 0 & 0\\
0 & 0 & 0 & 0 & 0 & 0 & 0 & 0\\
0 & 0 & 0 & 0 & 0 & 0 & 0 & 0\\
0 & 0 & 0 & 0 & 0 & 0 & 0 & 0\\
0 & -1 & 0 & 0 & 0 & 0 & 0 & 0
\end{pmatrix}\!.
\end{equation}
For a system consisting of $N$ unit cells we can set we define a composite vector:
\begin{eqnarray}
\vec{\underline{d}}=\begin{pmatrix}\vec{d}_{1} & \vec{d}_{2} & \dots & \vec{d}_{N}\end{pmatrix}
\end{eqnarray}
and the Hamiltonian becomes:
\begin{eqnarray}
{\cal H}&=&\vec{\underline{d}}^{\dagger}\left(\hat{T}_{0}\otimes\hat{P}+\hat{T}_{1}\otimes\hat{R}+\hat{T}_{-1}\otimes\hat{R}^{\dagger}\right)\vec{\underline{d}}
+\vec{\underline{d}}\left(\hat{T}_{0}\otimes\hat{Q}\right)\vec{\underline{d}}+\vec{\underline{d}}^{\dagger}\left(\hat{T}_{0}\otimes\hat{Q}^{\star}\right)\vec{\underline{d}}^{\dagger}\nonumber\\
&=&\vec{\underline{d}}^{\dagger}\hat{H}\vec{\underline{d}}+
\vec{\underline{d}}\hat{K}\vec{\underline{d}}+\vec{\underline{d}}^{\dagger}\hat{K}^{\star}\vec{\underline{d}}^{\dagger}
\end{eqnarray}
with $\hat{T}_{0}$ being $N\times N$ identity matrix and $\hat{T}_{1}$ being a matrix describing shift by one of all unit cells, i.e.,
\begin{equation}
\hat{T}_{1}=
\begin{pmatrix}0 & 1\\
 & 0 & 1\\
 &  & \ddots & \ddots\\
 &  &  & 0 & 1\\
x &  &  &  & 0
\end{pmatrix},\quad
\hat{T}_{-1}:=\hat{T}_{1}^T.
\end{equation}
Here $x=1$ for periodic and $x=0$ for open boundary conditions.

\subsection{Liouvillian}
Following the third quantization procedure by Prosen and Seligman \cite{prosen_quantization_2010,Brzezicki.2023} we find that the Liouvillian can be written in the form of
\begin{equation}
{\cal L}=-2\begin{pmatrix}\vec{\underline{d}}_{0}^{\,\prime} & \vec{\underline{d}}_{1}^{\,\prime}\end{pmatrix}\hat{X}^{T}\begin{pmatrix}\vec{\underline{d}}_{0} & \vec{\underline{d}}_{1}\end{pmatrix}+\begin{pmatrix}\vec{\underline{d}}_{0}^{\,\prime} & \vec{\underline{d}}_{1}^{\,\prime}\end{pmatrix}\hat{Y}\begin{pmatrix}\vec{\underline{d}}_{0}^{\,\prime} & \vec{\underline{d}}_{1}^{\,\prime}
\end{pmatrix}
\label{eq:Lmodel}
\end{equation}
with operators $d_{i,\mu}$ satisfying almost canonical commutation relations for bosons:
\begin{equation}
[d_{i,\mu},d_{j,\nu}^{\prime}]=\delta_{ij}\delta_{\mu\nu}
\end{equation}
where $d_{j,\nu}^{\prime}$ is not Hermitian conjugate of $d_{j,\nu}$. In the Liouvillan the matrices defining Hamiltonian enter as blocks of $\hat X$ and $\hat Y$:
\begin{equation}
\hat{X}=\frac{1}{2}\begin{pmatrix}i\hat{H}^{\star}+\hat{M} & -2i\hat{K}\\
2i\hat{K}^{\star} & -i\hat{H}+\hat{M}^{\star}
\end{pmatrix},
\,\hat{Y}=\begin{pmatrix}-i\hat{K}^{\star} & 0\\
0 & i\hat{K}
\end{pmatrix},
\label{eq:XY}
\end{equation}
where $\hat M$ is matrix describing dissipation. In our case it takes a diagonal form of:
\begin{eqnarray}
\hat{M}=\hat{T}_{0}\otimes\frac{1}{2}{\rm diag}\begin{pmatrix}\kappa & \gamma & \kappa & \gamma & \kappa & \gamma & \kappa & \gamma\end{pmatrix}.
\end{eqnarray}
Liouvillan is a quadratic form of boson-like operators and can be brought to a diagonal form of
\begin{equation}
{\cal L} = -2 \sum_{i=1}^{16N}\beta_{i}\xi_i^{\prime}\xi_i,
\end{equation}
by a suitable symplectic transformation. One finds that the rapidities $\beta_{i}$ are the eigenvalues of the matrix $\hat X$ so that one can identify $X$ to be proportional to the non-Hermitian Hamiltonian of the system. Strictly speaking we can identify:
\begin{equation}
\hat{H}_{\rm NH} = -2i\hat{X}.
\end{equation}

\subsection{Time-evolution}

To maximally simplify the notation let's write the Liouvillan as
\begin{equation}
{\cal L}=\vec{\alpha}^{\prime}\hat{A}\vec{\alpha}+\vec{\alpha}^{\prime}\hat{B}\vec{\alpha}^{\prime},
\label{eq:Lab}
\end{equation}
with $a_i$ and $a_i^{\prime}$ being almost canonical bosons and $\hat A$, $\hat B$ being $L\times L$ matrices. Now, the third-quantized Lindblad equation reads:
\begin{equation}
\frac{d}{dt}\rho(t)={\cal L}[\rho]
\end{equation}
thus the time-evolution of the density matrix is given by
\begin{equation}
\left|\rho(t)\right\rangle =\exp\left[t{\cal L}\right]\left|\rho(0)\right\rangle.
\end{equation}
The solution has a form of
\begin{equation}
\left|\rho(t)\right\rangle =
{\cal T}\exp\left[\int_0^t dt' {\cal L} (t') \right] \rho(0).
\end{equation}
Taking a coherent state as initial
\begin{equation}
\left|\rho(0)\right\rangle =\left|\vec{z}(0)\right\rangle= \exp[\vec{z}(0)\vec{\alpha}^{\,\prime}] \left|0\right\rangle,
\end{equation}
using Suzuki-Trotter decomposition and resolution of identity in coherent states we get:
\begin{equation}
\left|\rho(t)\right\rangle =\frac{1}{\pi^{2LN_t}}\int\left(\prod_{i=1}^{N_t}d\Im\vec{z}_{i}d\Re\vec{z}_{i}\right)\exp\left[\sum_{i=1}^{N_t}S_i\right]\left|\vec{z}_{1}\right\rangle,
\end{equation}
with
\begin{equation}
S_i=
\vec{z}_{i}^{\,\star}\left[1+\frac{t}{N_t}\hat{A}\right]\vec{z}_{i+1}+\vec{z}_{i}^{\,\star}\left[\frac{t}{N_t}\hat{B}\right]\vec{z}_{i}^{\,\star}-\vec{z}_{i}^{\,\star}\vec{z}_{i}
\end{equation}
and $\vec{z}_{N_t+1} \equiv \vec{z}(0)$ for number of time-steps $N_t\to\infty$. By calculating multi-dimensional Gaussian integral we can get:
\begin{widetext}
\begin{equation}
\left|\rho(t)\right\rangle \!=\!\!\frac{1}{\pi^{2L}}\!\int\! d\Im\vec{z}d\Re\vec{z}\exp\!\left[-\vec{z}^{\,\star}\vec{z}+\vec{z}^{\!\star}\exp\left[t\hat{A}\right]\vec{z}(0)+\frac{t}{N_t}\vec{z}^{\,\star}\left[\hat{B}+\hat{h}\hat{B}\hat{h}^{T}+\dots+\hat{h}^{N-1}\hat{B}(\hat{h}^T)^{N-1}\right]\vec{z}^{\,\star}\right]\left|\vec{z}\right\rangle.
\end{equation}
\end{widetext}
with
\begin{equation}
\hat{h}=1+\frac{t}{N}\hat{A}.
\end{equation}
The series in the exponent can be summed up in a similar way one is summing up a finite geometric series. Then we get for $N_t\to\infty$:
\begin{align}
\label{rho:integ}
\left|\rho(t)\right\rangle = \frac{1}{\pi^{2L}}\int d\Im\vec{z}d\Re\vec{z}
\exp\left[-\vec{z}^{\,\star}\vec{z}+\vec{z}^{\,\star}\exp\left[t\hat{A}\right]\vec{z}(0)+\vec{z}^{\,\star}\hat{C}(t)\vec{z}^{\,\star}\right]\left|\vec{z}\right\rangle
\end{align}
with $\hat{C}(t)$ being a matrix that satisfy Lyapunov equation of the form
\begin{equation}
\hat{A}\hat{C}(t)+\hat{C}(t)\hat{A}^{T}=\exp\left[t\hat{A}\right]\hat{B}\exp\left[t\hat{A}^{T}\right]-\hat{B}.
\end{equation}
After performing final Gaussian integration the time-evolved density matrix becomes:
\begin{equation}
\left|\rho(t)\right\rangle =\exp\left[\vec{\alpha}'\hat{C}(t)\vec{\alpha}'\right]\left|\exp\left[t\hat{A}\right]\vec{z}(0)\right\rangle.
\label{eq:rhot}
\end{equation}

\subsection{Solution of the Lyapunov equation}

The Lyapunov equation is known to have a solution of the closed form:
\begin{equation}
\hat{C}(t)=\int_{0}^{\;t}\exp\left[x\hat{A}\right]\hat{B}\exp\left[x\hat{A}^{T}\right]dx.
\end{equation}
Assuming that $\hat A$ can be diagonalized in the eigenbasis $\hat\beta$ we get:
\begin{equation}
\hat{\Lambda}=\hat{\beta}^{-1}\hat{A}\hat{\beta}=
{\rm diag}\begin{pmatrix}\lambda_{1} & \lambda_{2} & \dots& \lambda_{2L}\end{pmatrix}=
\hat{\beta}^{T}\hat{A}^T\hat{\beta}^{-1T}.
\end{equation}
Thus:
\begin{equation}
\hat{\beta}^{-1}\hat{C}(t)\hat{\beta}^{-1T}=\int_{0}^{\;t}\exp\left[x\hat{\Lambda}\right]\hat{\beta}^{-1}\hat{B}\hat{\beta}^{-1T}\exp\left[x\hat{\Lambda}\right]dx
\end{equation}
or:
\begin{align}
   &\left(\hat{\beta}^{-1}\hat{C}(t)\hat{\beta}^{-1T}\right)_{ij}=\left(\hat{\beta}^{-1}\hat{B}\hat{\beta}^{-1T}\right)_{ij}\nonumber\\
   &\times\int_{0}^{\;t}dx\exp\left[x(\lambda_i+\lambda_j)\right].
\end{align}
Then by explicitly calculating the integral we finally get:
\begin{equation}
\hat{C}(t)=\hat{\beta}\left[\left(\hat{\beta}^{-1}\hat{B}\hat{\beta}^{-1T}\right)\circ\hat{\Gamma}(t)\right]\hat{\beta}^{T},
\label{eq:Ct}
\end{equation}
with
\begin{equation}
\Gamma_{ij}(t)=\frac{\exp\left[t\left(\lambda_{i}+\lambda_{j}\right)\right]-1}{\lambda_{i}+\lambda_{j}}
\label{eq:Gij}
\end{equation}
and $\circ$ denoting Hadamard product of matrices.

\subsection{Correlation functions}

After starting from the Liouvillian of the general third-quantized form of Eq.~(\ref{eq:Lab}) and getting the time-evolved density matrix $\left|\rho(t)\right\rangle$ in the form of Eq.~(\ref{eq:rhot}) we can easily compute single or multi-boson correlation functions.
First, we need to make contact with our model Liouvillian of Eq.~(\ref{eq:Lmodel}).
We have:
\begin{equation}
\vec\alpha =\begin{pmatrix}\vec{\underline{d}}_{0} & \vec{\underline{d}}_{1}\end{pmatrix},\quad\hat A = -2\hat{X}^T,\quad\hat B = \hat{Y}.
\end{equation}
Here the 'flavor' index $\mu =0,1$ of $\vec{\underline{d}}_{\mu}$ is an effect of third quantization whereas all other indices of $\vec{\underline{d}}$ describe physical bosonic modes of the system. Therefore it makes sense to define a subblock structure of matrices $\hat A$, $\hat B$ and $\hat{C}(t)$, i.e.:
\begin{equation}
\hat A = \begin{pmatrix}
\hat{A}^{00} & \hat{A}^{01} \\
\hat{A}^{01} & \hat{A}^{11}
\end{pmatrix}\! , \,
\hat B = \begin{pmatrix}
\hat{B}^{00} & \hat{B}^{01} \\
\hat{B}^{01} & \hat{B}^{11}
\end{pmatrix}\! , \,
\hat C = \begin{pmatrix}
\hat{C}^{00} & \hat{C}^{01} \\
\hat{C}^{01} & \hat{C}^{11}
\end{pmatrix}\! .
\end{equation}
Now we can calculate some correlation function. Start with creator-annihilator
average. Following third quantization we get:
\begin{align}
    {\rm Tr}\left(d_{i}^{\dagger}d_{j}\rho(t)\right)=\left\langle 0\right|\left(d^{\prime}_{i,0}+d_{i,1}\right)d_{j,0}\left|\rho(t)\right\rangle
    =\left\langle 0\right|d_{i,1}d_{j,0}\left|\rho(t)\right\rangle.
\end{align}
The last expression can be easily calculate from the integral form of the density matrix $\left|\rho(t)\right\rangle$ of Eq.~(\ref{rho:integ}), namely:
\begin{align}
 {\rm Tr}\left(d_{i}^{\dagger}d_{j}\rho(t)\right)=\frac{1}{\pi^{2L}}\int d\Im\vec{z}d\Re\vec{z}\,\,\,z_{i,1}z_{j,0}
 \times\exp\left[-\vec{z}^{\,\star}\vec{z}+\vec{z}^{\,\star}\exp\left[t\hat{A}\right]\vec{z}(0)+\vec{z}^{\,\star}\hat{C}(t)\vec{z}^{\,\star}\right].
\end{align}
From the Gaussian integral we get:
\begin{align}
{\rm Tr}\left(d_{i}^{\dagger}d_{j}\rho(t)\right)=C(t)^{10}_{ij}+C(t)^{01}_{ji}
+\left(\exp\left[t\hat{A}\right]\vec{z}(0)\right)_{i,1}\left(\exp\left[t\hat{A}\right]\vec{z}(0)\right)_{j,0}.
\label{eq:ddd}
\end{align}
In analogical way we can get other single-boson correlators, i.e.,
\begin{align}
{\rm Tr}\left(d_{i}^{\dagger}d_{j}^{\dagger}\rho(t)\right)=C(t)^{11}_{ij}+C(t)^{11}_{ji}
+\left(\exp\left[t\hat{A}\right]\vec{z}(0)\right)_{i,1}\left(\exp\left[t\hat{A}\right]\vec{z}(0)\right)_{j,1}
\label{eq:dddd}
\end{align}
and
\begin{align}
{\rm Tr}\left(d_{i}d_{j}\rho(t)\right)=C(t)^{00}_{ij}+C(t)^{00}_{ji}
+\left(\exp\left[t\hat{A}\right]\vec{z}(0)\right)_{i,0}\left(\exp\left[t\hat{A}\right]\vec{z}(0)\right)_{j,0}.
\label{eq:dd}
\end{align}
If the starting state is not vacuum but a generic coherent state then also the linear averages of creation/annihilation operators are non-trivial. We get:
\begin{align}
{\rm Tr}\left(d_{i}^{\dagger}\rho(t)\right)=\left(\exp\left[t\hat{A}\right]\vec{z}(0)\right)_{i,1},\\
{\rm Tr}\left(d_{i}\rho(t)\right)=\left(\exp\left[t\hat{A}\right]\vec{z}(0)\right)_{i,0}.
\end{align}

\subsection{Non-zero temperature}

In presence of non-zero temperature the Liouvillan of Eq.~(\ref{eq:Lmodel}) gets modified by new form of the $\hat{X}$ and $\hat{Y}$ matrices, one gets:
\begin{equation}
\hat{X}=\frac{1}{2}\begin{pmatrix}i\hat{H}^{\star}+\hat{M} - \hat{N}^{\star} & -2i\hat{K}\\
2i\hat{K}^{\star} & -i\hat{H}+\hat{M}^{\star}-\hat{N}
\end{pmatrix},
\,\hat{Y}=\begin{pmatrix}-i\hat{K}^{\star} & \hat{N}\\
\hat{N}^{\star} & i\hat{K}
\end{pmatrix},
\end{equation}
where the definitions of the $\hat{M}$ and $\hat{N}$ block are given by:
\begin{eqnarray}
\hat{N}&=&\hat{T}_{0}\otimes\frac{1}{2}{\rm diag}\begin{pmatrix}\kappa n_a & \gamma n_b & \kappa n_a & \gamma n_b & \kappa n_a & \gamma n_b & \kappa n_a & \gamma n_b\end{pmatrix},\\
\hat{M}&=&\hat{T}_{0}\otimes\frac{1}{2}{\rm diag}\begin{pmatrix}\kappa (n_a+1) & \gamma (n_b+1) & \kappa (n_a+1) & \gamma (n_b+1) & \kappa (n_a+1) & \gamma (n_b+1) & \kappa (n_a+1) & \gamma (n_b+1)\end{pmatrix}.
\end{eqnarray}
Here $n_a$ and $n_b$ are thermal occupations of the $a$ and $b$ bosonic modes, so for zero temperature $\hat{N}=0$ and the $\hat{X}$ and $\hat{Y}$ matrices take the old form of Eq.~(\ref{eq:XY}). With new form of the Liouvillan the rest of derivation is done in the same manner.

\begin{figure}[ht]
  \includegraphics[width=0.95\columnwidth]{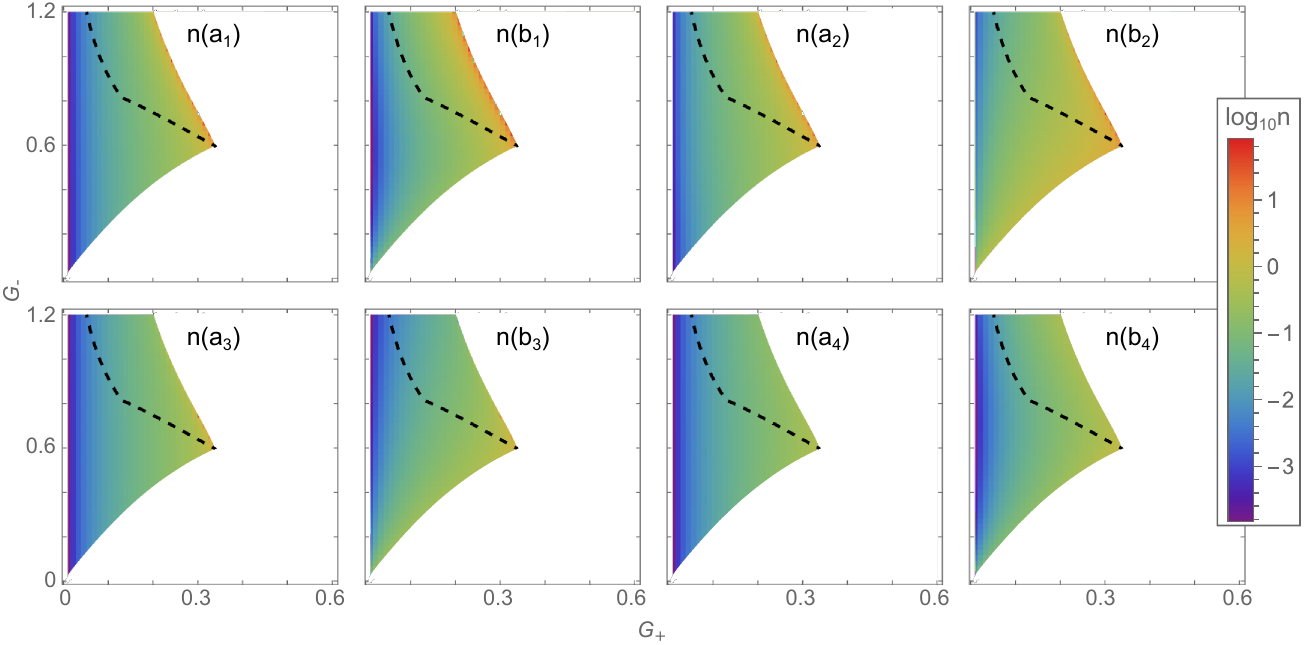}
  \caption{
Stationary-state densities for  $\gamma = 0.0001$ and $\kappa = 1.0$. Dashed line indicates boundary of end-states hidden in the bulk. Densities diverge in the white excluded region as there is no stationary state.
  }
  \label{fig:6}
\end{figure}

\subsection{Stationary correlation matrix}

Using equations (\ref{eq:ddd}-\ref{eq:dd}) it is simple to relate matrix $\hat C$ with the correlation matrix $\hat V$. First one has to notice that from the solution of the Lyapunov equation it follows that $\hat C$ is symmetric. Therefore taking initial condition as vacuum, i.e., $\vec{z}(0)=0$ we get for stationary state averages:
\begin{eqnarray}
    C_{ij}^{00}&=&\langle d_{i}d_{j}\rangle /2 \nonumber\\
    C_{ij}^{11}&=&\langle d_{i}^{\dagger}d_{j}^{\dagger}\rangle /2=C_{ij}^{00\star} \nonumber\\
    C_{ij}^{10}&=&\langle d_{i}^{\dagger}d_{j}\rangle /2 \nonumber\\
    C_{ij}^{01}&=&\langle d_{i}d_{j}^{\dagger}-\delta_{ij}\rangle /2=\left\langle a_{j}^{\dagger}a_{i}\right\rangle /2=C_{ji}^{10}
\end{eqnarray}
The correlation matrix can be written blockwise as:
\begin{equation}
\hat{V}=\begin{pmatrix}\hat{V}^{00} & \hat{V}^{01}\\
\hat{V}^{10} & \hat{V}^{11}
\end{pmatrix}
\end{equation}
with blocks given by elements: $V^{00}_{ij}=\langle d_{i}d_{j}^{\dagger}\rangle$, $V^{01}_{ij}=\langle d_{i}d_{j}\rangle$, $V^{10}_{ij}=\langle d_{i}^{\dagger}d_{j}^{\dagger}\rangle$ and $V^{11}_{ij}=\langle d_{i}^{\dagger}d_{j}\rangle$. Thus $\hat C$ matrix can be written as:
\begin{equation}
\hat{C}=\frac{1}{2}
\begin{pmatrix}\hat{V}^{01} & \hat{V}^{00}-1\\
\hat{V}^{11} & \hat{V}^{10}
\end{pmatrix}
=\frac{1}{2} \left[
\begin{pmatrix}\hat{V}^{00} & \hat{V}^{01}\\
\hat{V}^{10} & \hat{V}^{11}
\end{pmatrix}
-\begin{pmatrix}
1 & 0\\
0 & 0
\end{pmatrix}
\right]
\begin{pmatrix}
0 & 1\\
1 & 0
\end{pmatrix}.
\end{equation}
Now we can insert such form of matrix $\hat C$ into the stationary-state Lyapunov equation:
\begin{equation}
\hat{A}\hat{C}+\hat{C}\hat{A}^{T}=-\hat{B}
\end{equation}
we get an equation for the $\hat V$ matrix:
\begin{align}
\begin{pmatrix}2\hat{M} & 0\\
0 & 2\hat{N}
\end{pmatrix}=&
\left[
i\begin{pmatrix}\hat{H}-i(\hat{M}-\hat{N}) & 2\hat{K}\\
-2\hat{K} & -\hat{H}-i(\hat{M}-\hat{N})
\end{pmatrix}\right]
\begin{pmatrix}\hat{V}^{00} & \hat{V}^{01}\\
\hat{V}^{10} & \hat{V}^{11}
\end{pmatrix} \nonumber\\
+&
\begin{pmatrix}\hat{V}^{00} & \hat{V}^{01}\\
\hat{V}^{10} & \hat{V}^{11}
\end{pmatrix}
\left[
i\begin{pmatrix}\hat{H}-i(\hat{M}-\hat{N}) & 2\hat{K}\\
-2\hat{K} & -\hat{H}-i(\hat{M}-\hat{N})
\end{pmatrix}\right]^{\dagger},
\end{align}
or simply:
\begin{equation}
(i\hat{H}_{\rm NH})\hat{V} + \hat{V}(i\hat{H}_{\rm NH})^{\dagger} = \begin{pmatrix}2\hat{M} & 0\\
0 & 2\hat{N}
\end{pmatrix}.
\end{equation}
For this derivation we have taken advantage of the present current form of the Hamiltonian with $\hat{H}=\hat{H}^T$, $\hat{K}=\hat{K}^T=\hat{K}^{\star}$. Note that the non-Hermitian Hamiltonian $\hat{H}_{\rm NH}$ does not depend on temperature.

\begin{figure}[ht]
  \includegraphics[width=0.95\columnwidth]{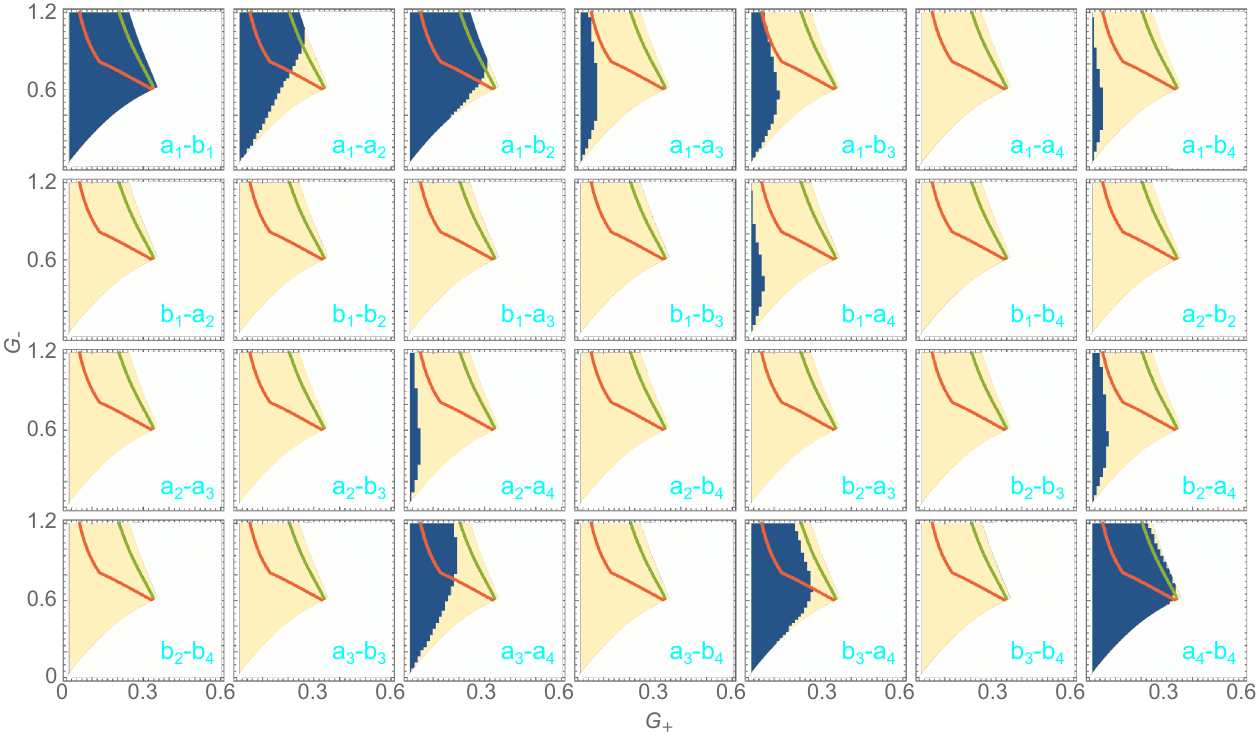}
  \caption{
Entanglement patterns for  $\gamma = 0.0001$
and $\kappa = 1.0$. Dark/bright color indicates non-vanishing/vanishing stationary state- or saturation-negativity
between two modes denoted in cyan in bottom right corner of each panel. Red and green lines indicates boundaries of:
end-states hidden in the bulk and end of the system having stationary state.
In the excluded white region the bulk has no stationary state. In the region between green line and white area the populations are diverging due to the instability of the end-states but negativity remains finite.
  }
  \label{fig:7}
\end{figure}

\subsection{Entanglement in the presence of unstable end modes}

Surprisingly, the entanglement can be calculated also when there is no stationary state because the end modes are unstable. In this case there are four degenerate end-modes which become unstable so the $\hat{C}(t)$ matrix can be evaluated from Eq. (\ref{eq:Ct}) in the limit of $t\to\infty$ by setting:
\begin{equation}
\Gamma_{ij}(t\to\infty)\to\frac{1}{\lambda_{i}+\lambda_{j}}\left[\begin{pmatrix}-1 & \cdots & -1\\
\vdots & \ddots & \vdots\\
-1 & \cdots & -1
\end{pmatrix}\oplus\begin{pmatrix}\xi & \xi & \xi & \xi\\
\xi & \xi & \xi & \xi\\
\xi & \xi & \xi & \xi\\
\xi & \xi & \xi & \xi
\end{pmatrix}\right]_{ij}
\end{equation}
with $\xi = \exp\left[2t\lambda_{{\rm end}}\right]$. Clearly $\xi$ diverges when $t\to\infty$. Using this form of $\hat{\Gamma}$ matrix and keeping $\xi$ as analytical variable we can express $\hat{C}(\xi\to\infty)$ and any two-point correlation function as analytical expression in $\xi$. Then we can quantify the entanglement for $t\to\infty$ by computing $\tilde{\nu}_-$ and taking analytical limit of $\tilde{\nu}_-(\xi\to\infty)$. Quite non-trivially this limit is well-defined and finite despite diverging populations of bosons.

\section{Entanglement measurement}

\begin{figure}[ht]
  \includegraphics[width=0.5\columnwidth]{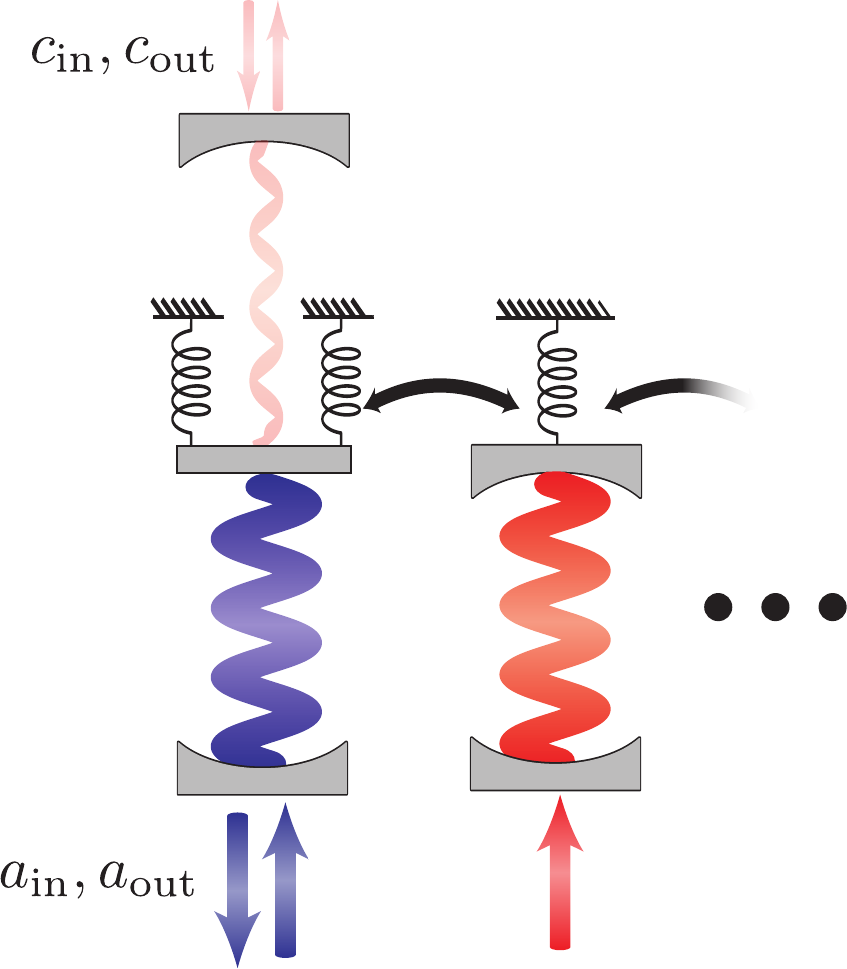}
  \caption{Cartoon depiction of the left end in the presence of an extra
    detection cavity weakly driven on the red sideband (intracavity mode $c$, input/output modes $c_{\rm in}$, $c_{\rm out}$).}
  \label{fig:8}
\end{figure}

We sketch here a possible detection scheme for the optomechanical entanglement
within the end sites of the chain. Following a known strategy
\cite{vitali_optomechanical_2007}, we suggest that the inclusion of an extra
cavity coupled to the mechanical resonators at each end of the chain can be
employed to monitor the mechanical motion of the extremal sites of the
superlattice. To this end, each detection cavity is weakly driven on the red
sideband, inducing an effective beam-splitter interaction between mechanical ($b_{\rm 1}$, $b_{\rm 4N}$) and readout ($c$, $d$) degrees of freedom (see Fig. \ref{fig:8}).

In the following we focus, without loss of generality on the first site of the
chain. The detection cavity described by bosonic operators $c$ and $c^\dagger$, is weakly driven on the red
sideband, giving rise (in the appropriate frame) to the following EOM for $c$ in Fourier space
\begin{equation}
  \label{eq:40}
  -i \omega c = \frac{\kappa}{2} c - i G_{\rm R} b_{\rm 1} +\sqrt{\kappa}\, c
\end{equation}
where, $G_{\rm R}$ and $\kappa_{\rm R}$ are the coupling induced by the
detection drive and the detection cavity linewidth, respectively. Hereafter, we will assume $\kappa_{\rm R}= \kappa$.
Taking into account the usual input-output relations
\begin{subequations}
  \begin{align}
    \label{eq:41}
    c_{\rm out}&=\sqrt{\kappa}\, c - c_{\rm in}\\
    \label{eq:42}
    a_{\rm out}&=\sqrt{\kappa}\, a - a_{\rm in}
  \end{align}
\end{subequations}
\cite{Walls.2008}, the output field for the readout cavity can be written as
\begin{eqnarray}
  \label{eq:43}
 c_{\rm out}& =\left(\chi_{\rm c} \kappa - 1\right)c_{\rm in} - i \chi_{\rm
              c}G_{\rm R}\sqrt{\kappa}\, b_{\rm 1}.
\end{eqnarray}
If we assume that the detection tone $G_{\rm R}$ is much weaker than the driving
tone $G_{+}$, the dynamics of the optomechanical system is
essentially unchanged by the detection process, simply leading to the
renormalization of $\gamma$ and $n_{\rm eff}$. As far as the superlattice is
concerned, we have demonstrated that our protocol is robust with respect to
changes in the thermal population and, in Section  \ref{sec:gammaVal} of the SI, we have numerically demonstrated the fact that, owing to the topological
features of our system, enlargement is robust also for large variations of the
mechanical linewidth on the end sites.

Assuming that the cavities are coupled to a zero-temperature bath, we can relate
the (normal-ordered) power spectrum of the correlations between the output fields
\begin{align}
  \label{eq:46}
  S^{\rm out}_{\rm 1,2}\left(\omega\right) = \int \frac{d \omega'}{2 \pi} \braket{O^\dagger_{\rm 1,out}(\omega) O_{\rm 2,out}(\omega')}
\end{align}
(where $O_{\rm 1,out},O_{\rm 2, out}=a_{\rm out}, c_{\rm out}$) to the power
spectrum of the correlations between  $a$ and $c$ (and, consequently, $b$)
\cite{Gardiner.2004,Walls.2008}, given that
\begin{align}
  \label{eq:48}
  S^{\rm out}_{\rm 1,2}\left(\omega\right) = \kappa  S_{\rm 1,2}\left(\omega\right).
\end{align}
\bibliography{OMtopo}